\documentclass{aa}  

\usepackage{graphicx}
\usepackage{txfonts}
\usepackage{soul}
\usepackage{xcolor}
\usepackage{multirow}
\usepackage{booktabs} 
\usepackage{tabularx} 
\usepackage{adjustbox} 
\usepackage{rotating}
\usepackage{lscape}
\usepackage{longtable}
\usepackage[flushleft]{threeparttable}
\usepackage{subfigure}
\usepackage{placeins}

\usepackage{ulem}
\begin{document}

   \title{Characterisation of magnetic activity of M dwarfs.}

   \subtitle{Possible impact on the surface brightness.}

   \author{R. V. Iba\~nez Bustos\inst{1}\thanks{Contact e-mail: romina.ibanez@oca.eu}
          \thanks{Based on data obtained at Complejo Astron\'omico El Leoncito, operated under the agreement between the Consejo Nacional de Investigaciones Cient\'\i ficas y T\'ecnicas de la Rep\'ublica Argentina and the National Universities of La Plata, C\'ordoba and San Juan.}
          \and
          A. P. Buccino\inst{2,3}
          \and
          N. Nardetto\inst{1}
          \and
          D. Mourard\inst{1}
          \and 
          M. Flores\inst{4,5}
          P. J. D. Mauas\inst{2,3}
}

   \institute{Laboratoire Lagrange, Université Côte d’Azur, Observatoire de la Côte d’Azur, CNRS, Boulevard de l’Observatoire, CS 34229, 06304 Nice Cedex 4, France
        \and
   Instituto de Astronom\'ia y F\'isica del Espacio (CONICET-UBA), C.C. 67 Sucursal 28, C1428EHA-Buenos Aires, Argentina.\\
        \and
            Departamento de F\'isica, Facultad de Ciencias Exactas y Naturales, Universidad de Buenos Aires, Buenos Aires, Argentina.\\
        \and
             Instituto de Ciencias Astron\'omicas, de la Tierra y del Espacio (ICATE-CONICET),  C.C 467, 5400 San Juan, Argentina.\\
        \and
            Facultad de Ciencias Exactas, F\'isicas y Naturales, Universidad Nacional de San Juan, San Juan, Argentina.          
            }
   \date{}

 
  \abstract
   {M dwarfs are an ideal laboratory for hunting Earth-like planets, and the study of chromospheric activity is an important part of this task. On the one hand, according to the study of short-term activity their high levels of magnetic activity can affect habitability and make it difficult to detect exoplanets orbiting around them. But on the other hand, long-term activity studies can show whether these stars exhibit cyclical behavior or not in their activity, facilitating the detection of planets in those periods of low magnetic activity.}
   {The long-term cyclical behavior of magnetic activity can be detected studying several spectral lines and explained by different stellar dynamo models (like $\alpha\Omega$ or $\alpha^2$ dynamos). 
   In the present work, we studied the Mount Wilson $S$-index to search for evidence of long-term activity possibly driven by a solar-type dynamo.}
   {We studied a sample of 35 M dwarfs with different levels of chromospheric activity and spectral classes ranging from dM0 to dM6. 
   To perform this, we used 2965 spectra in the optical range from different instruments installed in the Southern and Northern hemispheres to construct time series with extensions of up to 21 years. 
   We have analysed these time series with different time-domain techniques to detect cyclical patterns. 
   In addition, using 2MASS-$K_S$ and visible photometry we have also studied the potential impact of chromospheric activity on the surface brightness.}
   {Using the color index $(V-K_S)$, we have calculated the chromospheric emission levels and we found that most of the stars in the sample have low emission levels, indicating that most of them are inactive or very inactive stars.   
   For 31 stars over 35, we constructed time series using the $S$ indexes, and we detected 13 potential cycles of magnetic activity.
   These cycles have an approximate duration of between 3 and 19 years with false alarm probabilities (FAPs) less than 0.1\%.   
   For stars that do not show cyclic behavior, we have found that the mean value of the $S$-Index varies between 0.350 and 1.765 and its mean variability and chromospheric emission level are around 12\% and -5.110 dex, respectively.
   We do not find any impact of chromospheric activity on the surface brightness in the domain of $-5.6 < \log R'_{HK} < -4.5 $.}
   {}


   \maketitle
%

\section{Introduction}

Stellar magnetic activity is a phenomenon that has been extensively studied using several spectral lines sensitive to chromospheric activity (e.g. Mg \scriptsize{II}\normalsize\-, Ca \scriptsize{II}\normalsize\-, Na \scriptsize{I}\normalsize\-, and H$\alpha$) by integrating their fluxes in the line core (\citealt{Baliunas95}; \citealt{Diaz07}; \citealt{Lovis11, GomesdaSilva11, GomesdaSilva12}; \citealt{Buccino14, Flores16, Flores18, Ibanez18, Ibanez19, Ibanez20, Dimaio20, Lafarga21}; \citealt{Mignon23}).
The variability of each flux or equivalent width with time is related to stellar rotation and the evolution of the stellar active regions providing information on different regions at different height of the stellar atmosphere.
The stellar magnetic activity is assumed to be driven by a stellar dynamo that in the case of the Sun is called $\alpha\Omega$ dynamo and could be explained by the feedback between the differential rotation ($\Omega$-effect) of the Sun and the $\alpha$-effect related with the turbulent helical movements of the plasma in the solar convective zone \citep{Charbonneau10}.

The attempts to understand the stellar dynamo responsible for the magnetic activity cycles of the stars in our neighbourhood have focused on the relationship between stellar rotation and magnetic activity proxies.
It shows an increase in activity when the rotation period decreases, with a saturated activity regime for fast rotators. 
This relation has been widely studied in different works using different indicators: by \cite{Wright11}, \cite{Wright16}, and \cite{Wright18} in X-ray emission (using $L_X / L_{bol}$), and  by \cite{Astudillo17} and \cite{Newton17} in the optical range, employing the $\log(R'_{HK})$ and $L_{H\alpha} / L_{bol}$ indexes, respectively. 
These studies were especially focused on the study of M dwarfs, since these stars with masses between 0.1 and 0.5 M$_\odot $, constitute $\sim75 \%$ of the stars in the solar neighbourhood and also have a high occurrence rate of extrasolar planets orbiting in the habitable zone\footnote{The habitable zone is defined as the region in the equatorial plane of the star where water in the planet, if any, can be found in its liquid state.} surrounding the star (\citealt{Bonfils13, DressingCharbonneau15}).
They can exhibit high levels of chromospheric activity that can exceed the solar magnetic activity levels and several active stars are usually called flare stars due to the high frequency of these high-energy transient events (e.g., \citealt{Gunther20,Rodriguez20}).  
Therefore, a planet orbiting an M star could be frequently affected by small short flares and eventually by long high-energy flares, a fact which could constrain the exoplanet's habitability \citep{Buccino07, Vida17}, or even play an important role in the atmospheric chemistry of an orbiting planet (\citealt{Miguel15}).

On the other hand, the surface brightness colour relation (SBCR) is a tool of great interest for the study of stars hosting transiting exoplanets, especially M dwarfs, since due to their low size they represent an important laboratory for detecting Earth-like planets orbiting around them.
Such relation is usually calibrated by interferometric observations (\citealt{Kervella04, Dibenedetto05}) and recent studies based on a homogeneous methodology applied to a large number of interferometric data have shown that SBCRs depend not only on the temperature of the stars, but also on their luminosity class (\citealt{Salsi20, Salsi21}).
This has also been shown theoretically \citep{Salsi22}.
The SBCR is a fundamental tool that is used to easily and directly determine the angular diameter of any star from its photometric measurements, usually in two bands, in the visible and the near-infrared.
As an example, the SBCRs are currently implemented in the pipeline of the PLAnetary Transits and Oscillation of stars (PLATO) space mission (\citealt{Rauer14, Rauer24}), which provides an independent estimate of the stellar radius \citep{Gent22} and thus, the radius of the planet can be deduced directly from the transit \citep{Ligi16}. 
In order to characterise the stars in a large part of the Hertzsprung-Russell diagram and with the purpose of directly measuring the angular stellar diameters of about 800 stars, the Stellar Parameters and Images with a Cophased Array (SPICA) interferometer was installed at the Center for High Angular Resolution Astronomy (CHARA, see \citealt{Mourard18, Mourard22, Pannetier20}).
The SPICA instrument will help to improve SBCR also in M dwarfs in the near future. 

In the present study we have analysed the magnetic activity of a sample of 35 M dwarfs in order to characterise them according to chromospheric emission levels and analyse the possible impact that chromospheric activity could have on the surface brightness, as well as, to detect magnetic activity cycles.
To this end, we have organised this paper as follows: in Sect. \ref{sec.data} we describe the sample and the observations employed in this work.
In Sect. \ref{sec.analysis} we present our results from the chromospheric activity analysis and the surface brightness for the stars in the sample.
Finally, in Sect. \ref{sec.conclusions} we discuss the main conclusions of this analysis.

\section{Data} \label{sec.data}

\subsection{The stellar sample} \label{ssec.sample}

Our sample consists of 35 M dwarf stars, ranging from spectral class dM0 to dM6 and with V magnitude in the range 6.6 - 13.5. 
The selected stars come from a combination of M dwarfs studied under the HK$\alpha$ project using the REOSC spectrograph (see Section \ref{sec.observations}) and from the work of \cite{Salsi21} where they used a set of M dwarfs to calibrate the surface brightness-colour relation for these stars.

In the Table \ref{tab_sample} we show the fundamental parameters of our sample extracted from  Gaia DR3 \citep{GaiaDR323}, Simbad and CONCH-SHELL catalog \citep{Gaidos14} and also the angular diameters extracted from different references indicated in the last column of Table \ref{tab_sample}.

\begin{table*}[htb!]
\caption{Fundamental parameters of the stars in the sample. }
\begin{center}
\label{tab_sample}
\begin{tabular}{lccccccc}
\hline\hline\noalign{\smallskip}
Star             & SpT    & $T_{eff}$ [K] $^G$        & $B^S$ [mag]& $V^S$ [mag]& 2MASS-$K_S$ [mag]&  $\theta_{LD}$ [mas]        & ref. $\theta_{LD}$   \\
\hline\noalign{\smallskip}
GJ 1 *           & M2V    & 3355            & 10.02 & 8.61  & 4.50 & 0.812 $\pm$ 0.005    & 2                    \\
GJ 15 A *        & M2V    & 3352            & 10.12  & 8.13   & 4.02  & 1.005 $\pm$ 0.005    & 1                    \\
GJ 79            & M0V    & \multicolumn{1}{c}{--} & 10.32 & 8.88  & 5.18 & \multicolumn{1}{c}{--} & \multicolumn{1}{c}{--} \\
GJ 169           & M0.5V  & 3815            & 9.69  & 8.30    & 4.88 & \multicolumn{1}{c}{--} & \multicolumn{1}{c}{--} \\
GJ 176           & M2.5V  & \multicolumn{1}{c}{--} & 11.49  & 9.95  & 5.61 & 0.452 $\pm$ 0.020    & 2                    \\
GJ 205 *         & M1.5Ve & \multicolumn{1}{c}{--} & 9.44  & 7.84  & 4.04   & 0.943 $\pm$ 0.004    & 1                    \\
GJ 229           & M1V    & \multicolumn{1}{c}{--} & 9.61  & 8.13  & 4.17  & \multicolumn{1}{c}{--} & \multicolumn{1}{c}{--} \\
GJ 273 $^\dag$   & M3.5V  & 3024            & 11.44 & 9.87  & 4.86 & 0.763 $\pm$ 0.010    & 2                    \\
GJ 338 B $^\dag$ & M0V    & \multicolumn{1}{c}{--} & 9.05  & 7.63  & 4.14  &  0.856 $\pm$ 0.016    & 1                    \\
GJ 393 *         & M2V    & 3291            & 11.15 & 9.63   & 5.31 & 0.564 $\pm$ 0.021    & 4                    \\
GJ 406 *         & M6Ve   & --                   & 15.54 & 13.51 & 6.08 & 0.582 $\pm$ 0.020    & 2                    \\                   
GJ 411 $^\dag$   & M2+V   & 3511            & 8.96   & 7.52   & 3.34 $^S$  & 1.432 $\pm$ 0.013    & 1                    \\
GJ 412 A $^\dag$ & M1V  & 3382            & 10.27  & 8.80  & 4.77 & 0.764 $\pm$ 0.017    & 1                    \\
GJ 447 *         & M4V    & 3093            & 12.91 & 11.11 & 5.65 & 0.524 $\pm$ 0.029    & 2                    \\
GJ 488           & M0V    & 3738            & 9.89  & 8.47   & 4.88 & \multicolumn{1}{c}{--} & \multicolumn{1}{c}{--} \\
GJ 526 *         & M2V    & 3425            & 9.89  & 8.49  & 4.42 & 0.835 $\pm$ 0.014    & 1                    \\
GJ 536           & M0V    & 3638            & 11.18 & 9.71  & 5.68 & \multicolumn{1}{c}{--} & \multicolumn{1}{c}{--} \\
GJ 551 *         & M5.5Ve & 2829            & 12.95  & 11.01  & 4.38 & 1.103 $\pm$ 0.007    & 2                    \\
GJ 570 B         & M1.5V  & \multicolumn{1}{c}{--} & 9.56  & 8.07  & 3.80   & \multicolumn{1}{c}{--} & \multicolumn{1}{c}{--} \\
GJ 581 *         & M3V    & 3110            & 12.14  & 10.56  & 5.84 & 0.476 $\pm$ 0.007    & 2                    \\
GJ 628 $^\dag$   & M3V    & \multicolumn{1}{c}{--} & 11.64 & 10.07 & 5.08 & 0.661 $\pm$ 0.014    & 2                    \\
GJ 674 *         & M3V    & 3275            & 10.97 & 9.41  & 4.86 & 0.737 $\pm$ 0.037    & 2                    \\
GJ 687 *         & M3V  & 3192              & 10.65  & 9.15  & 4.55 & 0.859 $\pm$ 0.014    & 1                    \\
GJ 699 $^\dag$   & M4V    & 3099            & 11.24  & 9.51  & 4.52 & 1.004 $\pm$ 0.040    & 3                    \\
GJ 725 A $^\dag$ & M3V    & 3352            & 10.43  & 8.93   & 4.43 & 0.937 $\pm$ 0.008    & 1                    \\
GJ 725 B $^\dag$ & M3.5V  & 3180            & 11.32  & 9.78   & 5.00     & 0.851 $\pm$ 0.015    & 1                    \\
GJ 729 $^\dag$   & M3.5Ve & 3117            & 12.19  & 10.43  & 5.37  & 0.642 $\pm$ 0.020    & 2                    \\
GJ 752 A         & M3V    & 3236            & 10.63  & 9.12  & 4.67 & 0.836 $\pm$ 0.051    & 5                    \\
GJ 809 *         & M1V  & 3474            & 10.46 & 8.58  & 4.62 & 0.722 $\pm$ 0.008    & 1                    \\
GJ 825           & M1V    & \multicolumn{1}{c}{--} & 8.09   & 6.68   & 3.10  & \multicolumn{1}{c}{--} & \multicolumn{1}{c}{--} \\
GJ 832 *         & M2/3V  & 3310            & 10.18 & 8.69  & 4.50 & 0.814 $\pm$ 0.010    & 2                    \\
GJ 876 *         & M3.5V  & 3185            & 11.75 & 10.18 & 5.01  & 0.705 $\pm$ 0.009    & 2                    \\
GJ 880 $^\dag$   & M1.5V  & 3398            & 10.14 & 8.64  & 4.52 & 0.744 $\pm$ 0.004    & 1                    \\
GJ 887 $^\dag$   & M2V    & 3376            & 8.83   & 7.39   & 3.46  & 1.390 $\pm$ 0.040      & 3                    \\
GJ 908           & M1V    & 3619           & 10.43   & 8.99  & 5.04   & \multicolumn{1}{c}{--} & \multicolumn{1}{c}{--} \\
\hline
\end{tabular}
\tablefoot{In the last two columns we show the limb-darkened angular diameter of the stars extracted from the literature and its references. $^G$ values extracted from Gaia \textit{Data Release 3} \citep{GaiaDR323}. $^S$ values extracted from \textit{Simbad}. *, $^\dag$: stars considered and rejected by \citealt{Salsi21} (see Section \ref{ssec.SBCR}). 1: \citet{Boyajian12}; 2: \citet{Rabus19}; 3: \citet{Segransan03}; 4: \citet{Schaefer18}; 5: \citet{Berger06}.} 
\end{center}
\end{table*}

\subsection{Observations from the HK$\alpha$ Project} \label{sec.observations}

The HK$\alpha$ Project was started in 1999 with the aim to study the long-term chromospheric activity of southern cool stars.
In  this  program, we systematically observe late-type stars from dF5 to dM5.5 with the 2.15 m Jorge Sahade telescope at the CASLEO observatory, which is located at 2552 m above sea level in the Argentinian Andes.
The medium-resolution echelle spectra ($R\approx 13000$) were obtained with the REOSC\footnote{\textsf{http://www.casleo.gov.ar/instrumental/js-reosc.php}} spectrograph and they cover a maximum wavelength range between 386 and 669 nm.
We calibrated  all our echelle spectra in flux following the procedure described in \cite{Cincunegui04} which consists of two main phases based on main IRAF packages. 
First, we perform the standard echelle extraction and the proper wavelength calibration with the ThAr lamp.
Then, due to the strong CCD blaze, we perform an original calibration flux proccedure. 
For each science star, we obtain the long-slit spectrum in the same wavelength range calibrated in flux through standard procedures using standard spectrophotometric stars.
The flux calibrations of the long-slit spectra are reduced in order to increase the signal-to-noise ratio. 
Then, the corresponding sensitivity function is used for each order of the echelle spectrum.
Finally, each flux-calibrated order is combined to obtain a one-dimensional spectrum.

\subsection{Spectral monitoring surveys} 

We complemented our data with public observations from different spectrographs:
\textit{HARPS}, mounted at the 3.6 m telescope at the European Southern Observatory (ESO, Chile). 
It covers a wavelength range from 380 to 690 nm with a resolving power of R = 115000 in 72 echelle orders \citep{Mayor2003}.

\textit{HARPS - North} is the Northern Hemisphere counterpart of HARPS. HARPS-N is installed at the Italian Telescopio Nazionale Galileo, located at the Roque de los Muchachos Observatory on the island of La Palma, Canary Islands, Spain \citep{Cosentino2012}.

\textit{FEROS}, placed on the 2.2 m telescope in ESO, is equipped with two fibres and operates in a wavelength range of 360-920 nm with a resolution of $R \sim 48000$ \citep{Kaufer99}.

\textit{UVES}, attached to the Unit Telescope 2 (UT2) of the Very Large Telescope (VLT)  at ESO is designed to operate with high efficiency in its two arms covering the wavelength range 300 - 500 nm (blue) and 420 - 1100 nm (red). 
The maximum resolution is $R_b = $ 80000 or $R_r =$ 110000 in the blue and red arms, respectively \citep{Dekker2000}.

\textit{X-SHOOTER}, is a medium resolution spectrograph ($\sim 4000 - 17000$, depending on wavelength) mounted at the UT2 Cassegrain focus also at the VLT, it consists of three spectroscopic arms: UV-B (ultraviolet to blue range), VIS (visible range) and NIR (near infrared range) covering wavelength ranges 300 - 559.5 nm, 559.5 - 1024 nm and 1024 - 2480 nm, respectively \citep{Vernet2011}. 

\textit{HIRES}, attached at the Keck-I telescope has a limiting spectral resolution of $R =$ 67000 and a slit-width resolution of 39000 arcseconds operating between 300 and 1000 nm \citep{Vogt94}.

\textit{ESPRESSO}, installed on the VLT-ESO, is fed by the four Unit Telescopes (UTs) of the VLT and operates in a wavelength range of 380-686 nm with a resolution of $R \sim 200000$ \citep{Pepe2021}. 

All these spectra have been automatically processed by their respective pipelines\footnote{\textsf{https://www.eso.org/sci/facilities/lasilla/instruments.html}}$^,$\footnote{\textsf{https://www.eso.org/sci/facilities/paranal/instruments.html}} and all the information extracted and implemented in this work are reported in the Table B.1. 

\section{Analysis} \label{sec.analysis}

\subsection{Chromospheric activity levels}

To analyse the long-term magnetic activity of the stellar sample, we have focused our study on the emission of the Ca~ \scriptsize{II}\normalsize\ H and K resonance line-cores  at 3968 \AA\ and 3933 \AA\-, respectively.
The emission in these lines is related to the bright plages produced by non-thermal heating in the chromosphere, considered to be the source of these activity diagnostics.
We have calculated the well-known Mount Wilson $S$-Index for all collected spectroscopic data by measuring the ratio between the Ca \scriptsize{II}\normalsize\ emission lines integrated with a 1.09 \AA\ FWHM triangular profile and the photospheric continuum fluxes integrated in two 20 \AA\ passbands centred at 3891 and 4001 \AA\ \citep{Duncan91}.
We estimated a 4\% typical error of the $S$-Index derived from CASLEO (see the deduction in \citealt{Ibanez18}). 
The error of the $S$-Indexes derived from the other spectra, were calculated as the standard deviation of each monthly bin. For time intervals with only one observation in a month, we adopted the typical RMS dispersion of the bins. 

\begin{figure}[htb!]
\begin{center}
    \includegraphics[width=0.45\textwidth]{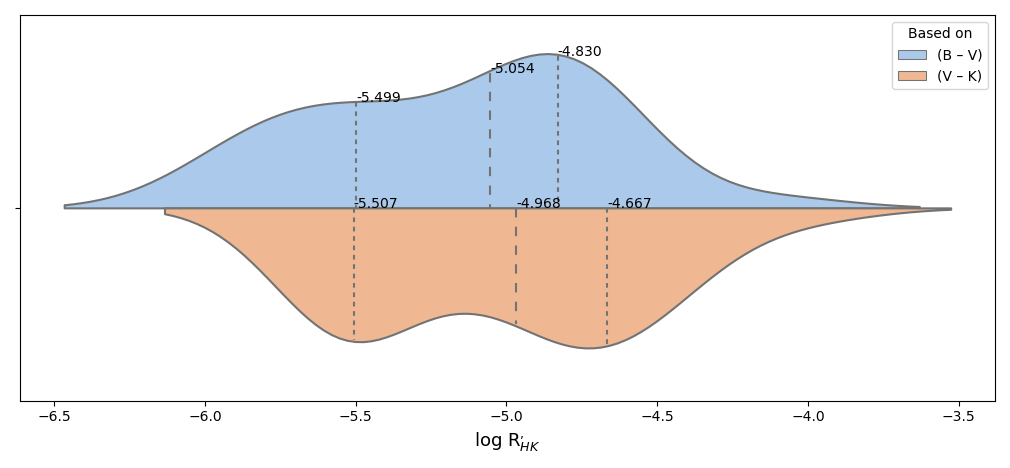}
\end{center}
\caption{Violin diagram of the distribution of chromospheric activity levels for the stars in the sample. The vertical lines with their respective numbers are indicating the values of the first, second and third quartile.
Most of the stars in the sample show low levels of chromospheric emission indicating that most of them are inactive or very inactive.}
\label{violin_logRHK}
\end{figure}

Using the calibration from \cite{Astudillo17}, we computed the chromospheric activity emission $\log R'_{HK}$ employing the colour indexes $(B-V)$ and $(V-K_S)$ using the values extracted from Simbad and 2MASS, both indicated in Table \ref{tab_sample}.
In Fig. \ref{violin_logRHK} we show a violin diagram of the obtained results where the vertical lines with their respective numbers correspond to the first, second and third quartile values.
Most of the stars in the sample show low levels of chromospheric emission indicating that most of them are inactive or very inactive.
The apparent bimodality of the distribution in Fig. \ref{violin_logRHK} is only due to the partial sample of stars analysed.
The left shift for the $\log R'_{HK}$ calibration using the colour index $(B-V)$ is probably due to the fact that M dwarfs emit little flux in the B-band. 
We therefore decided to use the values of $\log R'_{HK}$ based on $(V-K_S)$ colour indexes in our analysis.
The uncertainties in $\log R'_{HK}$, were computed using the standard deviation of the chromospheric emission levels for each star in the present analysis.

\subsection{Surface brightness study} \label{ssec.SBCR}

\begin{figure}[htb!]
    \centering
    \includegraphics[width=0.45\textwidth]{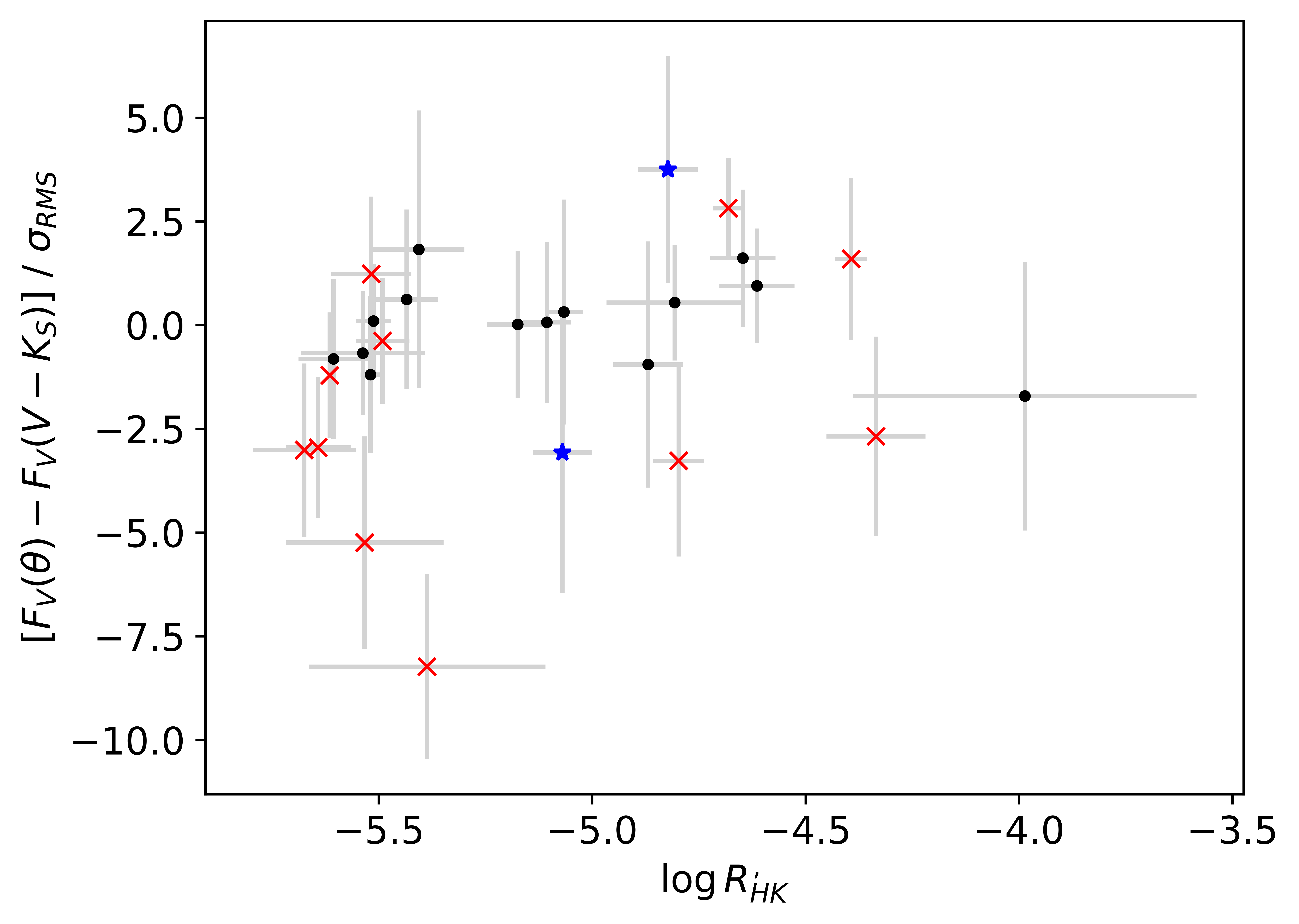}
    \caption{Difference between measured and computed surface brightness in units of $\sigma_{RMS}$ as a function of the $\log R'_{HK}$ index for the M dwarfs in the sample. In red crosses we represented the stars rejected  by \cite{Salsi21}, in black dots the targets considered by \cite{Salsi21} and in blue stars new targets added in this work. 
    In this plot, the uncertainty on the surface brightness of each measurement has also been divided by the rms.}
    \label{sbcr_v-k}
\end{figure}

To analyse the influence of stellar chromospheric activity on the surface brightness, we used the $\log R'_{HK}$ values obtained in the previous section and computed surface brightness following the work of \cite{Salsi21}.
To do this, we employed the values of the $V$ and $K_S$ from Simbad and 2MASS, respectively (see Table \ref{tab_sample}).

The surface brightness $F$, defined by $F = \log T_{eff} + 0.1BC$, is directly related to the bolometric correction ($BC$) and to the effective temperature ($T_{eff}$) of the star and therefore to its colour \citep{Wesselink69}. 
According to \cite{Barnes76}, the surface brightness in a given spectral band $F_\lambda$ can be found from their apparent magnitude and their true apparent limb-darkened angular diameter $\theta_{LD}$

\begin{equation}
    F_\lambda = C - 0.1 m_{\lambda 0} - 0.5 \log \theta_{LD}
    \label{eq.sbcr_1}
\end{equation}
where $m_{\lambda 0}$ is the apparent magnitude corrected from the interstellar extinction.
$C$ is a constant depending on the Sun's bolometric magnitude $M_{bol}$, its total flux $f_\odot$, and the Stefan-Boltzmann constant $\sigma$. 
For our study we adopted $C = 4.2196$ according to the more accurate estimations of solar parameters (\citealt{Mamajek15, Prsa16}).

On the other hand, the bolometric surface flux $f_{bol}$ of a star, which is expressed as the ratio between the bolometric flux $F_{bol}$ and the squared limb-darkened angular diameter $\theta_{LD}^2$, is linearly proportional to its effective temperature $T_{eff}^4$. 
It is thus also linearly linked to the colour $m_{\lambda 1} - m_{\lambda 2}$ . 
In this way, the surface brightness can be estimated by the following linear relation:
\begin{equation}
    F_{\lambda 1} = a (m_{\lambda 1} - m_{\lambda 2})_0 + b
    \label{eq.sbcr_2}
\end{equation}
where the subscript 0 refers to the magnitudes corrected for interstellar extinction.

The interstellar attenuation $A_V$ in the visible band is defined as $A_V = R_V \times E(B - V)$
where $R_V$ is the total-to-selective extinction ratio in the visible band and $E(B - V)$ is the $B - V$ color excess.
To compute this, we have used the Stilism\footnote{The online tool is available at \url{http://stilism.obspm.fr}.} online tool (\citealt{Lallement14, Capitanio17}) considering early Gaia DR3 distances \citep{GaiaDR323}. 
The interest of this tool lies in the three-dimensional maps of local interstellar matter it provides, based on measurements of the starlight absorption by dust or gaseous species. 

Taking all this into account and also the angular diameters extracted from the literature (see columns 7 and 8 of Table \ref{tab_sample}) together with the Eq. \ref{eq.sbcr_1}, we have computed $F_V(\theta)$ for the M dwarfs in the sample.
On the other hand, we have computed $F_V(V-K_S)$ employing Eq. \ref{eq.sbcr_2} and the colour indexes from Table \ref{tab_sample}.
Finally, in Fig. \ref{sbcr_v-k} we studied the possible influence of the chromospheric activity levels in the surface brightness plotting  the difference between measured and calculated surface brightness ($F_V (\theta) - F_V(V-K_S)$) divided by the $\sigma_{RMS}$ as a function of the chromospheric activity. 
We represented in red crosses the rejected stars by \cite{Salsi21}, in black dots the targets considered by \cite{Salsi21} and in blue stars the new targets added in this work.

From our results we can infer that there is no direct relationship between chromospheric activity and surface brightness in the domain of $-5.6 < \log R'_{HK} < -4.5 $.
For the stars that are further away from 5$\sigma_{RMS}$, we found GJ 725 B and GJ 699, two objects that were discarded by \cite{Salsi21}  because the interferometric observations to derive the angular diameters were not the most suitable.
However, there are other targets within the $\pm2.5 \sigma_{RMS}$ range that were also rejected by \cite{Salsi21}.
In their work they rejected stars using different approaches, one of them related to interferometric criteria but also due to stellar activity (fast rotators, spectroscopic binary, etc) and photometry found in the literature.
This preliminary study on the influence of chromospheric activity will be continued soon on the basis of new interferometric observations \citep{Mourard22}.

\subsection{Activity cycles}

\begin{table*}[htb!]
\caption{Synthesis of the characterisation and long-term activity analysis developed in this work.}
\begin{center}
\begin{tabular}{lcccccccc}
\label{tab_results}\\
\hline\hline\noalign{\smallskip}

Star     & \# Spectra & Time span & $\langle S \rangle$     & $\sigma_S /\langle S \rangle$ & $\log R'_{HK}$   & $\sigma_{\log R'_{HK}}$ & $P_{cyc}$                  & FAP   {[}\%{]}              \\
         &            & {[}d{]}   &              &                    &              &               & {[}d{]}                &  \scriptsize{IB19 | Bootstrap}\normalsize\            \\
\hline\noalign{\smallskip}
GJ 1     & 68         & 5238      & 0.399 & 0.140       & -5.537 & 0.145   & 4286 $\pm$ 479         & 0.01 | 0.04     \\
GJ 15 A  & 13         & 15        & 0.427 & 0.046       & -5.513 & 0.041 & --                     & --       \\
GJ 79    & 22         & 6589      & 1.703  & 0.064      & -4.497 & 0.032 & --                     & --        \\
GJ 169   & 3          & 9         & 1.287  & 0.036      & -4.524 & 0.018 & --                     & --        \\
GJ 176   & 112        & 4813      & 1.477  & 0.137       & -4.823 & 0.070 & --                     & --       \\
GJ 205   & 102        & 7675      & 1.801  & 0.175       & -4.614  & 0.088 & 5313 $\pm$ 862         & $<$ 0.1 | \ $<$ 0.01  \\
GJ 229 A & 86         & 7698      & 1.487  & 0.130       & -4.665 & 0.066 & 4949 $\pm$ 423         & $<$ 0.1 | $<$ 0.01  \\
GJ 273   & 172        & 4490      & 0.752 & 0.094       & -5.491 & 0.063 & --                     & --       \\
GJ 338 B & 7          & 2958      & 1.765  & 0.077      & -4.393  & 0.037 & --                     & --       \\
GJ 393   & 154        & 5121      & 0.939 & 0.081      & -5.066 & 0.045 & 1124 $\pm$ 29          & 0.003 | 0.02   \\
GJ 406   & 33         & 3733      & 155.893  & 0.902        & -3.986 & 0.402  & --                     & --     \\
GJ 411   & 1          & --        & 0.397    & --                 & -5.615 & --            & --                     & --       \\
GJ 412 A & 12         & 1742      & 0.350 & 0.115       & -5.675 & 0.121  & --                     & --        \\
GJ 447   & 65         & 4331      & 1.211  & 0.190       & -5.406  & 0.107  & 1956 $\pm$ 98 $^{IB}$  & 0.0002   \\
GJ 488   & 67         & 5232      & 1.603  & 0.097      & -4.480 & 0.048 & --                     & --        \\
GJ 526   & 47         & 6232      & 0.736 & 0.090      & -5.106 & 0.056 & 3739 $\pm$ 288         & 0.013 | 0.08   \\
GJ 536   & 216        & 5842      & 1.120  & 0.116       & -4.840 & 0.061 & 1643 $\pm$ 58          & 0.008 |  0.06   \\
         &            &           &              &                    &              &               & 4044 $\pm$ 443         & 0.05 | 0.1             \\
GJ 551   & 140        & 5495      & 11.994  & 0.399       & -4.807 & 0.160   & 4121 $\pm$ 321         & 0.02 | 0.05    \\
GJ 570 B & 129        & 1859      & 1.394  & 0.099      & -4.821 & 0.049 & --                     & --         \\
GJ 581   & 30         & 1883      & 0.537   & 0.106       & -5.606 & 0.082 & --                     & --     \\
GJ 628   & 154        & 4998      & 0.715 & 0.139       & -5.518 & 0.094 & 4498 $\pm$ 607         & 0.004  | 0.01  \\
GJ 674   & 74         & 5470      & 1.585  & 0.165       & -4.869 & 0.082 & --                     & --     \\
GJ 687   & 98         & 4         & 0.558 & 0.030      & -5.519 & 0.023 & --                     & --       \\
GJ 699   & 72         & 5834      & 0.695 & 0.319       & -5.533 & 0.185  & 3281 $\pm$ 255         & 0.001 | 0.02   \\
GJ 725 A & 21         & 3         & 0.454 & 0.088      & -5.642 & 0.076 & --                     & --       \\
GJ 725 B & 50         & 612       & 0.752   & 0.643       & -5.387 & 0.277  & --                     & --       \\
GJ 729   & 103        & 7603      & 7.418  & 0.274       & -4.335 & 0.116   & 1521 $\pm$ 20 $^{IB2}$ & $<$ 0.1  \\
GJ 752 A & 305        & 5892      & 1.004  & 0.124        & -5.070 & 0.069 & 4416 $\pm$ 462         & $<$ 0.01 | 0.02  \\
GJ 809   & 61         & 5152      & 1.545  & 0.148       & -4.647 & 0.076 & 2898 $\pm$ 273         & 0.001 | 0.03  \\
GJ 825   & 37         & 5474      & 1.094  & 0.101       & -4.673 & 0.051 & 2593 $\pm$ 137         & 0.04 | 0.1   \\
         &            &           &              &                    &              &               & 7038 $\pm$ 2700        & 0.0002 | 0.01  \\
GJ 832   & 172        & 5632      & 0.699 & 0.118       & -5.175 & 0.072 & 2297 $\pm$ 54          & $<$ 0.1 | $<$ 0.01 \\
GJ 876   & 96         & 3261      & 0.918 & 0.112       & -5.435 & 0.073 & --                     & --        \\
GJ 880   & 125        & 4450      & 1.635  & 0.072      & -4.681 & 0.036  & --                     & --        \\
GJ 887   & 79         & 2120      & 1.131  & 0.114       & -4.797 & 0.060 & --                     & --        \\
GJ 908   & 39         & 3525      & 0.490 & 0.124       & -5.326 & 0.095 & --                     & --       \\
\hline
\end{tabular}
\tablefoot{In the Col. 4 and Col. 5 we show, respectively, the $S$-Index obtained for each star in the sample and its variability. In Col. 6 and Col. 7 we show the chromospheric activity levels measured from the $\log R'_{HK}$ indicator and their standard deviation. Finally, we report the detected magnetic activity cycles with their respective FAPs in Col. 8 and Col. 9, respectively. With $^{IB}$ and $^{IB2}$ we highlighted the cycles extracted from \cite{Ibanez19} and \cite{Ibanez20}, respectively.}
\end{center}
\end{table*}

In order to look for some cyclical pattern in the magnetic activity of the stars in the sample, we have constructed the $S$ time series of each target containing at least 5 spectra.
We have implemented the same color code for all the time series in order to distinguish between the different instruments: blue for CASLEO spectra, orange for HARPS, green for FEROS, red for UVES, violet for XSHOOTER, brown for HIRES, yellow for HARPS-N and light blue for ESPRESSO spectra.
In this section, we show the time series for GJ 1 as an example of all the analysis implemented.
For the rest of the stars that present a cyclical behaviour in their magnetic activity, we have plotted the results in appendix \ref{ss.cycles} and those stars without a cyclical pattern are shown in appendix \ref{ss.noncycles}.
A summary of their characteristics as number of observations used ($N_{obs}$) or extension of the time series can be found in the second and third columns of Table \ref{tab_results}.
We have also cleaned them by discarding those spectra that are possibly contaminated by transient events, such as flares, using the same criteria discussed in \cite{Ibanez23} based on the distortion of the Balmer lines.
In some time series, we highlighted few points with red circles to indicate that the spectral line could be contaminated by flares and they were discarded from our analysis.

To characterise their activity based on decadal time-series of each individual star, we have also estimated the mean values of the Mount Wilson indexes ($\langle S \rangle$) and their dispersion $\sigma_S / \langle S \rangle$ as an estimation of the variability in the $S$-series (see Cols. 4 and 5 of Table \ref{tab_results}). 
For this purpose, we have considered the non-flaring time series so that the values found are purely due to due to non-transient variability of the magnetic field.

To study the long-term magnetic activity and to know if it exhibits a cyclical behaviour or not, we have searched for significant signals in the periodograms of the $S$-series.
For all of them we employed the generalized Lomb-Scargle (GLS) periodogram \citep{Zechmeister09}.
To estimate the significance of the signals found in the periodograms we have taken into account the false alarm probability (FAP) as we expressed in \cite{Ibanez18}. 
To complement this, for those stars with a significant FAP we have also computed the bootstrap randomisation of the data following \cite{Endl01} (see column 9 of Table \ref{tab_results}).
We define a significant signal as those frequencies whose power shows a FAP $\leq$ 0.1\% in both methods above-mentioned.

Due to inhomogeneity and differences in the cadence of the time series or also due to the difference in sampling, several significant frequencies might appear in the periodograms.
When the periodograms have displayed numerous significant peaks, we have proceeded in different ways: by binning the time series if the number of points is not well distributed over its time span or by subtracting the harmonic function fit associated with the most significant period/s from the time series.
We have also considered two complementary tools: the Bayesian generalized Lomb-Scargle (BGLS) periodogram described in \cite{Mortier15} and the CLEAN deconvolution algorithm developed by \cite{Roberts87}.
When we needed to estimate the probability that a given frequency is related to an activity cycle, we calculated the BGLS, while when we needed to discard frequencies from our periodograms due to, for example, aliasing induced by the spectral window function, we used the CLEAN algorithm.

We have added the word ‘potential/possible’ to distinguish that this is a first detection and not a confirmation of the activity cycle already reported in the literature.

We have detected 13 periodic signals that correspond to the first evidence of potential magnetic activity cycles\footnote{We have added the word `potential/possible' to distinguish that this is a first detection and not a confirmation of the activity cycle already reported in the literature.} and others already reported in the literature that we have been able to confirm in this work.. 
The error of the detected periods depends on the finite frequency resolution of their corresponding periodogram $\delta\nu$ as given by equation (2) in \cite{Lamm04}, $\delta P = \frac{\delta\nu P^2}{2}$.
For those stars where cyclical behaviour due to stellar magnetic activity was detected, we have plotted the folded phases with their respective periods detected by the aforementioned periodograms. 
In each of these plots, we have plotted the least-square fit in red dashed line and the red shadow represents the $\pm 3\sigma$ deviation.

We summarize the results of this section in columns 4 to 9 of Table \ref{tab_results}. 
Here we provide the analysis carried out for the different time series of the stars in the sample and a detailed discussion over our results are shown in the Sect. \ref{sec.conclusions}.

\label{ssec.ciclos.gl1}

\begin{figure}[htb!]
\begin{center}
    \includegraphics[width=0.4\textwidth]{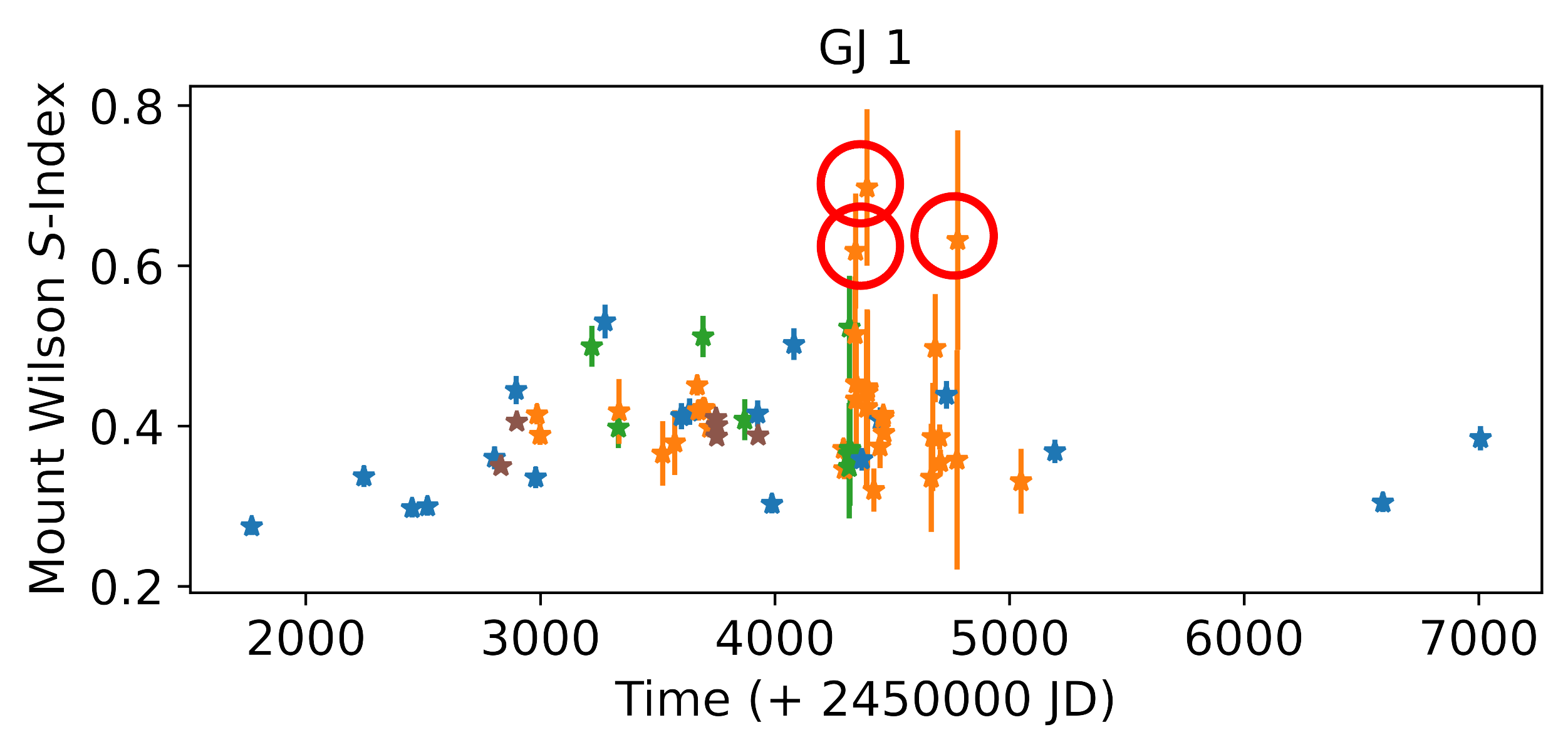}
    \includegraphics[width=0.45\textwidth]{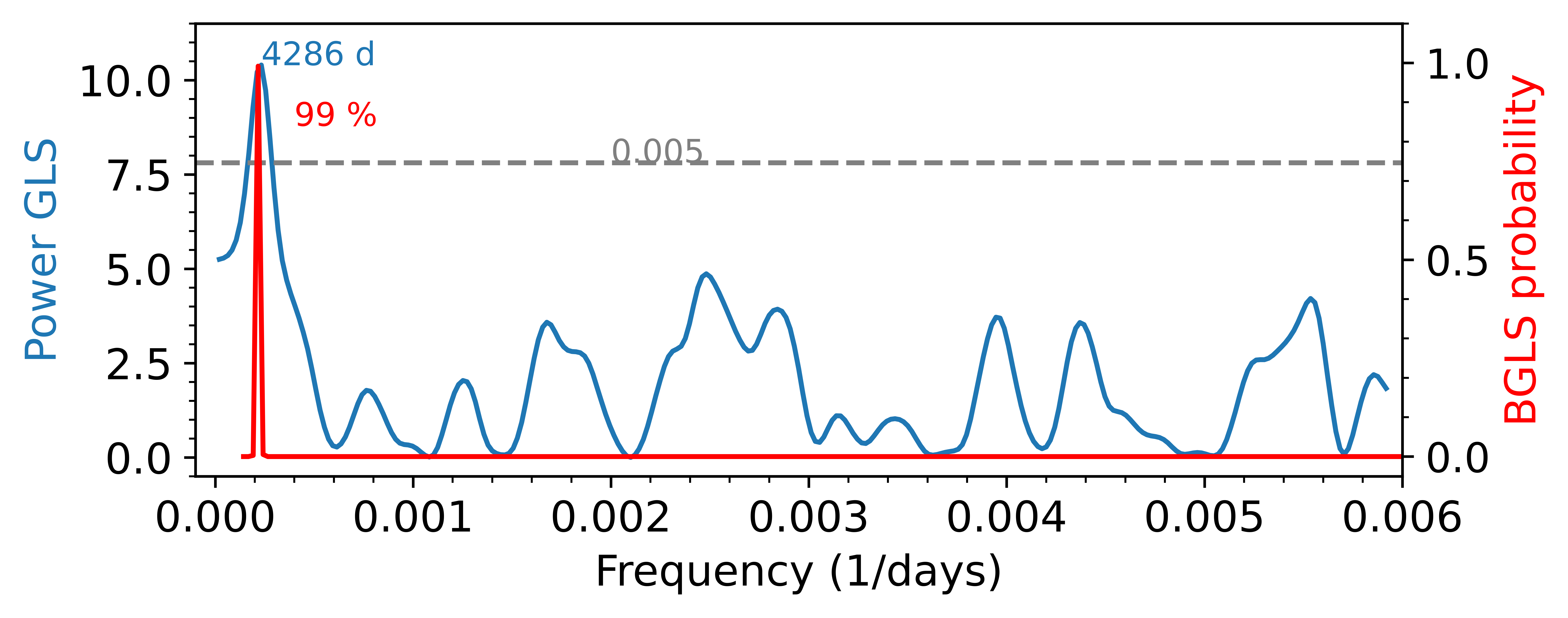}
    \includegraphics[width=0.25\textwidth]{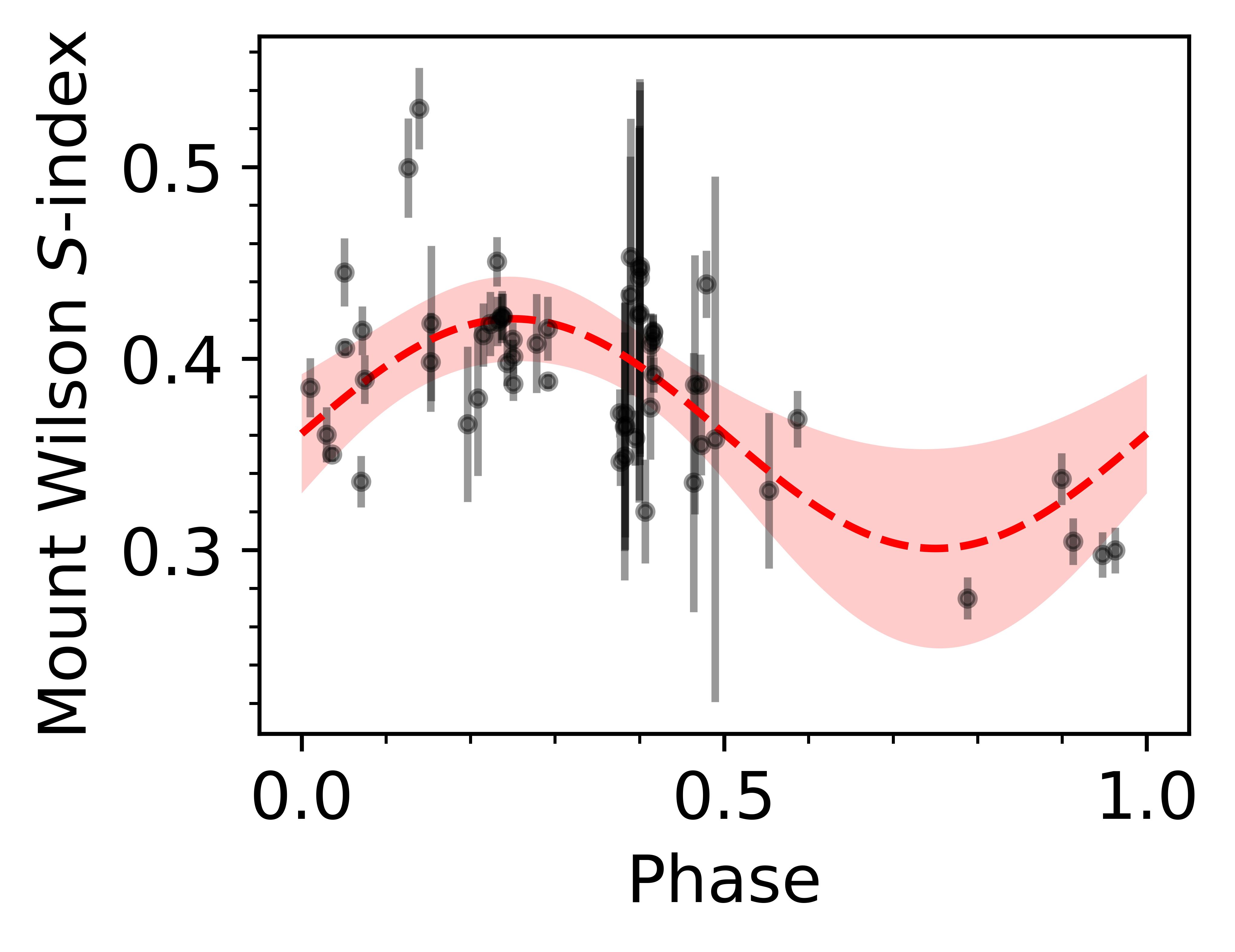}
\end{center}
\caption{\textbf{GJ 1.} \textit{Top.} $S$-Index time series. \textit{Middle.} GLS periodogram (blue): $(4286 \pm 479)$ days with FAP = 0.01\%; FAP$_{bootstrap} = 0.04$ \%. BGLS periodogram (red): there is a probability of 99\% that the peak of $\sim4200$ days corresponds with a magnetic activity cycle for this star. \textit{Bottom.} Phase folded time series with a $\sim4200$ day period.} 
\label{per_gl1}
\end{figure}

\noindent
\textit{-- GJ 1 - HD 225213 - HIP 439}.
In Fig. \ref{per_gl1}, we show the time series and the results obtained from the GLS and BGLS periodograms.
We found that the most significant peak corresponds to a period $P_{GLS} = (4286 \pm 479)$ days with a FAP $\sim 0.0001$. 
To corroborate that the 4286-day peak corresponds to an activity cycle and not an artifact or alias, we subtracted the harmonic function fit associated with this period from its time series and executed the GLS periodogram for the new residual series.
We did not find any significant peaks in this step.
On the other hand, to find the probability that this peak of $\sim$ 4200 days corresponds to an activity cycle, we studied the BGLS periodogram (red curve in Fig. \ref{per_gl1}). 
From this analysis, we conclude that the peak associated with the 4286-day period corresponds to a potential activity cycle with a 99\% probability for this partially convective star.
\\

\noindent
\textit{-- GJ 205 - HD 36395 - LHS 30}.
For this star we have executed GLS and CLEAN periodograms (see middle panel of Fig. \ref{per_gj205}). 
For the $S$-Index time series, a single significant peak stands out with a period of $P_{GLS} = 5313 \pm 862$ days and a FAP < 0.1\%; FAP$_{bootstrap} < 0.01$ \% for the GLS periodogram and $P_{CLEAN} = 5479 \pm 196$ days for the CLEAN.
Both peaks would be associated with an activity cycle between 14.5 and 15 years for GJ 205.
Subtracting the period of $\sim$5400 days from our time series and running the periodograms again, we found no significant peaks.
Thus, we conclude that the quiescent star GJ 205 has a potential magnetic activity cycle of $\sim$5400 days. \\

\noindent
\textit{-- GJ 229 A - HD 42581 - HIP 29295}. 
In this study, performing the GLS periodogram, we obtained two significant peaks of $P_{GLS; 1} = 4949 \pm 423$ days and $P_{GLS; 2} = 123.2 \pm 0.4$ days, with FAPs less than 0.1\% (see middle panel of Fig. \ref{per_gj229}). 
To analyse whether the $P_{GLS; 2}$ peak is a sampling effect, we executed the CLEAN periodogram where we find the only significant signal at a period of $P_{CLEAN} = 4811 \pm 150$ days.
Therefore, we have discarded that the 123-day peak is associated with modulations related to the activity of the star.
We have also analysed our time series with the BGLS periodogram and we found that there is a 99\% probability that the 4900-day peak is related to the activity cycle for our $S$-series.
Subtracting the 4900-day cycle from our series, we found no other significant peaks during the execution of the periodograms. 
Then, we have concluded that a cycle of $\sim$13 yr is possible modulating the magnetic activity cycle for this star.
This cycle does not agree with those already reported in the literature, we discussed our results in Sect. \ref{ssec.activityCycles}. \\

\noindent
\textit{-- GJ 393 - BD+01 2447 - HIP 51317}.
To study its long-term activity, we have binned the time series every 30 days for those points in HJD $>$ 6000 d in order to analyse a more homogeneous series. We calculated the GLS and CLEAN periodograms for the time series and we obtained only two significant peaks with the GLS: $P_{GLS; 1}$ = (1124 $\pm$ 29) d and $P_{GLS; 2}$ = (274 $\pm$ 2) d with FAPs $\sim 0.003$ \% and $\sim 0.04$ \%, respectively (see Fig. middle panel of \ref{per_gl393}).
To know if the 274-peak is an alias we executed the CLEAN periodogram.
Only one significant frequency appear with a period of $P_{CLEAN}$ = (1112 $\pm$ 25) d with a FAP$_{bootstrap} = 0.02$ \%. 
For this star, we concluded that GJ 393 presents a possible activity cycle around 1100 days. \\

\noindent
\textit{-- GJ 526 - HD 119850 - HIP 67155}.
In the middle panel of Fig. \ref{per_gl526} we show the GLS and CLEAN periodograms where it is observed that there is a cyclic behaviour in the activity with a period of $(3739 \pm 288)$ days with a FAP of 0. 013\%; FAP$_{bootstrap} = 0.08$ \%  for the GLS case (blue line), while for CLEAN (orange line) we obtained a period of $(3115 \pm 1336)$ days\footnote{Here the error was computed as the HWHM$\times P^2$, where $P$ is the period found.}.
These results are in agreement with those reported by \cite{SuarezMascareno16}.
Then, we concluded that GJ 526 exhibit a magnetic activity cycle of $\sim$10 yr.\\

\noindent
\textit{-- GJ 536 - HD 122303 - 	HIP 68469}.
We analysed the binned series with the GLS and CLEAN periodograms in Fig. \ref{per_gl536}.
In the case of the GLS periodogram, we found a peak corresponding to a period of (1643 $\pm$ 58) days with a FAP of $8\times10^{-5}$; FAP$_{bootstrap} = 0.06$ \%,  while for the CLEAN case we found a period of (1667 $\pm$ 24) days.
To search for other potential magnetic activity cycles, we analyse the residuals of the series by subtracting the harmonic signal corresponding to the $\sim$1600-day period.
In this new residual series we found a significant period peak ($4044 \pm 443$) days with a FAP of 0.05\% (see bottom panel of Fig. \ref{per_gl536}) and FAP$_{bootstrap} = 0.1$ \%.
Subtracting this new signal from our series, we did not detect any significant period.
Finally, to corroborate our results, we have worked with the original series without binning.
Both cyclical signals are present in the unbinned series where we have detected periods of $1643 \pm 51$ days and when subtracting this signal, one of $4382 \pm 469$ days, both with FAPs < 0.1\%.
We therefore conclude that in this spectroscopic series, the period of $\sim$1600 days is modulated by a larger activity cycle of period $\sim$4000 days. 
This period is about twice as long as that reported by \cite{SuarezMascareno17}. 
We discussed our results in Sect. \ref{ssec.activityCycles}.\\

\noindent
\textit{-- GJ 551 - HIP 70890 - Proxima Centauri}.
In the top panel of Fig. \ref{per_gj551}, we show the $S$-Index time series followed by the GLS and CLEAN periodograms where several peaks with high significance appear. 
GLS: $P_{GLS; 1} = (4946 \pm 522)$ d, $P_{GLS; 2} = (110.6 \pm 0.3)$ d, $P_{GLS; 3} = (81.5 \pm 0.3)$ d con FAPs $<0.01$ \%).
CLEAN: $P_{CLEAN; 1} = (4226 \pm 325)$ d and $P_{CLEAN; 2} = (81.5 \pm 0.2)$ d. 

As the rotational modulation of GJ 551 ($\sim$82 d, \citealt{Kiraga07}) is present in the periodograms performed on our series, we subtract the harmonic signal of $\sim$82 days and run the GLS periodogram to the new residual series. 
In this analysis, we obtained as a result a single significant peak of $P_{GLS} = 4121 \pm 321$ days with a FAP of 0.0002  (see third panel of Fig. \ref{per_gj551}) and FAP$_{bootstrap} = 0.05$ \% . 
When we subtracted this signal from our series and ran the GLS periodogram, we did not obtain any other significant peaks.
The magnetic activity of GJ 551 has already been studied and its activity cycles have been reported in several works (\citealt{Cincunegui07, Wargelin17}). 
Our cycle of $\sim$4000 days differs from those reported in the literature. 
We discussed our results in Sect. \ref{ssec.activityCycles}.\\

\noindent
\textit{-- GJ 628 - BD-12 4523 - V* V2306 Oph}.
As the GLS periodogram displayed multiple peaks of high significance, we used the series binned every 30 days to study the long-term magnetic activity of this star.
Using GLS and CLEAN we have obtained only one significant peak: $P_{GLS}$ = (4498 $\pm$ 607) d with FAP of $\sim 0.004$ \%; FAP$_{bootstrap} = 0.01$ \%  and $P_{CLEAN}$ = (4153 $\pm$ 173) d, respectively (see middle panel of  Fig. \ref{per_gl628}). 
Finally, to corroborate our result we used the $S$ series without binning and with the GLS periodogram we obtained as the most significant peak a signal of $P_{GLS}$ = (4090 $\pm$ 363) d with a FAP $<<$ 0.001\%.
We conclude that GJ 628 present a potential activity cycle of $\sim$4100 days.
\\

\noindent
\textit{-- GJ 699 - HIP 87937 - Barnard's star}.
In Fig. \ref{per_gl699}, we show the time series, the GLS and BGLS periodograms and the phase for this target.
The only significant period $P_{cyc} = (3281 \pm 255)$ d with a FAP of $1\times10^{-5}$; FAP$_{bootstrap} = 0.02$ \% would be related to the activity cycle with a probability of 96\%.
\cite{Toledo19} reported the rotation period for this star ($P_{rot} = $ 145 days) with an activity cycle of $P_{m_V}$ = 3846.15 days using the ASAS photometric database, and a cycle of $P_S$ = 3225.81 days using spectroscopic data from 7 different spectrographs.
Our result is in line with \cite{Toledo19}. 
However, we would like to point out that despite this, the $S$-index time series constructed in the present work does not show clear evidence of a periodic signal, as can also be seen in the plotted phase in Fig. \ref{per_gl699}. 
In the work of \cite{Toledo19}, they have used a modified S-index constructed from the Ca \scriptsize{II}\normalsize\- lines, so it is logical that the time series are not similar. 
Furthermore, it must be mentioned that the data used are not the same and  we did not find the peak related to the rotational modulation of $\sim$ 145 days. 
Therefore, in this work we conclude that despite having reproduced the potential cycle already detected by the aforementioned authors ($\sim$9 yr), this star should continue to be monitored for future analysis of the long-term magnetic activity.\\

\noindent
\textit{-- GJ 752 A - HD 180617 - V$^*$ V1428 Aql}. 
To analyse its long-term magnetic activity, we have implemented the GLS and CLEAN periodograms for the binned time series.
The results are shown in the middle panel of Fig. \ref{per_gj752A}.
We obtained only one significant peak: $P_{GLS}$ = (4416 $\pm$ 462) d with FAP < 0.01\%; FAP$_{bootstrap} = 0.02$ \% and $P_{CLEAN}$ = (4527 $\pm$ 174) d, respectively. 
When we removed this signal from our binned time series, we did not find another significant peak. 
We discussed our results in Sect. \ref{ssec.activityCycles}. 
As follows, we concluded that the magnetic activity of this star is possible modulated by a cycle of $\sim$12 years.\\

\noindent
\textit{-- GJ 809 - HD 199305 - HIP 103096}.
To study its long-term activity, we calculated the GLS and CLEAN periodograms for the time series (see middle panel of Fig. \ref{per_gl809} ) and we obtained only one significant peak: $P_{GLS}$ = (2898 $\pm$ 273) d with FAP of $\sim 0.001$ \%; FAP$_{bootstrap} = 0.03$ \%  and $P_{CLEAN}$ = (2861 $\pm$ 79) d, respectively. 
The 418-day peak present in the GLS periodogram disappears in the CLEAN periodogram because it is probably related to an alias induced by the spectral window function.
Then, we concluded that this star exhibit a potential activity cycle of $\sim$2800 days.\\

\noindent
\textit{-- GJ 825 - HD 202560 - V$^*$ AX Mic}.
To study its long-term activity, we calculated the GLS periodogram where we obtained two significant peaks: $P_{GLS;1}$ = (7038 $\pm$ 2700) d and $P_{GLS;2}$ = (365 $\pm$ 1) d with FAPs of $2\times10^{-6}$ and $1\times10^{-4}$, respectively (see second panel of Fig. \ref{per_gl825}, blue line).
To investigate whether the second peak associated to 1 year is related to an activity cycle or to a sampling effect, we computed the CLEAN periodogram.
In the second panel of Fig. \ref{per_gl825}, we show this periodogram in orange line.
Since the 365-day peak is not present in the CLEAN periodogram, we conclude that this peak is due to an alias induced by the spectral window function.
On the other hand, we can observe that the largest peak still appears and in this case its value is $P_{CLEAN}$ = (6078 $\pm$ 338) d.

Since the period of $\sim$7000 days is larger than the time span of the sample ($\sim$5400 days), we studied the residual series obtained by subtracting the peak of $\sim$7000 days from our $S$ series.
We show the results in the fourth panel of Fig. \ref{per_gl825}, where we found a single significant peak of $P_{GLS}$ = (2593 $\pm$ 137) d with a FAP of 0.04\%; FAP$_{bootstrap} = 0.1$ \%  and $P_{CLEAN}$ = (2378 $\pm$ 104) days.
Thus, we conclude that GJ 825 present a possible activity cycle of $\sim$2500 days that could be modulated by a higher peak of $\sim$7000 days. \\

\noindent
\textit{-- GJ 832 - HD 204961 - HIP 106440}.
To study its long-term activity, we calculated the GLS and CLEAN periodograms for the binned time series and we obtained only one significant peak: $P_{GLS}$ = (2297 $\pm$ 54) d with FAP of $3\times10^{-7}$; FAP$_{bootstrap} < 0.01$ \%  and $P_{CLEAN}$ = (2333 $\pm$ 49) d, respectively (see middle panel of Fig. \ref{per_gl832}). 
We concluded that a cycle of $\sim$6 years is potentially modulating the magnetic activity of this star.\\

\subsubsection{Other M dwarfs} \label{ssec.ciclos.gl908}

We did not detect any evidence of cyclical activities in the time series for the stars GJ 169, GJ 338 B, GJ 488, GJ 79, GJ 412 A, GJ 880, GJ 570 B, GJ 908, GJ 411, GJ 887, GJ 15 A, GJ 687, GJ 725 B, GJ 273, GJ 876, GJ 725 A, GJ 674, GJ 581, GJ 176 and GJ 406.
The time series of these non-cyclical stars are plotted in the Appendix \ref{aa.timeseries}.
Probably, they could be non-cyclic stars or, in several cases, the time-span or the sampling of the dataset could be insufficient to detect a significant period. 
Thus, we need to keep observing this sample to build more reliable time series.
It may also be due to the poor distribution and inhomogeneity of the points in the time series or simply because the stars have very low levels of chromospheric emission, which classifies them as very inactive and in this work also without magnetic activity cycles.
This is the case of GJ 273 ($\log R'_{HK} = -5.491 \pm 0.063$) which, with a large number of spectra approximately well distributed throughout the time series, does not present a cyclical behaviour in its magnetic activity due to the low variability of the star's $S$-Index (see its time series in the Appendix \ref{aa.timeseries}). 

Discarding GJ 406, in this stellar sample the mean value of the $S$-Index varies between 0.350 and 1.765 and their average variability and chromospheric emission level are around 12\% and -5.110 dex, respectively.
For the particular case of GJ 406, we found a variability of 90\% in the $S$ time series and the chromospheric emission level is the highest value in all the sample ($\log R'_{HK} = -3.986 \pm 0.402$).
These values are indicating that this star is a very active star with high chromospheric emission and therefore, considering the percentage of variability found possibly most of the spectra are contaminated by flares. 
The way to improve this last issue in order to study the magnetic activity of this star, is observing it with a short cadence in both, photometry and spectroscopy techniques.

For the particular case of GJ 581, the activity cycle for this star has already been reported by \cite{GomesdaSilva12}. 
They found a period of 3.85 years using the Na \scriptsize{I}\normalsize\- spectral lines of HARPS spectra.
In \cite{Robertson14b} they also studied the magnetic activity for this star and they concluded that the $\sim 1300$ d-peak is the beat frequency of the 125- and 138-day signals associated with the rotation period but alternatively, the cycle may exist.
In the present study, we used HARPS spectra with defined Ca \scriptsize{II}\normalsize\- line emission where spectra with poor signal-to-noise ratios were discarded.
Although, Na \scriptsize{I}\normalsize\- and Ca \scriptsize{II}\normalsize\- indexes are not strictly correlated, due to different atmospheric sources responsible for the emission \citep{Meunier24},  we also computed the Na \scriptsize{I}\normalsize\- D index for our data set. 
Nevertheless, the data set used in this work is not exactly the same as the one used in \citeauthor{GomesdaSilva12}'s work.
From our analysis, we were unable to find the 1300 day-signal by using similar dataset through the $S$-index and Na \scriptsize{I}\normalsize\- time series. 
Therefore, we conclude that this star should continue to be monitored for the analysis of its long-term magnetic activity and the possible confirmation of the activity cycle reported by the aforementioned works.

\section{Discussions and conclusions} \label{sec.conclusions}

\subsection{Surface brightness study} \label{ssec.sbcr.concl}

In this work we have carried out an analysis of how magnetic activity in the stellar chromosphere of M dwarfs might have an impact on the surface brightness.
As we mentioned in Sect. \ref{ssec.SBCR}, it seems that there is no direct relationship between the two quantities, but this result is based on a small sample of targets.

The impact of the magnetic activity on surface brightness and on SBC relation is important, but fundamentally, improving the SBCR for M dwarfs (even with for those stars with low magnetic activity levels) is even now necessary, as refinements are still required as \cite{Kiman24} have recently quoted. 

Our goal in the future is to improve the calibration of the surface brightness - colour relation in M dwarfs by using a larger sample. 
First of all, we must include more active stars in the sample since we currently only have 3 targets (GJ 338 B, GJ 729 and GJ 406) with values higher than $\log R'_{HK} = -4.5$. 
To do so, we will have to observe these stars by interferometry to obtain their angular diameters directly, as only a few of them have been reported for M dwarfs. 
Once the CHARA/SPICA interferometer commissioning tests have been completed, we will be able to observe the M dwarfs, re-calibrate their SBCR and, in particular, improve its  robustness.

\subsection{Activity cycles} \label{ssec.activityCycles}

From 8 different spectroscopic databases, we have built $S$-index time series for 31 over 35 dM stars in the sample presented in Appendix \ref{aa.timeseries}. 
For those series of time-span longer than 4300 days, we have computed GLS periodograms complemented by the BGLS and CLEAN periodograms and we have detected decadal and yearly cyclical patterns in 13 stars. 
These cycles have a duration between approximately 3 and 19 years.
We have also detected double cycles in three partially convective stars: GJ 536, GJ 825 and GJ 752 A.
However, the periods detected for GJ 536, GJ 551, GJ 752 A and GJ 229 A show discrepancies with those already reported in the literature.
In the following paragraphs, we discuss a specific analysis on these targets.

\textit{GJ 536.} \cite{SuarezMascareno17} reported a period of 824.9 $\pm$ 1.7 days which is about half what we found in this paper.
\cite{SuarezMascareno17} have concluded that the periodicities detected when they analysed the Mount Wilson index and the H$\alpha$ index, may be apparent and close to the real periodicities caused by sampling, given that the series studied do not have an optimal extension to detect signals of prolonged periods.
In this work we use 216 spectra covering a time span of $\sim$5800 days implying that our $S$-series is the longest series studied until now for this star, which shows a higher fidelity when searching for signals related to the magnetic activity cycle.
In addition, in this work we have also studied the GJ 536 time series using the CLEAN algorithm that allows us to detect possible spurious signals due to sampling.

\textit{GJ 551.} In \cite{Wargelin17} they reported a $\sim$7-year magnetic activity cycle for Proxima Centauri using ASAS photometry. 
It is important to note the different contributions depending on the observational technique used. On the one hand, photometry tracks starspots and filaments while spectroscopy tracks plages related to local magnetic field concentrations. 
The intensity of the emission of Ca \scriptsize{II}\normalsize\- spectral lines increases with the fraction of non-thermal chromospheric heating that can be produced, for example, by local magnetic inhomogeneities, thus providing a useful spectroscopic indicator to measure the intensity and area covered by magnetic fields.
Recently, in \cite{SuarezMascareno20} they combined HARPS, UVES and ESPRESSO spectroscopy where they detected no signals related to the possible magnetic activity cycle concluding that any possible long-term variation is suppressed by the floating means of the different data sets.
In our case, using only the 40 CASLEO spectra we obtain a cycle of $P_{GLS-CASLEO} = (4121 \pm 816)$ days with a FAP of 0.006 compatible with the result obtained in our previous analysis. 
Thus we can conclude that our result is not affected by the different data sets used in our study, but rather they complement each other.
Finally, it is important to note the rellevance of the discrimination in our study of those spectra that may be contaminated by flares in this type of highly active stars where intensifications and broadening in the Balmer lines and in some cases in the Ca \scriptsize{II}\normalsize\- lines can be observed.

\textit{GJ 752.} \cite{Buccino11} have reported a magnetic activity cycle of 7 years employing ASAS photometric data and the CASLEO spectroscopic data.
On the other hand, \cite{SuarezMascareno16} have detected an activity cycle of 9.3 years using a similar ASAS database.
Our results are in disagreement with those reported in literature.
However, when we work with the series without binning, the most significant peak still appears to be $\sim$4400 d. 
When this signal is discarded, the most significant period appears to be $\sim$2500 d, similar to that reported in \cite{Buccino11}. 
We have not found this signal in our binned series despite the fact that our time series has an extension of more than 16 years. 
Therefore, we conclude that this star presents a magnetic activity modulated by a $\sim$12 yr cycle in the time series of the chromospheric $S$-Index.
It is important to emphasise that the cycles detected by photometry may not agree with those reported by chromospheric indicators for this kind of stars. 
We therefore highlight that it is of great importance to pursue a more detailed study of the photospheric (spot) and chromospheric (plages) correlation in M stars.

\textit{GJ 229.} \cite{Buccino11} using the Ca K-line flux measured on flux-calibrated CASLEO spectra, obtained a magnetic activity cycle of 1649 days.
On the other hand, \cite{SuarezMascareno16} and \cite{DiezAlonso19} using ASAS photometric data, reported an activity cycle of 8.3 years for this star (almost twice as long as the cycle found by \citealt{Buccino11}).
Recently, \cite{Mignon23}, measuring the $H\alpha$ line from HARPS spectra, detected a cycle around $\sim700$ days.
The time series used in these works have shorter extensions than the cycle we reported in the present work ($\sim$4900 days). 
Therefore, our results do not indicate incompatibility with what is reported in the literature, but we have the advantage of having studied a longer time series ($\sim$7700 d) with a larger number of spectra (86) than that of both authors.

There are other incompatibilities between our study and those from other authors, for example in the case of GJ 1.
It is important to note that its rotation period ($P_{rot} = 60$ d) was reported by \cite{SuarezMascareno15} employing HARPS spectra by analysing the Ca \scriptsize{II}\normalsize\- and H$\alpha$ lines.
More recently, \cite{Mignon23} have reported a rotation period of 91 days studying the variability of Ca \scriptsize{II}\normalsize, Na \scriptsize{I}\normalsize\ and H$\alpha$ lines.
In the present study, we did not find such modulation (see Fig. \ref{per_gl1}, \textit{Middle}) and therefore we conclude that a photometric analysis to corroborate the GJ 1 rotation period reported by \citeauthor{SuarezMascareno15} or \citeauthor{Mignon23} is of great importance.

In conclusion, even for the study of long-term magnetic activity it is necessary to highlight the importance of high cadence photometric or spectroscopic studies to detect rotation periods in M dwarfs.
Regarding the building of the time series, it is essential to have reliable series with a good temporal extension, taking into account that the discrimination of spectra that may be contaminated by transient events in this type of stars is crucial.
Finally, it is of great importance to study them with different techniques to mitigate sampling effects such as spurious signals or aliasing due to the spectral window function.

\section{Data Availability}

Table B.1 is only available in electronic form at the CDS via anonymous ftp to cdsarc.u-strasbg.fr (130.79.128.5) or via http://cdsweb.u-strasbg.fr/cgi-bin/qcat?J/A+A/.

\begin{acknowledgements}
      This project has received funding from the European Research Council (ERC) under the European Union’s Horizon 2020 research and innovation programme (Grant agreement No. 101019653).
\end{acknowledgements}

%
%

\bibliographystyle{aa}
\small
\bibliography{biblio}

\begin{thebibliography}{80}
\expandafter\ifx\csname natexlab\endcsname\relax\def\natexlab#1{#1}\fi

\bibitem[{{Astudillo-Defru} {et~al.}(2017){Astudillo-Defru}, {Delfosse},
  {Bonfils}, {Forveille}, {Lovis}, \& {Rameau}}]{Astudillo17}
{Astudillo-Defru}, N., {Delfosse}, X., {Bonfils}, X., {et~al.} 2017, \aap, 600,
  A13

\bibitem[{{Baliunas} {et~al.}(1995){Baliunas}, {Donahue}, {Soon}, {Horne},
  {Frazer}, {Woodard-Eklund}, {Bradford}, {Rao}, {Wilson}, {Zhang}, \&
  {Bennett}}]{Baliunas95}
{Baliunas}, S.~L., {Donahue}, R.~A., {Soon}, W.~H., {et~al.} 1995, ApJ, 438,
  269

\bibitem[{Barnes \& Evans(1976)}]{Barnes76}
Barnes, T.~G. \& Evans, D.~S. 1976, \mnras, 174, 489

\bibitem[{{Berger} {et~al.}(2006){Berger}, {Gies}, {McAlister}, {ten
  Brummelaar}, {Henry}, {Sturmann}, {Sturmann}, {Turner}, {Ridgway},
  {Aufdenberg}, \& {M{\'e}rand}}]{Berger06}
{Berger}, D.~H., {Gies}, D.~R., {McAlister}, H.~A., {et~al.} 2006, \apj, 644,
  475

\bibitem[{{Bonfils} {et~al.}(2013){Bonfils}, {Delfosse}, {Udry}, {Forveille},
  {Mayor}, {Perrier}, {Bouchy}, {Gillon}, {Lovis}, {Pepe}, {Queloz}, {Santos},
  {S{\'e}gransan}, \& {Bertaux}}]{Bonfils13}
{Bonfils}, X., {Delfosse}, X., {Udry}, S., {et~al.} 2013, \aap, 549, A109

\bibitem[{{Boyajian} {et~al.}(2012){Boyajian}, {von Braun}, {van Belle},
  {McAlister}, {ten Brummelaar}, {Kane}, {Muirhead}, {Jones}, {White},
  {Schaefer}, {Ciardi}, {Henry}, {L{\'o}pez-Morales}, {Ridgway}, {Gies}, {Jao},
  {Rojas-Ayala}, {Parks}, {Sturmann}, {Sturmann}, {Turner}, {Farrington},
  {Goldfinger}, \& {Berger}}]{Boyajian12}
{Boyajian}, T.~S., {von Braun}, K., {van Belle}, G., {et~al.} 2012, \apj, 757,
  112

\bibitem[{{Buccino} {et~al.}(2011){Buccino}, {D{\'{\i}}az}, {Luoni},
  {Abrevaya}, \& {Mauas}}]{Buccino11}
{Buccino}, A.~P., {D{\'{\i}}az}, R.~F., {Luoni}, M.~L., {Abrevaya}, X.~C., \&
  {Mauas}, P.~J.~D. 2011, \aj, 141, 34

\bibitem[{{Buccino} {et~al.}(2007){Buccino}, {Lemarchand}, \&
  {Mauas}}]{Buccino07}
{Buccino}, A.~P., {Lemarchand}, G.~A., \& {Mauas}, P.~J.~D. 2007, \icarus, 192,
  582

\bibitem[{{Buccino} {et~al.}(2014){Buccino}, {Petrucci}, {Jofr{\'e}}, \&
  {Mauas}}]{Buccino14}
{Buccino}, A.~P., {Petrucci}, R., {Jofr{\'e}}, E., \& {Mauas}, P.~J.~D. 2014,
  \apjl, 781, L9

\bibitem[{{Capitanio} {et~al.}(2017){Capitanio}, {Lallement}, {Vergely},
  {Elyajouri}, \& {Monreal-Ibero}}]{Capitanio17}
{Capitanio}, L., {Lallement}, R., {Vergely}, J.~L., {Elyajouri}, M., \&
  {Monreal-Ibero}, A. 2017, \aap, 606, A65

\bibitem[{{Charbonneau}(2010)}]{Charbonneau10}
{Charbonneau}, P. 2010, Living Reviews in Solar Physics, 7, 3

\bibitem[{{Cincunegui} {et~al.}(2007){Cincunegui}, {D{\'{\i}}az}, \&
  {Mauas}}]{Cincunegui07}
{Cincunegui}, C., {D{\'{\i}}az}, R.~F., \& {Mauas}, P.~J.~D. 2007, \aap, 461,
  1107

\bibitem[{{Cincunegui} \& {Mauas}(2004)}]{Cincunegui04}
{Cincunegui}, C. \& {Mauas}, P.~J.~D. 2004, A\&A, 414, 699

\bibitem[{{Cosentino} {et~al.}(2012){Cosentino}, {Lovis}, {Pepe}, {Collier
  Cameron}, {Latham}, {Molinari}, {Udry}, {Bezawada}, {Black}, {Born},
  {Buchschacher}, {Charbonneau}, {Figueira}, {Fleury}, {Galli}, {Gallie},
  {Gao}, {Ghedina}, {Gonzalez}, {Gonzalez}, {Guerra}, {Henry}, {Horne},
  {Hughes}, {Kelly}, {Lodi}, {Lunney}, {Maire}, {Mayor}, {Micela}, {Ordway},
  {Peacock}, {Phillips}, {Piotto}, {Pollacco}, {Queloz}, {Rice}, {Riverol},
  {Riverol}, {San Juan}, {Sasselov}, {Segransan}, {Sozzetti}, {Sosnowska},
  {Stobie}, {Szentgyorgyi}, {Vick}, \& {Weber}}]{Cosentino2012}
{Cosentino}, R., {Lovis}, C., {Pepe}, F., {et~al.} 2012, in Society of
  Photo-Optical Instrumentation Engineers (SPIE) Conference Series, Vol. 8446,
  Ground-based and Airborne Instrumentation for Astronomy IV, ed. I.~S.
  {McLean}, S.~K. {Ramsay}, \& H.~{Takami}, 84461V

\bibitem[{{Dekker} {et~al.}(2000){Dekker}, {D'Odorico}, {Kaufer}, {Delabre}, \&
  {Kotzlowski}}]{Dekker2000}
{Dekker}, H., {D'Odorico}, S., {Kaufer}, A., {Delabre}, B., \& {Kotzlowski}, H.
  2000, in Society of Photo-Optical Instrumentation Engineers (SPIE) Conference
  Series, Vol. 4008, Optical and IR Telescope Instrumentation and Detectors,
  ed. M.~{Iye} \& A.~F. {Moorwood}, 534--545

\bibitem[{{Di Benedetto}(2005)}]{Dibenedetto05}
{Di Benedetto}, G.~P. 2005, \mnras, 357, 174

\bibitem[{{Di Maio} {et~al.}(2020){Di Maio}, {Argiroffi}, {Micela}, {Benatti},
  {Lanza}, {Scandariato}, {Maldonado}, {Maggio}, {Affer}, \&
  {Claudi}}]{Dimaio20}
{Di Maio}, C., {Argiroffi}, C., {Micela}, G., {et~al.} 2020, \aap, 642, A53

\bibitem[{{D{\'{\i}}az} {et~al.}(2007){D{\'{\i}}az}, {Gonz{\'a}lez},
  {Cincunegui}, \& {Mauas}}]{Diaz07}
{D{\'{\i}}az}, R.~F., {Gonz{\'a}lez}, J.~F., {Cincunegui}, C., \& {Mauas},
  P.~J.~D. 2007, \aap, 474, 345

\bibitem[{{D{\'\i}ez Alonso} {et~al.}(2019){D{\'\i}ez Alonso}, {Caballero},
  {Montes}, {de Cos Juez}, {Dreizler}, {Dubois}, {Jeffers}, {Lalitha}, {Naves},
  {Reiners}, {Ribas}, {Vanaverbeke}, {Amado}, {B{\'e}jar},
  {Cort{\'e}s-Contreras}, {Herrero}, {Hidalgo}, {K{\"u}rster}, {Logie},
  {Quirrenbach}, {Rau}, {Seifert}, {Sch{\"o}fer}, \& {Tal-Or}}]{DiezAlonso19}
{D{\'\i}ez Alonso}, E., {Caballero}, J.~A., {Montes}, D., {et~al.} 2019, \aap,
  621, A126

\bibitem[{{Dressing} \& {Charbonneau}(2015)}]{DressingCharbonneau15}
{Dressing}, C.~D. \& {Charbonneau}, D. 2015, \apj, 807, 45

\bibitem[{{Duncan} {et~al.}(1991){Duncan}, {Vaughan}, {Wilson}, {Preston},
  {Frazer}, {Lanning}, {Misch}, {Mueller}, {Soyumer}, {Woodard}, {Baliunas},
  {Noyes}, {Hartmann}, {Porter}, {Zwaan}, {Middelkoop}, {Rutten}, \&
  {Mihalas}}]{Duncan91}
{Duncan}, D.~K., {Vaughan}, A.~H., {Wilson}, O.~C., {et~al.} 1991, \apjs, 76,
  383

\bibitem[{{Endl} {et~al.}(2001){Endl}, {K{\"u}rster}, {Els}, {Hatzes}, \&
  {Cochran}}]{Endl01}
{Endl}, M., {K{\"u}rster}, M., {Els}, S., {Hatzes}, A.~P., \& {Cochran}, W.~D.
  2001, \aap, 374, 675

\bibitem[{{Flores} {et~al.}(2016){Flores}, {Gonz{\'a}lez}, {Jaque Arancibia},
  {Buccino}, \& {Saffe}}]{Flores16}
{Flores}, M., {Gonz{\'a}lez}, J.~F., {Jaque Arancibia}, M., {Buccino}, A., \&
  {Saffe}, C. 2016, \aap, 589, A135

\bibitem[{{Flores} {et~al.}(2018){Flores}, {Gonz{\'a}lez}, {Jaque Arancibia},
  {Saffe}, {Buccino}, {L{\'o}pez}, {Iba{\~n}ez Bustos}, \&
  {Miquelarena}}]{Flores18}
{Flores}, M., {Gonz{\'a}lez}, J.~F., {Jaque Arancibia}, M., {et~al.} 2018,
  \aap, 620, A34

\bibitem[{{Gaia Collaboration} {et~al.}(2023){Gaia Collaboration}, {Vallenari},
  {Brown}, {Prusti}, {de Bruijne}, {Arenou}, {Babusiaux}, {Biermann},
  {Creevey}, {Ducourant}, {Evans}, {Eyer}, {Guerra}, {Hutton}, {Jordi},
  {Klioner}, {Lammers}, {Lindegren}, {Luri}, {Mignard}, {Panem}, {Pourbaix},
  {Randich}, {Sartoretti}, {Soubiran}, {Tanga}, {Walton}, {Bailer-Jones},
  {Bastian}, {Drimmel}, {Jansen}, {Katz}, {Lattanzi}, {van Leeuwen}, {Bakker},
  {Cacciari}, {Casta{\~n}eda}, {De Angeli}, {Fabricius}, {Fouesneau},
  {Fr{\'e}mat}, {Galluccio}, {Guerrier}, {Heiter}, {Masana}, {Messineo},
  {Mowlavi}, {Nicolas}, {Nienartowicz}, {Pailler}, {Panuzzo}, {Riclet}, {Roux},
  {Seabroke}, {Sordo}, {Th{\'e}venin}, {Gracia-Abril}, {Portell}, {Teyssier},
  {Altmann}, {Andrae}, {Audard}, {Bellas-Velidis}, {Benson}, {Berthier},
  {Blomme}, {Burgess}, {Busonero}, {Busso}, {C{\'a}novas}, {Carry}, {Cellino},
  {Cheek}, {Clementini}, {Damerdji}, {Davidson}, {de Teodoro}, {Nu{\~n}ez
  Campos}, {Delchambre}, {Dell'Oro}, {Esquej}, {Fern{\'a}ndez-Hern{\'a}ndez},
  {Fraile}, {Garabato}, {Garc{\'\i}a-Lario}, {Gosset}, {Haigron}, {Halbwachs},
  {Hambly}, {Harrison}, {Hern{\'a}ndez}, {Hestroffer}, {Hodgkin}, {Holl},
  {Jan{\ss}en}, {Jevardat de Fombelle}, {Jordan}, {Krone-Martins}, {Lanzafame},
  {L{\"o}ffler}, {Marchal}, {Marrese}, {Moitinho}, {Muinonen}, {Osborne},
  {Pancino}, {Pauwels}, {Recio-Blanco}, {Reyl{\'e}}, {Riello}, {Rimoldini},
  {Roegiers}, {Rybizki}, {Sarro}, {Siopis}, {Smith}, {Sozzetti}, {Utrilla},
  {van Leeuwen}, {Abbas}, {{\'A}brah{\'a}m}, {Abreu Aramburu}, {Aerts},
  {Aguado}, {Ajaj}, {Aldea-Montero}, {Altavilla}, {{\'A}lvarez}, {Alves},
  {Anders}, {Anderson}, {Anglada Varela}, {Antoja}, {Baines}, {Baker},
  {Balaguer-N{\'u}{\~n}ez}, {Balbinot}, {Balog}, {Barache}, {Barbato},
  {Barros}, {Barstow}, {Bartolom{\'e}}, {Bassilana}, {Bauchet}, {Becciani},
  {Bellazzini}, {Berihuete}, {Bernet}, {Bertone}, {Bianchi}, {Binnenfeld},
  {Blanco-Cuaresma}, {Blazere}, {Boch}, {Bombrun}, {Bossini}, {Bouquillon},
  {Bragaglia}, {Bramante}, {Breedt}, {Bressan}, {Brouillet}, {Brugaletta},
  {Bucciarelli}, {Burlacu}, {Butkevich}, {Buzzi}, {Caffau}, {Cancelliere},
  {Cantat-Gaudin}, {Carballo}, {Carlucci}, {Carnerero}, {Carrasco},
  {Casamiquela}, {Castellani}, {Castro-Ginard}, {Chaoul}, {Charlot}, {Chemin},
  {Chiaramida}, {Chiavassa}, {Chornay}, {Comoretto}, {Contursi}, {Cooper},
  {Cornez}, {Cowell}, {Crifo}, {Cropper}, {Crosta}, {Crowley}, {Dafonte},
  {Dapergolas}, {David}, {David}, {de Laverny}, {De Luise}, {De March}, {De
  Ridder}, {de Souza}, {de Torres}, {del Peloso}, {del Pozo}, {Delbo},
  {Delgado}, {Delisle}, {Demouchy}, {Dharmawardena}, {Di Matteo}, {Diakite},
  {Diener}, {Distefano}, {Dolding}, {Edvardsson}, {Enke}, {Fabre}, {Fabrizio},
  {Faigler}, {Fedorets}, {Fernique}, {Fienga}, {Figueras}, {Fournier},
  {Fouron}, {Fragkoudi}, {Gai}, {Garcia-Gutierrez}, {Garcia-Reinaldos},
  {Garc{\'\i}a-Torres}, {Garofalo}, {Gavel}, {Gavras}, {Gerlach}, {Geyer},
  {Giacobbe}, {Gilmore}, {Girona}, {Giuffrida}, {Gomel}, {Gomez},
  {Gonz{\'a}lez-N{\'u}{\~n}ez}, {Gonz{\'a}lez-Santamar{\'\i}a},
  {Gonz{\'a}lez-Vidal}, {Granvik}, {Guillout}, {Guiraud},
  {Guti{\'e}rrez-S{\'a}nchez}, {Guy}, {Hatzidimitriou}, {Hauser}, {Haywood},
  {Helmer}, {Helmi}, {Sarmiento}, {Hidalgo}, {Hilger}, {H{\l}adczuk}, {Hobbs},
  {Holland}, {Huckle}, {Jardine}, {Jasniewicz}, {Jean-Antoine Piccolo},
  {Jim{\'e}nez-Arranz}, {Jorissen}, {Juaristi Campillo}, {Julbe}, {Karbevska},
  {Kervella}, {Khanna}, {Kontizas}, {Kordopatis}, {Korn}, {K{\'o}sp{\'a}l},
  {Kostrzewa-Rutkowska}, {Kruszy{\'n}ska}, {Kun}, {Laizeau}, {Lambert},
  {Lanza}, {Lasne}, {Le Campion}, {Lebreton}, {Lebzelter}, {Leccia}, {Leclerc},
  {Lecoeur-Taibi}, {Liao}, {Licata}, {Lindstr{\o}m}, {Lister}, {Livanou},
  {Lobel}, {Lorca}, {Loup}, {Madrero Pardo}, {Magdaleno Romeo}, {Managau},
  {Mann}, {Manteiga}, {Marchant}, {Marconi}, {Marcos}, {Marcos Santos},
  {Mar{\'\i}n Pina}, {Marinoni}, {Marocco}, {Marshall}, {Martin Polo},
  {Mart{\'\i}n-Fleitas}, {Marton}, {Mary}, {Masip}, {Massari},
  {Mastrobuono-Battisti}, {Mazeh}, {McMillan}, {Messina}, {Michalik}, {Millar},
  {Mints}, {Molina}, {Molinaro}, {Moln{\'a}r}, {Monari}, {Mongui{\'o}},
  {Montegriffo}, {Montero}, {Mor}, {Mora}, {Morbidelli}, {Morel}, {Morris},
  {Muraveva}, {Murphy}, {Musella}, {Nagy}, {Noval}, {Oca{\~n}a}, {Ogden},
  {Ordenovic}, {Osinde}, {Pagani}, {Pagano}, {Palaversa}, {Palicio},
  {Pallas-Quintela}, {Panahi}, {Payne-Wardenaar}, {Pe{\~n}alosa Esteller},
  {Penttil{\"a}}, {Pichon}, {Piersimoni}, {Pineau}, {Plachy}, {Plum}, {Poggio},
  {Pr{\v{s}}a}, {Pulone}, {Racero}, {Ragaini}, {Rainer}, {Raiteri}, {Rambaux},
  {Ramos}, {Ramos-Lerate}, {Re Fiorentin}, {Regibo}, {Richards}, {Rios Diaz},
  {Ripepi}, {Riva}, {Rix}, {Rixon}, {Robichon}, {Robin}, {Robin}, {Roelens},
  {Rogues}, {Rohrbasser}, {Romero-G{\'o}mez}, {Rowell}, {Royer}, {Ruz Mieres},
  {Rybicki}, {Sadowski}, {S{\'a}ez N{\'u}{\~n}ez}, {Sagrist{\`a} Sell{\'e}s},
  {Sahlmann}, {Salguero}, {Samaras}, {Sanchez Gimenez}, {Sanna},
  {Santove{\~n}a}, {Sarasso}, {Schultheis}, {Sciacca}, {Segol}, {Segovia},
  {S{\'e}gransan}, {Semeux}, {Shahaf}, {Siddiqui}, {Siebert}, {Siltala},
  {Silvelo}, {Slezak}, {Slezak}, {Smart}, {Snaith}, {Solano}, {Solitro},
  {Souami}, {Souchay}, {Spagna}, {Spina}, {Spoto}, {Steele},
  {Steidelm{\"u}ller}, {Stephenson}, {S{\"u}veges}, {Surdej}, {Szabados},
  {Szegedi-Elek}, {Taris}, {Taylor}, {Teixeira}, {Tolomei}, {Tonello}, {Torra},
  {Torra}, {Torralba Elipe}, {Trabucchi}, {Tsounis}, {Turon}, {Ulla}, {Unger},
  {Vaillant}, {van Dillen}, {van Reeven}, {Vanel}, {Vecchiato}, {Viala},
  {Vicente}, {Voutsinas}, {Weiler}, {Wevers}, {Wyrzykowski}, {Yoldas}, {Yvard},
  {Zhao}, {Zorec}, {Zucker}, \& {Zwitter}}]{GaiaDR323}
{Gaia Collaboration}, {Vallenari}, A., {Brown}, A.~G.~A., {et~al.} 2023, \aap,
  674, A1

\bibitem[{{Gaidos} {et~al.}(2014){Gaidos}, {Mann}, {L{\'e}pine}, {Buccino},
  {James}, {Ansdell}, {Petrucci}, {Mauas}, \& {Hilton}}]{Gaidos14}
{Gaidos}, E., {Mann}, A.~W., {L{\'e}pine}, S., {et~al.} 2014, \mnras, 443, 2561

\bibitem[{{Gent} {et~al.}(2022){Gent}, {Bergemann}, {Serenelli}, {Casagrande},
  {Gerber}, {Heiter}, {Kovalev}, {Morel}, {Nardetto}, {Adibekyan}, {Silva
  Aguirre}, {Asplund}, {Belkacem}, {del Burgo}, {Bigot}, {Chiavassa},
  {Rodr{\'\i}guez D{\'\i}az}, {Goupil}, {Gonz{\'a}lez Hern{\'a}ndez},
  {Mourard}, {Merle}, {M{\'e}sz{\'a}ros}, {Marshall}, {Ouazzani}, {Plez},
  {Reese}, {Trampedach}, \& {Tsantaki}}]{Gent22}
{Gent}, M.~R., {Bergemann}, M., {Serenelli}, A., {et~al.} 2022, \aap, 658, A147

\bibitem[{{Gomes da Silva} {et~al.}(2011){Gomes da Silva}, {Santos}, {Bonfils},
  {Delfosse}, {Forveille}, \& {Udry}}]{GomesdaSilva11}
{Gomes da Silva}, J., {Santos}, N.~C., {Bonfils}, X., {et~al.} 2011, \aap, 534,
  A30

\bibitem[{{Gomes da Silva} {et~al.}(2012){Gomes da Silva}, {Santos}, {Bonfils},
  {Delfosse}, {Forveille}, {Udry}, {Dumusque}, \& {Lovis}}]{GomesdaSilva12}
{Gomes da Silva}, J., {Santos}, N.~C., {Bonfils}, X., {et~al.} 2012, \aap, 541,
  A9

\bibitem[{{G{\"u}nther} {et~al.}(2020){G{\"u}nther}, {Zhan}, {Seager},
  {Rimmer}, {Ranjan}, {Stassun}, {Oelkers}, {Daylan}, {Newton}, {Kristiansen},
  {Olah}, {Gillen}, {Rappaport}, {Ricker}, {Vanderspek}, {Latham}, {Winn},
  {Jenkins}, {Glidden}, {Fausnaugh}, {Levine}, {Dittmann}, {Quinn},
  {Krishnamurthy}, \& {Ting}}]{Gunther20}
{G{\"u}nther}, M.~N., {Zhan}, Z., {Seager}, S., {et~al.} 2020, \aj, 159, 60

\bibitem[{{Iba{\~n}ez Bustos} {et~al.}(2023){Iba{\~n}ez Bustos}, {Buccino},
  {Flores}, {Martinez}, \& {Mauas}}]{Ibanez23}
{Iba{\~n}ez Bustos}, R.~V., {Buccino}, A.~P., {Flores}, M., {Martinez}, C.~F.,
  \& {Mauas}, P.~J.~D. 2023, \aap, 672, A37

\bibitem[{{Iba{\~n}ez Bustos} {et~al.}(2019{\natexlab{a}}){Iba{\~n}ez Bustos},
  {Buccino}, {Flores}, {Martinez}, {Maizel}, {Messina}, \& {Mauas}}]{Ibanez18}
{Iba{\~n}ez Bustos}, R.~V., {Buccino}, A.~P., {Flores}, M., {et~al.}
  2019{\natexlab{a}}, \mnras, 483, 1159

\bibitem[{{Iba{\~n}ez Bustos} {et~al.}(2019{\natexlab{b}}){Iba{\~n}ez Bustos},
  {Buccino}, {Flores}, \& {Mauas}}]{Ibanez19}
{Iba{\~n}ez Bustos}, R.~V., {Buccino}, A.~P., {Flores}, M., \& {Mauas},
  P.~J.~D. 2019{\natexlab{b}}, \aap, 628, L1

\bibitem[{{Iba{\~n}ez Bustos} {et~al.}(2020){Iba{\~n}ez Bustos}, {Buccino},
  {Messina}, {Lanza}, \& {Mauas}}]{Ibanez20}
{Iba{\~n}ez Bustos}, R.~V., {Buccino}, A.~P., {Messina}, S., {Lanza}, A.~F., \&
  {Mauas}, P.~J.~D. 2020, \aap, 644, A2

\bibitem[{{Kaufer} {et~al.}(1999){Kaufer}, {Stahl}, {Tubbesing},
  {N{\o}rregaard}, {Avila}, {Francois}, {Pasquini}, \& {Pizzella}}]{Kaufer99}
{Kaufer}, A., {Stahl}, O., {Tubbesing}, S., {et~al.} 1999, The Messenger, 95, 8

\bibitem[{{Kervella} {et~al.}(2004){Kervella}, {Th{\'e}venin}, {Di Folco}, \&
  {S{\'e}gransan}}]{Kervella04}
{Kervella}, P., {Th{\'e}venin}, F., {Di Folco}, E., \& {S{\'e}gransan}, D.
  2004, \aap, 426, 297

\bibitem[{{Kiman} {et~al.}(2024){Kiman}, {Brandt}, {Faherty}, \&
  {Popinchalk}}]{Kiman24}
{Kiman}, R., {Brandt}, T.~D., {Faherty}, J.~K., \& {Popinchalk}, M. 2024, \aj,
  168, 126

\bibitem[{{Kiraga} \& {Stepien}(2007)}]{Kiraga07}
{Kiraga}, M. \& {Stepien}, K. 2007, Acta Astron., 57, 149

\bibitem[{{Lafarga} {et~al.}(2021){Lafarga}, {Ribas}, {Reiners}, {Quirrenbach},
  {Amado}, {Caballero}, {Azzaro}, {B{\'e}jar}, {Cort{\'e}s-Contreras},
  {Dreizler}, {Hatzes}, {Henning}, {Jeffers}, {Kaminski}, {K{\"u}rster},
  {Montes}, {Morales}, {Oshagh}, {Rodr{\'\i}guez-L{\'o}pez}, {Sch{\"o}fer},
  {Schweitzer}, \& {Zechmeister}}]{Lafarga21}
{Lafarga}, M., {Ribas}, I., {Reiners}, A., {et~al.} 2021, \aap, 652, A28

\bibitem[{{Lallement} {et~al.}(2014){Lallement}, {Vergely}, {Valette},
  {Puspitarini}, {Eyer}, \& {Casagrande}}]{Lallement14}
{Lallement}, R., {Vergely}, J.~L., {Valette}, B., {et~al.} 2014, \aap, 561, A91

\bibitem[{{Lamm} {et~al.}(2004){Lamm}, {Bailer-Jones}, {Mundt}, {Herbst}, \&
  {Scholz}}]{Lamm04}
{Lamm}, M.~H., {Bailer-Jones}, C.~A.~L., {Mundt}, R., {Herbst}, W., \&
  {Scholz}, A. 2004, \aap, 417, 557

\bibitem[{{Ligi} {et~al.}(2016){Ligi}, {Creevey}, {Mourard}, {Crida},
  {Lagrange}, {Nardetto}, {Perraut}, {Schultheis}, {Tallon-Bosc}, \& {ten
  Brummelaar}}]{Ligi16}
{Ligi}, R., {Creevey}, O., {Mourard}, D., {et~al.} 2016, \aap, 586, A94

\bibitem[{{Lovis} {et~al.}(2011){Lovis}, {Dumusque}, {Santos}, {Bouchy},
  {Mayor}, {Pepe}, {Queloz}, {S{\'e}gransan}, \& {Udry}}]{Lovis11}
{Lovis}, C., {Dumusque}, X., {Santos}, N.~C., {et~al.} 2011, arXiv e-prints,
  arXiv:1107.5325

\bibitem[{{Mamajek} {et~al.}(2015){Mamajek}, {Torres}, {Prsa}, {Harmanec},
  {Asplund}, {Bennett}, {Capitaine}, {Christensen-Dalsgaard}, {Depagne},
  {Folkner}, {Haberreiter}, {Hekker}, {Hilton}, {Kostov}, {Kurtz}, {Laskar},
  {Mason}, {Milone}, {Montgomery}, {Richards}, {Schou}, \&
  {Stewart}}]{Mamajek15}
{Mamajek}, E.~E., {Torres}, G., {Prsa}, A., {et~al.} 2015, arXiv e-prints,
  arXiv:1510.06262

\bibitem[{{Mayor} {et~al.}(2003){Mayor}, {Pepe}, {Queloz}, {Bouchy},
  {Rupprecht}, {Lo Curto}, {Avila}, {Benz}, {Bertaux}, {Bonfils}, {Dall},
  {Dekker}, {Delabre}, {Eckert}, {Fleury}, {Gilliotte}, {Gojak}, {Guzman},
  {Kohler}, {Lizon}, {Longinotti}, {Lovis}, {Megevand}, {Pasquini}, {Reyes},
  {Sivan}, {Sosnowska}, {Soto}, {Udry}, {van Kesteren}, {Weber}, \&
  {Weilenmann}}]{Mayor2003}
{Mayor}, M., {Pepe}, F., {Queloz}, D., {et~al.} 2003, The Messenger, 114, 20

\bibitem[{{Meunier} {et~al.}(2024){Meunier}, {Mignon}, {Kretzschmar}, \&
  {Delfosse}}]{Meunier24}
{Meunier}, N., {Mignon}, L., {Kretzschmar}, M., \& {Delfosse}, X. 2024, \aap,
  684, A106

\bibitem[{{Mignon} {et~al.}(2023){Mignon}, {Meunier}, {Delfosse}, {Bonfils},
  {Santos}, {Forveille}, {Gaisn{\'e}}, {Astudillo-Defru}, {Lovis}, \&
  {Udry}}]{Mignon23}
{Mignon}, L., {Meunier}, N., {Delfosse}, X., {et~al.} 2023, \aap, 675, A168

\bibitem[{{Miguel} {et~al.}(2015){Miguel}, {Kaltenegger}, {Linsky}, \&
  {Rugheimer}}]{Miguel15}
{Miguel}, Y., {Kaltenegger}, L., {Linsky}, J.~L., \& {Rugheimer}, S. 2015,
  \mnras, 446, 345

\bibitem[{{Mortier} {et~al.}(2015){Mortier}, {Faria}, {Correia}, {Santerne}, \&
  {Santos}}]{Mortier15}
{Mortier}, A., {Faria}, J.~P., {Correia}, C.~M., {Santerne}, A., \& {Santos},
  N.~C. 2015, A\&A, 573, A101

\bibitem[{{Mourard} {et~al.}(2022){Mourard}, {Berio}, {Pannetier}, {Nardetto},
  {Allouche}, {Bailet}, {Dejonghe}, {Geneslay}, {Jacqmart}, {Lagarde},
  {Lecron}, {Morand}, {Rousseau}, {Salabert}, {Spang}, {Albrecht}, {Anugu},
  {Bourg{\`e}s}, {ten Brummelaar}, {Creevey}, {Deheuvels}, {Domiciano de
  Souza}, {Gies}, {Ligi}, {Mella}, {Perraut}, {Schaefer}, \&
  {Wittkowski}}]{Mourard22}
{Mourard}, D., {Berio}, P., {Pannetier}, C., {et~al.} 2022, in Society of
  Photo-Optical Instrumentation Engineers (SPIE) Conference Series, Vol. 12183,
  Optical and Infrared Interferometry and Imaging VIII, ed. A.~{M{\'e}rand},
  S.~{Sallum}, \& J.~{Sanchez-Bermudez}, 1218308

\bibitem[{{Mourard} {et~al.}(2018){Mourard}, {Nardetto}, {ten Brummelaar},
  {Bailet}, {Berio}, {Bresson}, {Cassaing}, {Clausse}, {Dejonghe}, {Lagarde},
  {Martinod}, {Michau}, {Perraut}, {Petit}, {Tallon}, \&
  {Tallon-Bosc}}]{Mourard18}
{Mourard}, D., {Nardetto}, N., {ten Brummelaar}, T., {et~al.} 2018, in Society
  of Photo-Optical Instrumentation Engineers (SPIE) Conference Series, Vol.
  10701, Optical and Infrared Interferometry and Imaging VI, ed. M.~J.
  {Creech-Eakman}, P.~G. {Tuthill}, \& A.~{M{\'e}rand}, 1070120

\bibitem[{{Newton} {et~al.}(2017){Newton}, {Irwin}, {Charbonneau}, {Berlind},
  {Calkins}, \& {Mink}}]{Newton17}
{Newton}, E.~R., {Irwin}, J., {Charbonneau}, D., {et~al.} 2017, \apj, 834, 85

\bibitem[{{Pannetier} {et~al.}(2020){Pannetier}, {Mourard}, {Berio},
  {Cassaing}, {Allouche}, {Anugu}, {Bailet}, {ten Brummelaar}, {Dejonghe},
  {Gies}, {Jocou}, {Kraus}, {Lacour}, {Lagarde}, {Le Bouquin}, {Lecron},
  {Monnier}, {Nardetto}, {Patru}, {Perraut}, {Petrov}, {Rousseau}, {Stee},
  {Sturmann}, \& {Sturmann}}]{Pannetier20}
{Pannetier}, C., {Mourard}, D., {Berio}, P., {et~al.} 2020, in Society of
  Photo-Optical Instrumentation Engineers (SPIE) Conference Series, Vol. 11446,
  Optical and Infrared Interferometry and Imaging VII, ed. P.~G. {Tuthill},
  A.~{M{\'e}rand}, \& S.~{Sallum}, 114460T

\bibitem[{{Pepe} {et~al.}(2021){Pepe}, {Cristiani}, {Rebolo}, {Santos},
  {Dekker}, {Cabral}, {Di Marcantonio}, {Figueira}, {Lo Curto}, {Lovis},
  {Mayor}, {M{\'e}gevand}, {Molaro}, {Riva}, {Zapatero Osorio}, {Amate},
  {Manescau}, {Pasquini}, {Zerbi}, {Adibekyan}, {Abreu}, {Affolter}, {Alibert},
  {Aliverti}, {Allart}, {Allende Prieto}, {{\'A}lvarez}, {Alves}, {Avila},
  {Baldini}, {Bandy}, {Barros}, {Benz}, {Bianco}, {Borsa}, {Bourrier},
  {Bouchy}, {Broeg}, {Calderone}, {Cirami}, {Coelho}, {Conconi}, {Coretti},
  {Cumani}, {Cupani}, {D'Odorico}, {Damasso}, {Deiries}, {Delabre},
  {Demangeon}, {Dumusque}, {Ehrenreich}, {Faria}, {Fragoso}, {Genolet},
  {Genoni}, {G{\'e}nova Santos}, {Gonz{\'a}lez Hern{\'a}ndez}, {Hughes},
  {Iwert}, {Kerber}, {Knudstrup}, {Landoni}, {Lavie}, {Lillo-Box}, {Lizon},
  {Maire}, {Martins}, {Mehner}, {Micela}, {Modigliani}, {Monteiro}, {Monteiro},
  {Moschetti}, {Murphy}, {Nunes}, {Oggioni}, {Oliveira}, {Oshagh}, {Pall{\'e}},
  {Pariani}, {Poretti}, {Rasilla}, {Rebord{\~a}o}, {Redaelli}, {Santana
  Tschudi}, {Santin}, {Santos}, {S{\'e}gransan}, {Schmidt}, {Segovia},
  {Sosnowska}, {Sozzetti}, {Sousa}, {Span{\`o}}, {Su{\'a}rez Mascare{\~n}o},
  {Tabernero}, {Tenegi}, {Udry}, \& {Zanutta}}]{Pepe2021}
{Pepe}, F., {Cristiani}, S., {Rebolo}, R., {et~al.} 2021, \aap, 645, A96

\bibitem[{{Pr{\v{s}}a} {et~al.}(2016){Pr{\v{s}}a}, {Harmanec}, {Torres},
  {Mamajek}, {Asplund}, {Capitaine}, {Christensen-Dalsgaard}, {Depagne},
  {Haberreiter}, {Hekker}, {Hilton}, {Kopp}, {Kostov}, {Kurtz}, {Laskar},
  {Mason}, {Milone}, {Montgomery}, {Richards}, {Schmutz}, {Schou}, \&
  {Stewart}}]{Prsa16}
{Pr{\v{s}}a}, A., {Harmanec}, P., {Torres}, G., {et~al.} 2016, \aj, 152, 41

\bibitem[{{Rabus} {et~al.}(2019){Rabus}, {Lachaume}, {Jord{\'a}n}, {Brahm},
  {Boyajian}, {von Braun}, {Espinoza}, {Berger}, {Le Bouquin}, \&
  {Absil}}]{Rabus19}
{Rabus}, M., {Lachaume}, R., {Jord{\'a}n}, A., {et~al.} 2019, \mnras, 484, 2674

\bibitem[{{Rauer} {et~al.}(2024){Rauer}, {Aerts}, {Cabrera}, {Deleuil},
  {Erikson}, {Gizon}, {Goupil}, {Heras}, {Lorenzo-Alvarez}, {Marliani},
  {Martin-Garcia}, {Mas-Hesse}, {O'Rourke}, {Osborn}, {Pagano}, {Piotto},
  {Pollacco}, {Ragazzoni}, {Ramsay}, {Udry}, {Appourchaux}, {Benz},
  {Brandeker}, {G{\"u}del}, {Janot-Pacheco}, {Kabath}, {Kjeldsen}, {Min},
  {Santos}, {Smith}, {Suarez}, {Werner}, {Aboudan}, {Abreu}, {Acu a}, {Adams},
  {Adibekyan}, {Affer}, {Agneray}, {Agnor}, {Aguirre B{\o}rsen-Koch}, {Ahmed},
  {Aigrain}, {Al-Bahlawan}, {Alcacera Gil}, {Alei}, {Alencar}, {Alexander},
  {Alfonso-Garz{\'o}n}, {Alibert}, {Allende Prieto}, {Almeida}, {Alonso
  Sobrino}, {Altavilla}, {Althaus}, {Alonso Alvarez Trujillo}, {Amarsi},
  {Ammler-von Eiff}, {Am{\^o}res}, {Andrade}, {Antoniadis-Karnavas},
  {Ant{\'o}nio}, {Aparicio del Moral}, {Appolloni}, {Arena}, {Armstrong},
  {Aroca Aliaga}, {Asplund}, {Audenaert}, {Auricchio}, {Avelino}, {Baeke},
  {Bailli{\'e}}, {Balado}, {Ballber Balaguer{\'o}}, {Balestra}, {Ball},
  {Ballans}, {Ballot}, {Barban}, {Barbary}, {Barbieri}, {Barcel{\'o} Forteza},
  {Barker}, {Barklem}, {Barnes}, {Barrado Navascues}, {Barragan}, {Baruteau},
  {Basu}, {Baudin}, {Baumeister}, {Bayliss}, {Bazot}, {Beck}, {Bedding},
  {Belkacem}, {Bellinger}, {Benatti}, {Benomar}, {B{\'e}rard}, {Bergemann},
  {Bergomi}, {Bernardo}, {Biazzo}, {Bignamini}, {Bigot}, {Billot}, {Binet},
  {Biondi}, {Biondi}, {Birch}, {Bitsch}, {Bluhm Ceballos}, {B{\'o}di},
  {Bogn{\'a}r}, {Boisse}, {Bolmont}, {Bonanno}, {Bonavita}, {Bonfanti},
  {Bonfils}, {Bonito}, {Bonomo}, {B{\"o}rner}, {Boro Saikia}, {Borreguero
  Mart{\'\i}n}, {Borsa}, {Borsato}, {Bossini}, {Bouchy}, {Bou{\'e}},
  {Boufleur}, {Boumier}, {Bourrier}, {Bowman}, {Bozzo}, {Bradley}, {Bray},
  {Bressan}, {Breton}, {Brienza}, {Brito}, {Brogi}, {Brown}, {Brown}, {Brun},
  {Bruno}, {Bruns}, {Buchhave}, {Bugnet}, {Buldgen}, {Burgess}, {Busatta},
  {Busso}, {Buzasi}, {Caballero}, {Cabral}, {Cabrero Gomez}, {Calderone},
  {Cameron}, {Cameron}, {Campante}, {Campos Gestal}, {Canto Martins}, {Cara},
  {Carone}, {Carrasco}, {Casagrande}, {Casewell}, {Cassisi}, {Castellani},
  {Castro}, {Catala}, {Catal{\'a}n Fern{\'a}ndez}, {Catelan}, {Cegla},
  {Cerruti}, {Cessa}, {Chadid}, {Chaplin}, {Charpinet}, {Chiappini},
  {Chiarucci}, {Chiavassa}, {Chinellato}, {Chirulli}, {Christensen-Dalsgaard},
  {Church}, {Claret}, {Clarke}, {Claudi}, {Clermont}, {Coelho}, {Coelho},
  {Cogato}, {Colom{\'e}}, {Condamin}, {Conde Garc{\'\i}a}, \&
  {Conseil}}]{Rauer24}
{Rauer}, H., {Aerts}, C., {Cabrera}, J., {et~al.} 2024, arXiv e-prints,
  arXiv:2406.05447

\bibitem[{{Rauer} {et~al.}(2014){Rauer}, {Catala}, {Aerts}, {Appourchaux},
  {Benz}, {Brandeker}, {Christensen-Dalsgaard}, {Deleuil}, {Gizon}, {Goupil},
  {G{\"u}del}, {Janot-Pacheco}, {Mas-Hesse}, {Pagano}, {Piotto}, {Pollacco},
  {Santos}, {Smith}, {Su{\'a}rez}, {Szab{\'o}}, {Udry}, {Adibekyan}, {Alibert},
  {Almenara}, {Amaro-Seoane}, {Eiff}, {Asplund}, {Antonello}, {Barnes},
  {Baudin}, {Belkacem}, {Bergemann}, {Bihain}, {Birch}, {Bonfils}, {Boisse},
  {Bonomo}, {Borsa}, {Brand{\~a}o}, {Brocato}, {Brun}, {Burleigh}, {Burston},
  {Cabrera}, {Cassisi}, {Chaplin}, {Charpinet}, {Chiappini}, {Church},
  {Csizmadia}, {Cunha}, {Damasso}, {Davies}, {Deeg}, {D{\'\i}az}, {Dreizler},
  {Dreyer}, {Eggenberger}, {Ehrenreich}, {Eigm{\"u}ller}, {Erikson}, {Farmer},
  {Feltzing}, {de Oliveira Fialho}, {Figueira}, {Forveille}, {Fridlund},
  {Garc{\'\i}a}, {Giommi}, {Giuffrida}, {Godolt}, {Gomes da Silva}, {Granzer},
  {Grenfell}, {Grotsch-Noels}, {G{\"u}nther}, {Haswell}, {Hatzes},
  {H{\'e}brard}, {Hekker}, {Helled}, {Heng}, {Jenkins}, {Johansen},
  {Khodachenko}, {Kislyakova}, {Kley}, {Kolb}, {Krivova}, {Kupka}, {Lammer},
  {Lanza}, {Lebreton}, {Magrin}, {Marcos-Arenal}, {Marrese}, {Marques},
  {Martins}, {Mathis}, {Mathur}, {Messina}, {Miglio}, {Montalban}, {Montalto},
  {Monteiro}, {Moradi}, {Moravveji}, {Mordasini}, {Morel}, {Mortier},
  {Nascimbeni}, {Nelson}, {Nielsen}, {Noack}, {Norton}, {Ofir}, {Oshagh},
  {Ouazzani}, {P{\'a}pics}, {Parro}, {Petit}, {Plez}, {Poretti}, {Quirrenbach},
  {Ragazzoni}, {Raimondo}, {Rainer}, {Reese}, {Redmer}, {Reffert},
  {Rojas-Ayala}, {Roxburgh}, {Salmon}, {Santerne}, {Schneider}, {Schou},
  {Schuh}, {Schunker}, {Silva-Valio}, {Silvotti}, {Skillen}, {Snellen}, {Sohl},
  {Sousa}, {Sozzetti}, {Stello}, {Strassmeier}, {{\v{S}}vanda}, {Szab{\'o}},
  {Tkachenko}, {Valencia}, {Van Grootel}, {Vauclair}, {Ventura}, {Wagner},
  {Walton}, {Weingrill}, {Werner}, {Wheatley}, \& {Zwintz}}]{Rauer14}
{Rauer}, H., {Catala}, C., {Aerts}, C., {et~al.} 2014, Experimental Astronomy,
  38, 249

\bibitem[{{Roberts} {et~al.}(1987){Roberts}, {Lehar}, \& {Dreher}}]{Roberts87}
{Roberts}, D.~H., {Lehar}, J., \& {Dreher}, J.~W. 1987, \aj, 93, 968

\bibitem[{{Robertson} {et~al.}(2014){Robertson}, {Mahadevan}, {Endl}, \&
  {Roy}}]{Robertson14b}
{Robertson}, P., {Mahadevan}, S., {Endl}, M., \& {Roy}, A. 2014, Science, 345,
  440

\bibitem[{{Rodr{\'\i}guez Mart{\'\i}nez} {et~al.}(2020){Rodr{\'\i}guez
  Mart{\'\i}nez}, {Lopez}, {Shappee}, {Schmidt}, {Jayasinghe}, {Kochanek},
  {Auchettl}, \& {Holoien}}]{Rodriguez20}
{Rodr{\'\i}guez Mart{\'\i}nez}, R., {Lopez}, L.~A., {Shappee}, B.~J., {et~al.}
  2020, \apj, 892, 144

\bibitem[{{Salsi} {et~al.}(2020){Salsi}, {Nardetto}, {Mourard}, {Creevey},
  {Huber}, {White}, {Hocd{\'e}}, {Morand}, {Tallon-Bosc}, {Farrington},
  {Chelli}, \& {Duvert}}]{Salsi20}
{Salsi}, A., {Nardetto}, N., {Mourard}, D., {et~al.} 2020, \aap, 640, A2

\bibitem[{{Salsi} {et~al.}(2021){Salsi}, {Nardetto}, {Mourard}, {Graczyk},
  {Taormina}, {Creevey}, {Hocd{\'e}}, {Morand}, {Perraut}, {Pietrzynski}, \&
  {Schaefer}}]{Salsi21}
{Salsi}, A., {Nardetto}, N., {Mourard}, D., {et~al.} 2021, \aap, 652, A26

\bibitem[{{Salsi} {et~al.}(2022){Salsi}, {Nardetto}, {Plez}, \&
  {Mourard}}]{Salsi22}
{Salsi}, A., {Nardetto}, N., {Plez}, B., \& {Mourard}, D. 2022, \aap, 662, A120

\bibitem[{{Schaefer} {et~al.}(2018){Schaefer}, {White}, {Baines}, {Boyajian},
  {ten Brummelaar}, {Farrington}, {Sturmann}, {Sturmann}, \&
  {Turner}}]{Schaefer18}
{Schaefer}, G.~H., {White}, R.~J., {Baines}, E.~K., {et~al.} 2018, \apj, 858,
  71

\bibitem[{{S{\'e}gransan} {et~al.}(2003){S{\'e}gransan}, {Kervella},
  {Forveille}, \& {Queloz}}]{Segransan03}
{S{\'e}gransan}, D., {Kervella}, P., {Forveille}, T., \& {Queloz}, D. 2003,
  \aap, 397, L5

\bibitem[{{Su{\'a}rez Mascare{\~n}o} {et~al.}(2020){Su{\'a}rez Mascare{\~n}o},
  {Faria}, {Figueira}, {Lovis}, {Damasso}, {Gonz{\'a}lez Hern{\'a}ndez},
  {Rebolo}, {Cristiani}, {Pepe}, {Santos}, {Zapatero Osorio}, {Adibekyan},
  {Hojjatpanah}, {Sozzetti}, {Murgas}, {Abreu}, {Affolter}, {Alibert},
  {Aliverti}, {Allart}, {Allende Prieto}, {Alves}, {Amate}, {Avila}, {Baldini},
  {Bandi}, {Barros}, {Bianco}, {Benz}, {Bouchy}, {Broeng}, {Cabral},
  {Calderone}, {Cirami}, {Coelho}, {Conconi}, {Coretti}, {Cumani}, {Cupani},
  {D'Odorico}, {Deiries}, {Delabre}, {Di Marcantonio}, {Dumusque},
  {Ehrenreich}, {Fragoso}, {Genolet}, {Genoni}, {G{\'e}nova Santos}, {Hughes},
  {Iwert}, {Kerber}, {Knusdstrup}, {Landoni}, {Lavie}, {Lillo-Box}, {Lizon},
  {Lo Curto}, {Maire}, {Manescau}, {Martins}, {M{\'e}gevand}, {Mehner},
  {Micela}, {Modigliani}, {Molaro}, {Monteiro}, {Monteiro}, {Moschetti},
  {Mueller}, {Nunes}, {Oggioni}, {Oliveira}, {Pall{\'e}}, {Pariani},
  {Pasquini}, {Poretti}, {Rasilla}, {Redaelli}, {Riva}, {Santana Tschudi},
  {Santin}, {Santos}, {Segovia}, {Sosnowska}, {Sousa}, {Span{\`o}}, {Tenegi},
  {Udry}, {Zanutta}, \& {Zerbi}}]{SuarezMascareno20}
{Su{\'a}rez Mascare{\~n}o}, A., {Faria}, J.~P., {Figueira}, P., {et~al.} 2020,
  \aap, 639, A77

\bibitem[{{Su{\'a}rez Mascare{\~n}o} {et~al.}(2017){Su{\'a}rez Mascare{\~n}o},
  {Gonz{\'a}lez Hern{\'a}ndez}, {Rebolo}, {Astudillo-Defru}, {Bonfils},
  {Bouchy}, {Delfosse}, {Forveille}, {Lovis}, {Mayor}, {Murgas}, {Pepe},
  {Santos}, {Udry}, {W{\"u}nsche}, \& {Velasco}}]{SuarezMascareno17}
{Su{\'a}rez Mascare{\~n}o}, A., {Gonz{\'a}lez Hern{\'a}ndez}, J.~I., {Rebolo},
  R., {et~al.} 2017, \aap, 597, A108

\bibitem[{{Su{\'a}rez Mascare{\~n}o} {et~al.}(2016){Su{\'a}rez Mascare{\~n}o},
  {Rebolo}, \& {Gonz{\'a}lez Hern{\'a}ndez}}]{SuarezMascareno16}
{Su{\'a}rez Mascare{\~n}o}, A., {Rebolo}, R., \& {Gonz{\'a}lez Hern{\'a}ndez},
  J.~I. 2016, \aap, 595, A12

\bibitem[{{Su{\'a}rez Mascare{\~n}o} {et~al.}(2015){Su{\'a}rez Mascare{\~n}o},
  {Rebolo}, {Gonz{\'a}lez Hern{\'a}ndez}, \& {Esposito}}]{SuarezMascareno15}
{Su{\'a}rez Mascare{\~n}o}, A., {Rebolo}, R., {Gonz{\'a}lez Hern{\'a}ndez},
  J.~I., \& {Esposito}, M. 2015, \mnras, 452, 2745

\bibitem[{{Toledo-Padr{\'o}n} {et~al.}(2019){Toledo-Padr{\'o}n}, {Gonz{\'a}lez
  Hern{\'a}ndez}, {Rodr{\'\i}guez-L{\'o}pez}, {Su{\'a}rez Mascare{\~n}o},
  {Rebolo}, {Butler}, {Ribas}, {Anglada-Escud{\'e}}, {Johnson}, {Reiners},
  {Caballero}, {Quirrenbach}, {Amado}, {B{\'e}jar}, {Morales}, {Perger},
  {Jeffers}, {Vogt}, {Teske}, {Shectman}, {Crane}, {D{\'\i}az}, {Arriagada},
  {Holden}, {Burt}, {Rodr{\'\i}guez}, {Herrero}, {Murgas}, {Pall{\'e}},
  {Morales}, {L{\'o}pez-Gonz{\'a}lez}, {D{\'\i}ez Alonso}, {Tuomi}, {Kiraga},
  {Engle}, {Guinan}, {Strachan}, {Aceituno}, {Aceituno}, {Casanova},
  {Mart{\'\i}n-Ruiz}, {Montes}, {Ortiz}, {Sota}, {Briol}, {Barbieri},
  {Cervini}, {Deldem}, {Dubois}, {Hambsch}, {Harris}, {Kotnik}, {Logie},
  {Lopez}, {McNeely}, {Ogmen}, {P{\'e}rez}, {Rau}, {Rodr{\'\i}guez}, {Urquijo},
  \& {Vanaverbeke}}]{Toledo19}
{Toledo-Padr{\'o}n}, B., {Gonz{\'a}lez Hern{\'a}ndez}, J.~I.,
  {Rodr{\'\i}guez-L{\'o}pez}, C., {et~al.} 2019, \mnras, 488, 5145

\bibitem[{{Vernet} {et~al.}(2011){Vernet}, {Dekker}, {D'Odorico}, {Kaper},
  {Kjaergaard}, {Hammer}, {Randich}, {Zerbi}, {Groot}, {Hjorth}, {Guinouard},
  {Navarro}, {Adolfse}, {Albers}, {Amans}, {Andersen}, {Andersen}, {Binetruy},
  {Bristow}, {Castillo}, {Chemla}, {Christensen}, {Conconi}, {Conzelmann},
  {Dam}, {de Caprio}, {de Ugarte Postigo}, {Delabre}, {di Marcantonio},
  {Downing}, {Elswijk}, {Finger}, {Fischer}, {Flores}, {Fran{\c{c}}ois},
  {Goldoni}, {Guglielmi}, {Haigron}, {Hanenburg}, {Hendriks}, {Horrobin},
  {Horville}, {Jessen}, {Kerber}, {Kern}, {Kiekebusch}, {Kleszcz}, {Klougart},
  {Kragt}, {Larsen}, {Lizon}, {Lucuix}, {Mainieri}, {Manuputy}, {Martayan},
  {Mason}, {Mazzoleni}, {Michaelsen}, {Modigliani}, {Moehler}, {M{\o}ller},
  {Norup S{\o}rensen}, {N{\o}rregaard}, {P{\'e}roux}, {Patat}, {Pena}, {Pragt},
  {Reinero}, {Rigal}, {Riva}, {Roelfsema}, {Royer}, {Sacco}, {Santin},
  {Schoenmaker}, {Spano}, {Sweers}, {Ter Horst}, {Tintori}, {Tromp}, {van
  Dael}, {van der Vliet}, {Venema}, {Vidali}, {Vinther}, {Vola}, {Winters},
  {Wistisen}, {Wulterkens}, \& {Zacchei}}]{Vernet2011}
{Vernet}, J., {Dekker}, H., {D'Odorico}, S., {et~al.} 2011, \aap, 536, A105

\bibitem[{{Vida} {et~al.}(2017){Vida}, {K{\H o}v{\'a}ri}, {P{\'a}l},
  {Ol{\'a}h}, \& {Kriskovics}}]{Vida17}
{Vida}, K., {K{\H o}v{\'a}ri}, Z., {P{\'a}l}, A., {Ol{\'a}h}, K., \&
  {Kriskovics}, L. 2017, ApJ, 841, 124

\bibitem[{Vogt {et~al.}(1994)Vogt, Allen, Bigelow, Bresee, Brown, Cantrall,
  Conrad, Couture, Delaney, Epps, Hilyard, Hilyard, Horn, Jern, Kanto, Keane,
  Kibrick, Lewis, Osborne, Pardeilhan, Pfister, Ricketts, Robinson, Stover,
  Tucker, Ward, \& Wei}]{Vogt94}
Vogt, S.~S., Allen, S.~L., Bigelow, B.~C., {et~al.} 1994, in Instrumentation in
  Astronomy VIII, ed. D.~L. Crawford \& E.~R. Craine, Vol. 2198, International
  Society for Optics and Photonics (SPIE), 362 -- 375

\bibitem[{{Wargelin} {et~al.}(2017){Wargelin}, {Saar}, {Pojma{\'n}ski},
  {Drake}, \& {Kashyap}}]{Wargelin17}
{Wargelin}, B.~J., {Saar}, S.~H., {Pojma{\'n}ski}, G., {Drake}, J.~J., \&
  {Kashyap}, V.~L. 2017, \mnras, 464, 3281

\bibitem[{{Wesselink}(1969)}]{Wesselink69}
{Wesselink}, A.~J. 1969, \mnras, 144, 297

\bibitem[{{Wright} \& {Drake}(2016)}]{Wright16}
{Wright}, N.~J. \& {Drake}, J.~J. 2016, \nat, 535, 526

\bibitem[{{Wright} {et~al.}(2011){Wright}, {Drake}, {Mamajek}, \&
  {Henry}}]{Wright11}
{Wright}, N.~J., {Drake}, J.~J., {Mamajek}, E.~E., \& {Henry}, G.~W. 2011,
  \apj, 743, 48

\bibitem[{{Wright} {et~al.}(2018){Wright}, {Newton}, {Williams}, {Drake}, \&
  {Yadav}}]{Wright18}
{Wright}, N.~J., {Newton}, E.~R., {Williams}, P.~K.~G., {Drake}, J.~J., \&
  {Yadav}, R.~K. 2018, \mnras, 479, 2351

\bibitem[{{Zechmeister} \& {K{\"u}rster}(2009)}]{Zechmeister09}
{Zechmeister}, M. \& {K{\"u}rster}, M. 2009, A\&A, 496, 577

\end{thebibliography}

\begin{appendix} 

\section{Time series} \label{aa.timeseries}

In this section we show the time series constructed in the present work for the stars in the sample. 
In Section \ref{ss.cycles} we show the time series together with their periodograms and phases folded time series for those stars that show a cyclical behaviour in their activity that is possibly associated with a potential magnetic activity cycle.
In Section \ref{ss.noncycles} we plot the time series for those stars that do not show a cyclical behaviour in their magnetic activity and we also present the time series for the stars GJ 729 and GJ 447 already reported in \cite{Ibanez20} and \cite{Ibanez19}, respectively.

The time series were constructed using our own observations from CASLEO and those public domain observations that we found in 7 different databases.
To distinguish between the different instruments, we employed the same color code for all the time series constructed: blue for CASLEO spectra, orange for HARPS, green for FEROS, red for UVES, violet for XSHOOTER, brown for HIRES, yellow for HARPS-N and light blue for ESPRESSO spectra.

We have studied each spectrum in detail analysing the different spectral lines as we already suggest in \cite{Ibanez23}.
In some of the following time series, we highlighted some points with red circles to indicate that the spectral line could be contaminated by flares and they were discarded from our analysis.

\subsection{Stars with potential magnetic activity cycles} \label{ss.cycles}

\begin{figure}[htb!]
    \centering
    \includegraphics[width=0.4\textwidth]{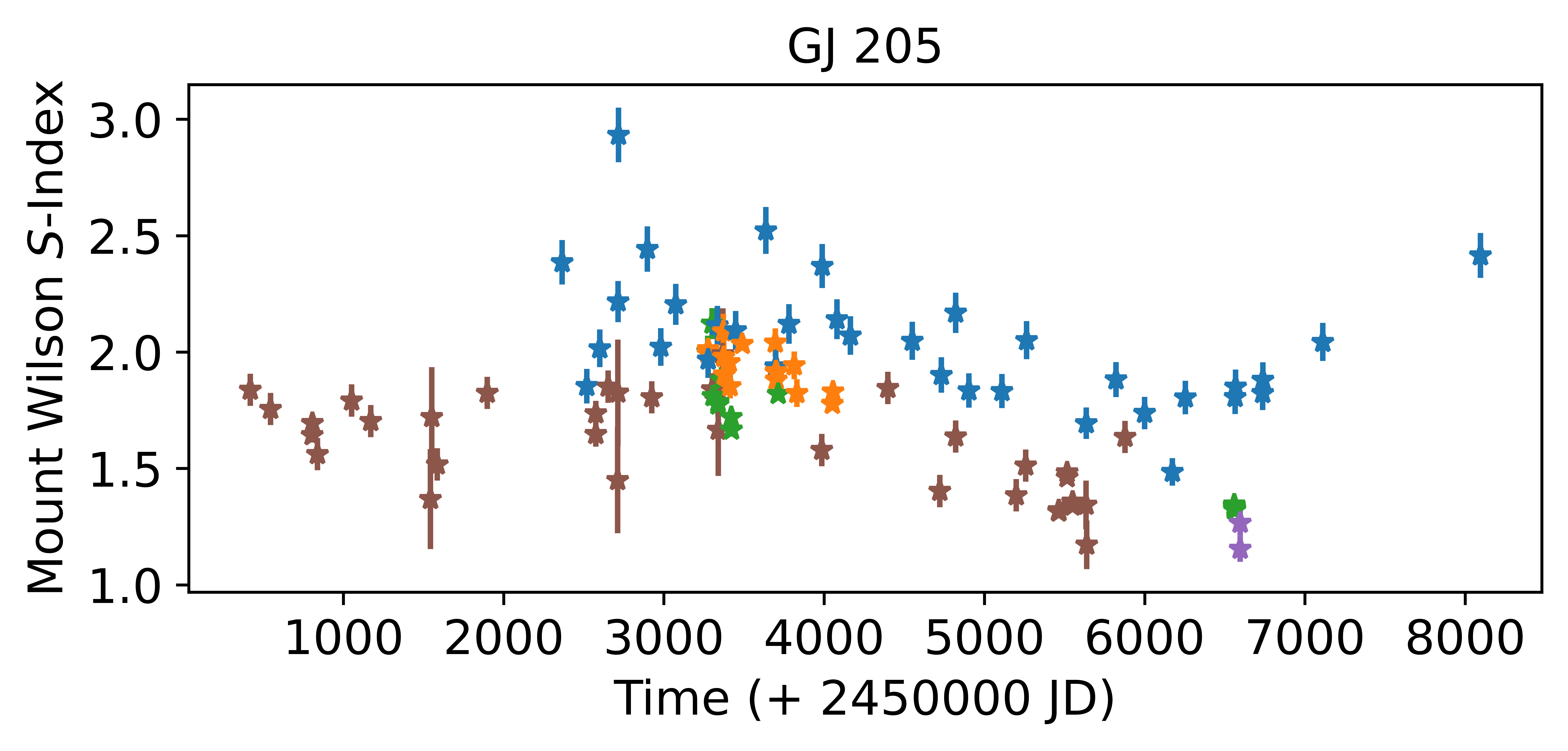}
    \includegraphics[width=0.45\textwidth]{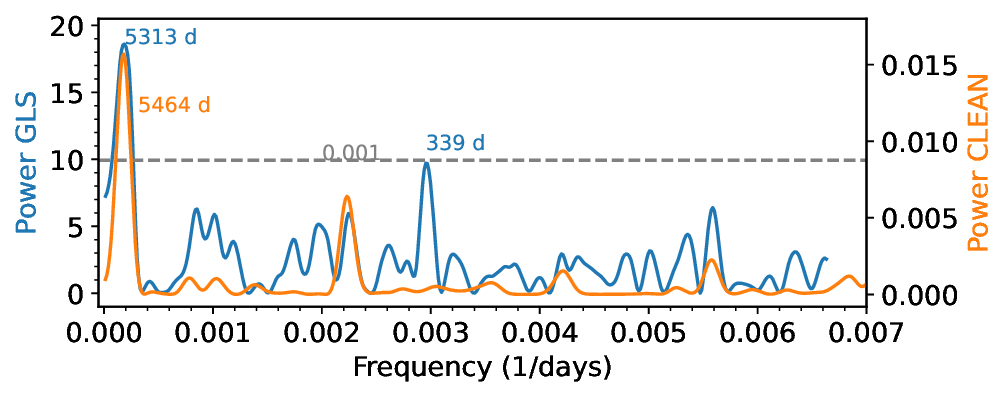}
    \includegraphics[width=0.25\textwidth]{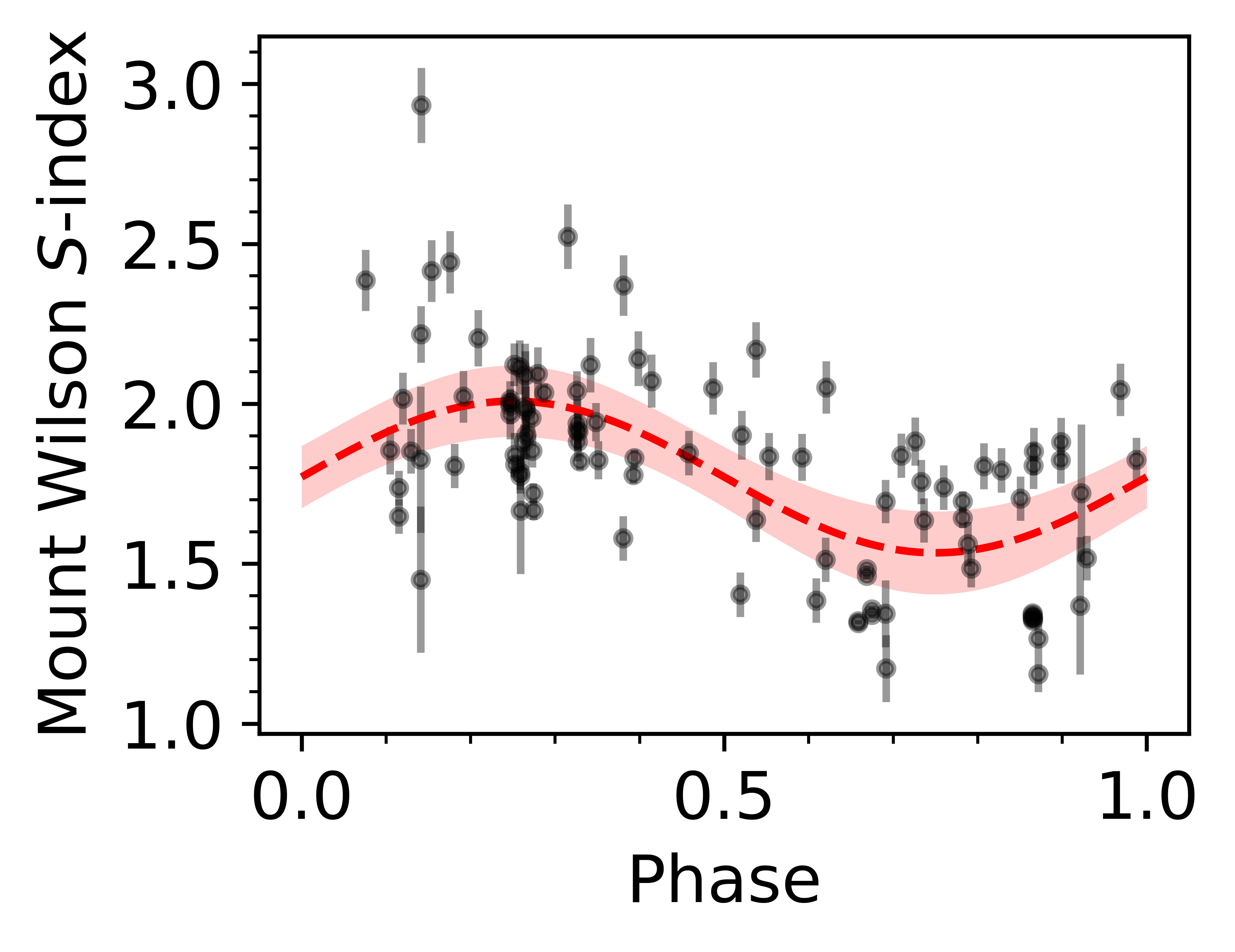}
    \caption{\textbf{GJ 205.} \textit{Top.} $S$-Index time series. \textit{Middle.} GLS and CLEAN periodograms: $P_{GLS; 1} = (5313 \pm  862)$ d and $P_{GLS; 2} = 339 \pm 3$ d with a  FAP < 0.01\% and FAP $\sim$ 0.001, respectively. From the CLEAN periodogram we have obtained as significant period $P_{CLEAN} = (5479 \pm  196)$ days. \textit{Bottom.} Phase folded time series with a $\sim5300$ day period.}
    \label{per_gj205}
\end{figure}

\begin{figure}[htb!]
\begin{center}
    \includegraphics[width=0.4\textwidth]{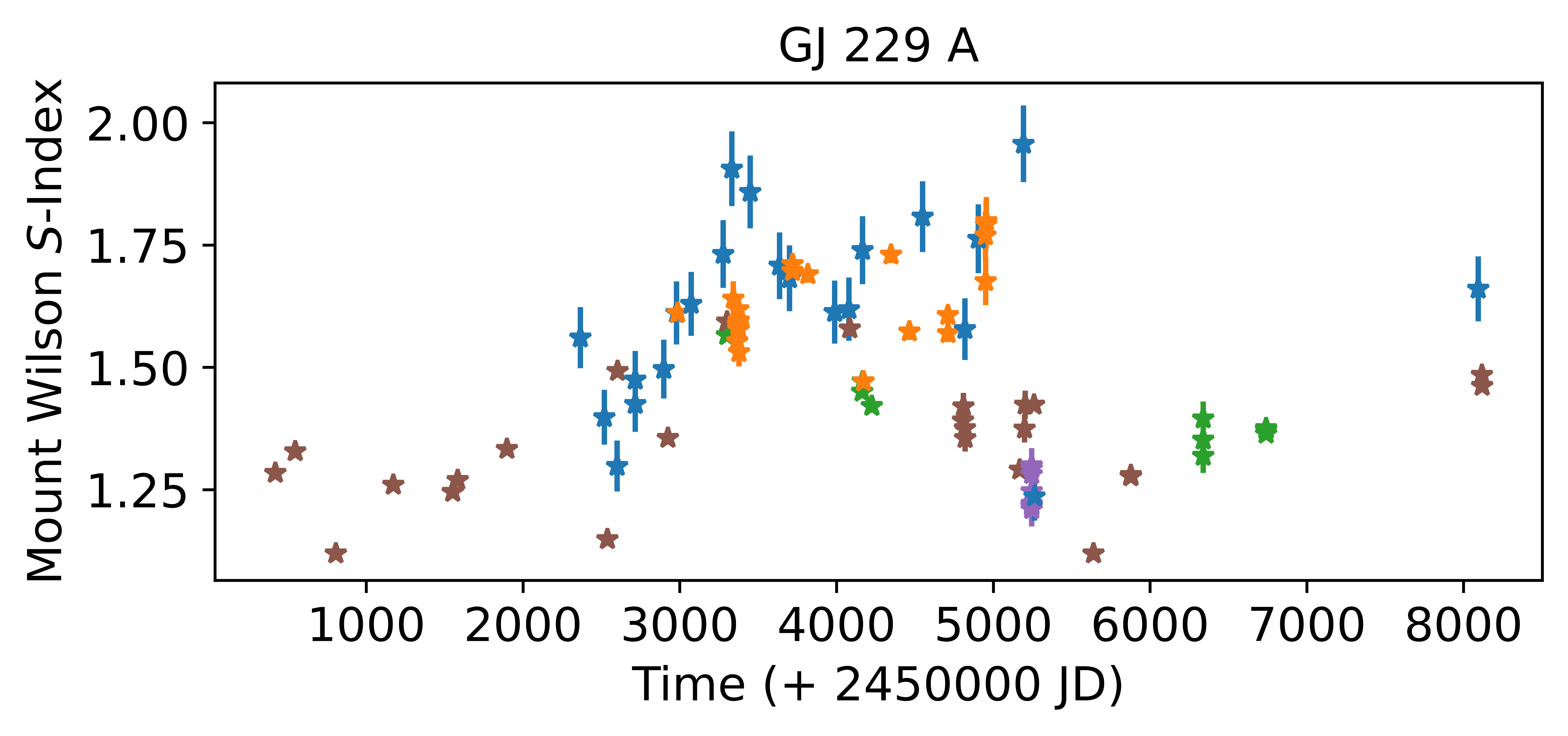}
    \includegraphics[width=0.45\textwidth]{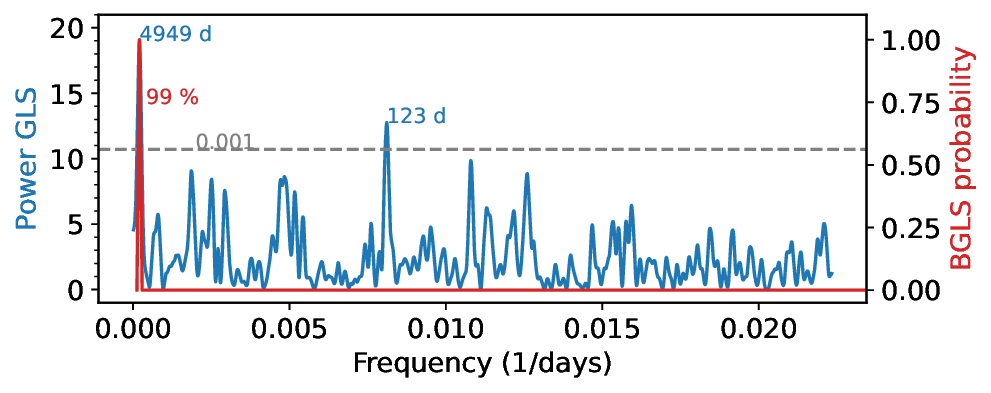}
    \includegraphics[width=0.25\textwidth]{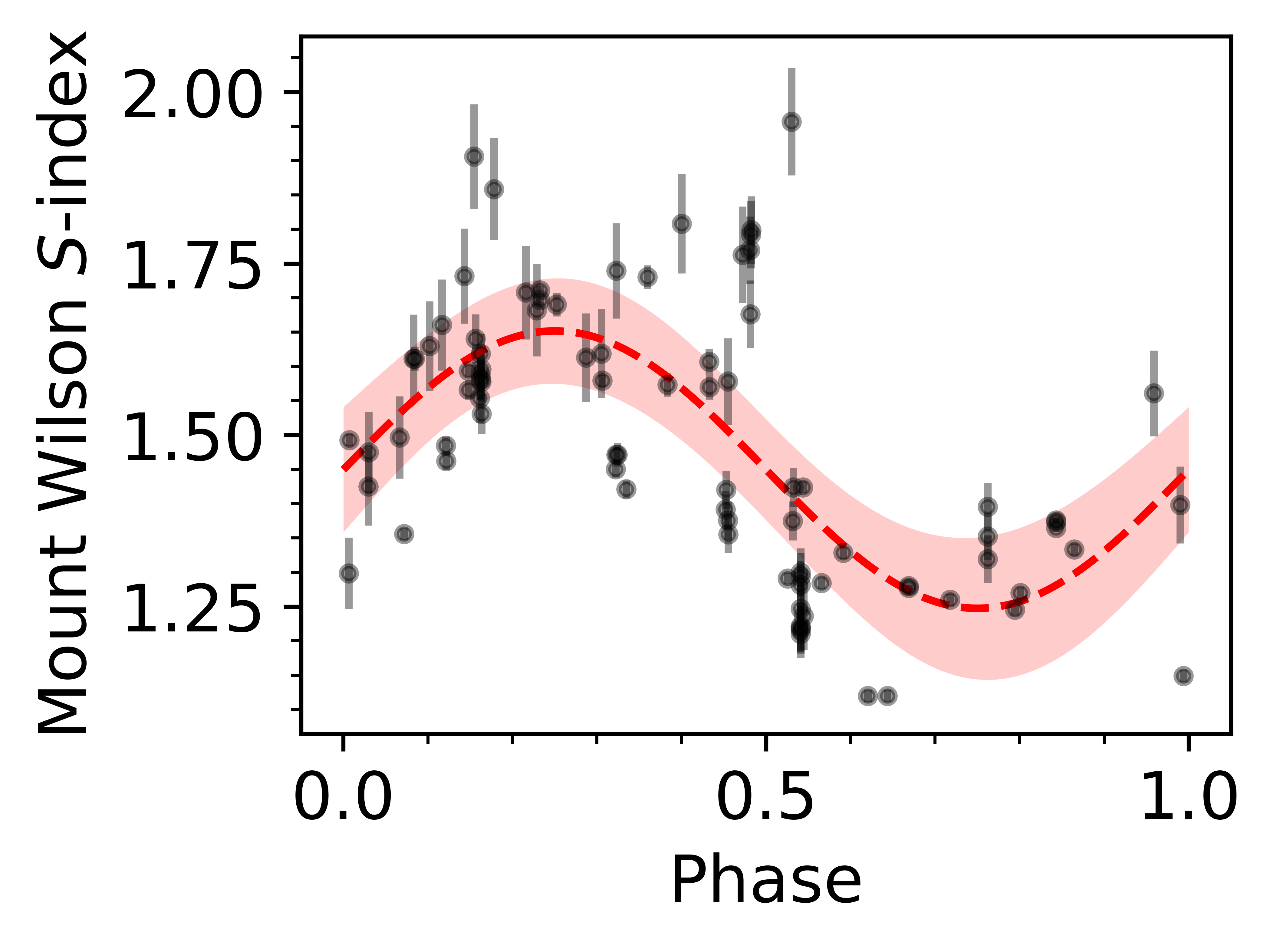}
\end{center}
\caption{Same as Fig. \ref{per_gl1} but  for \textbf{GJ 229 A.} $P_{GLS; 1} = (4949 \pm  423)$ d and $P_{GLS; 2} = (123.2 \pm 0.4)$ d with FAPs < 0.1\%.
Performing the BGLS periodogram we obtained a 99\% probability that the $\sim 4900$ d peak is present in our sample. \textit{Bottom.} Phase folded time series with a $\sim4900$ day period.}
\label{per_gj229}
\end{figure}

\begin{figure}[htb!]
\begin{center}
    \includegraphics[width=0.4\textwidth]{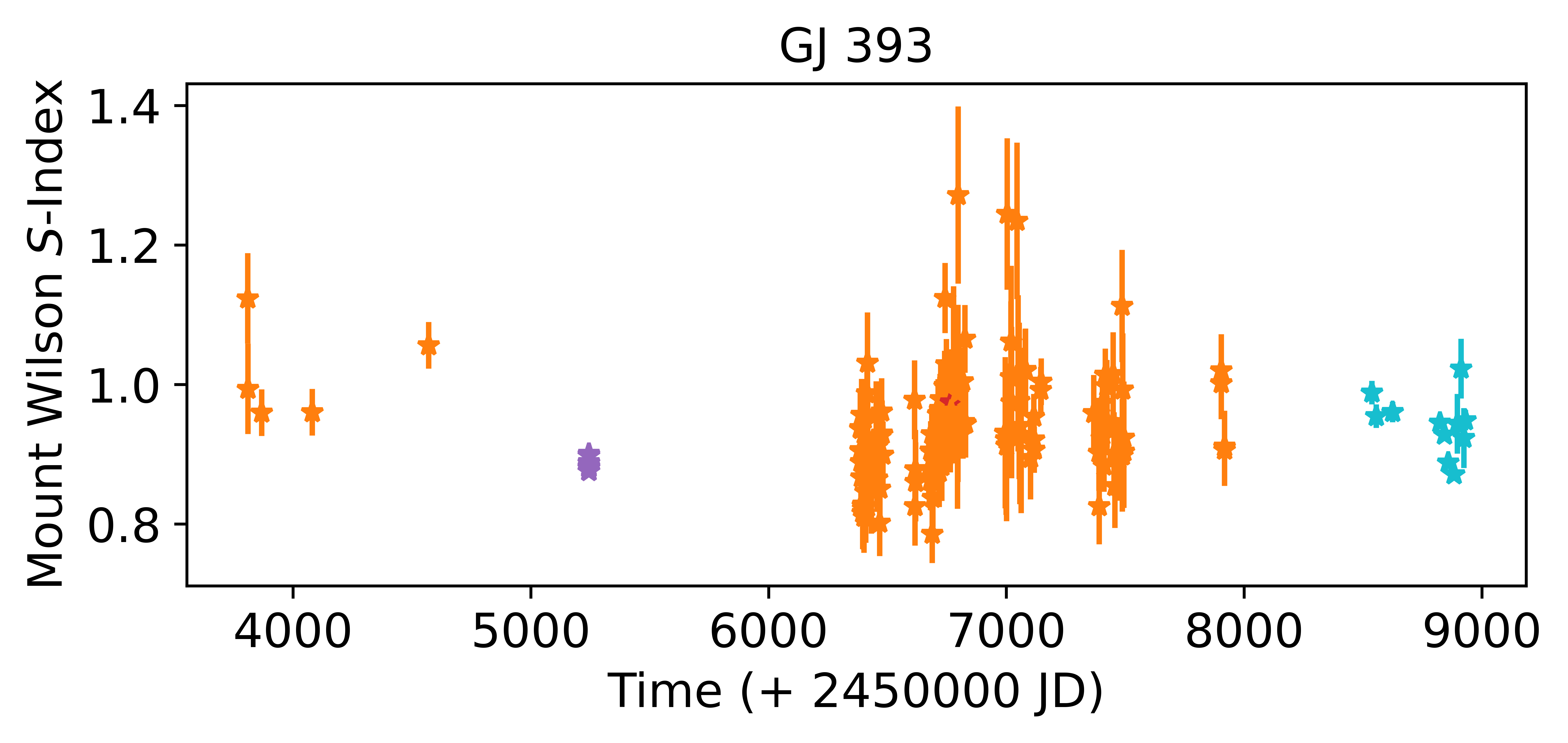}
    \includegraphics[width=0.45\textwidth]{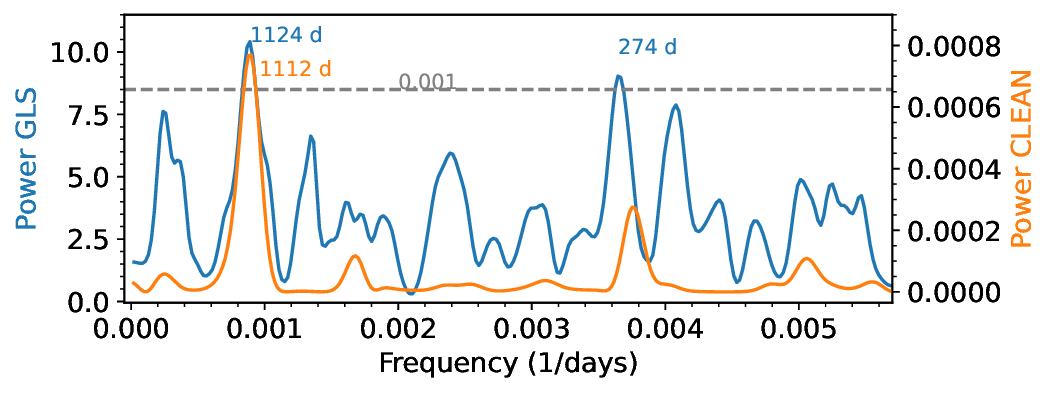}
    \includegraphics[width=0.25\textwidth]{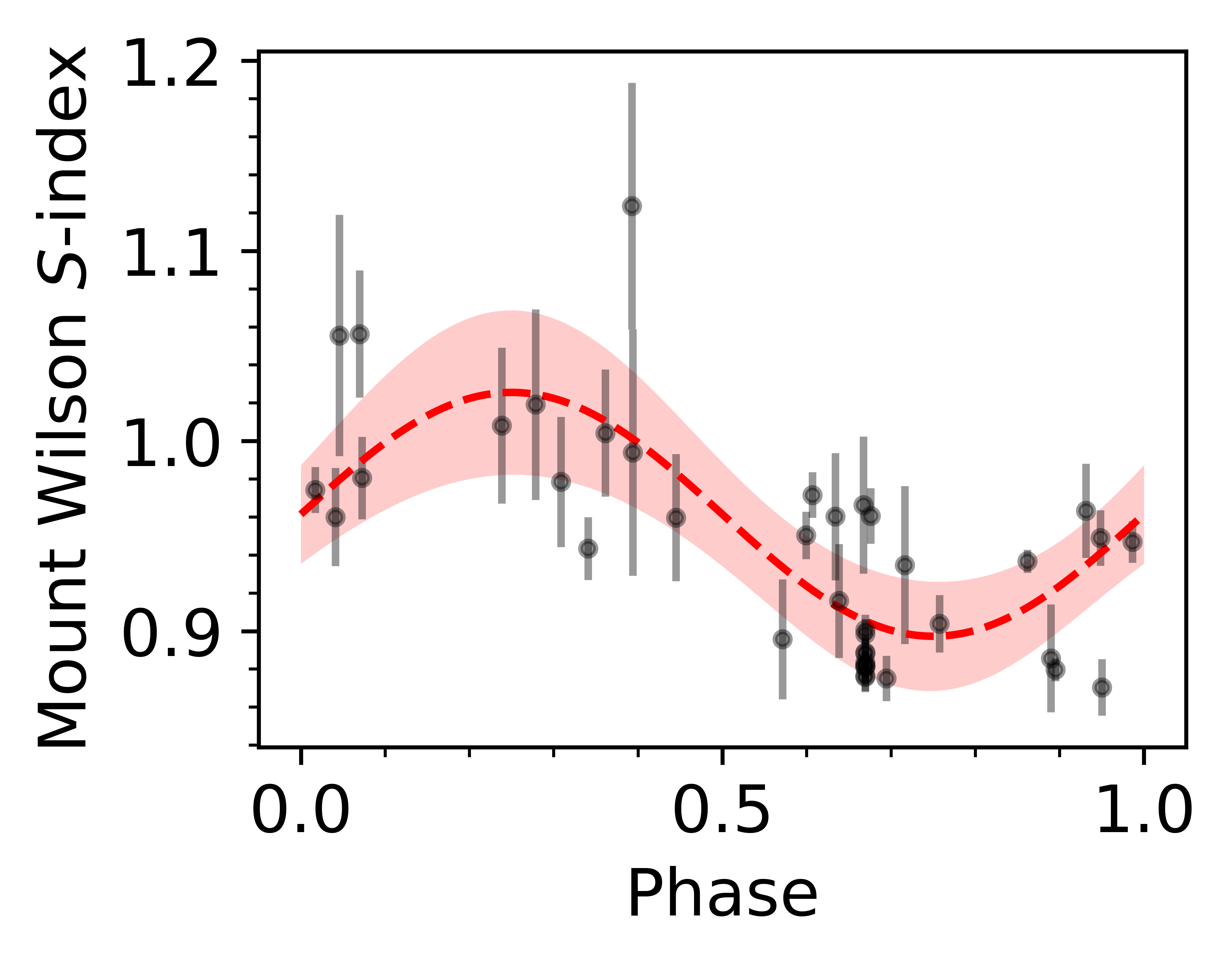}
\end{center}
\caption{Same as Fig. \ref{per_gj205} but for \textbf{GJ 393.} $P_{GLS; 1}$ = (1124 $\pm$ 29) d and $P_{GLS; 2}$ = (274 $\pm$ 2) d with FAPs $\sim 0.003$ \% and $\sim 0.04$ \%, respectively. From the CLEAN periodogram we obtained a significant period of $P_{CLEAN}$ = (1112 $\pm$ 25) d. \textit{Bottom.} Phase folded time series with a $\sim1100$ day period.}
\label{per_gl393}
\end{figure}

\begin{figure}[]
\begin{center}
    \includegraphics[width=0.4\textwidth]{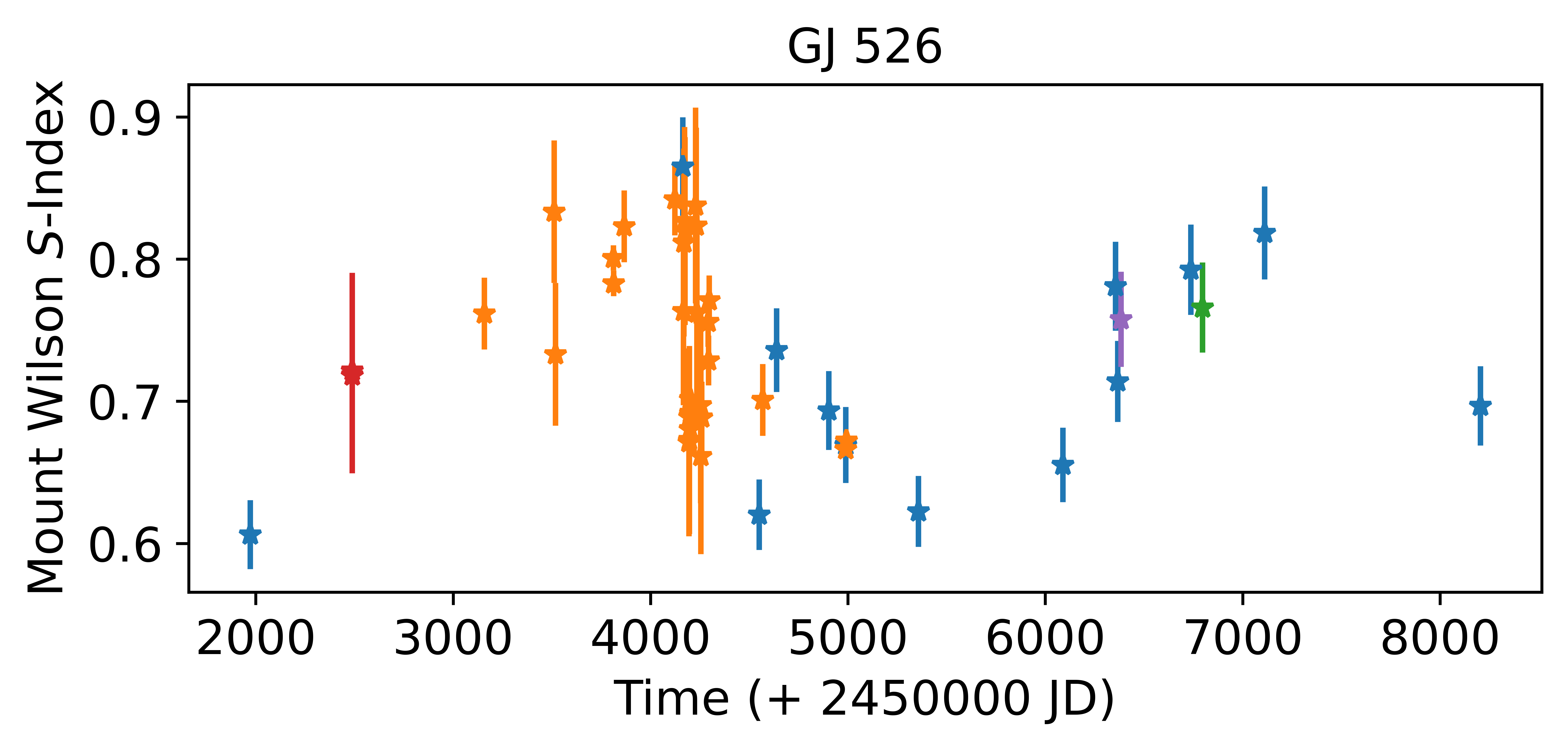}
    \includegraphics[width=0.45\textwidth]{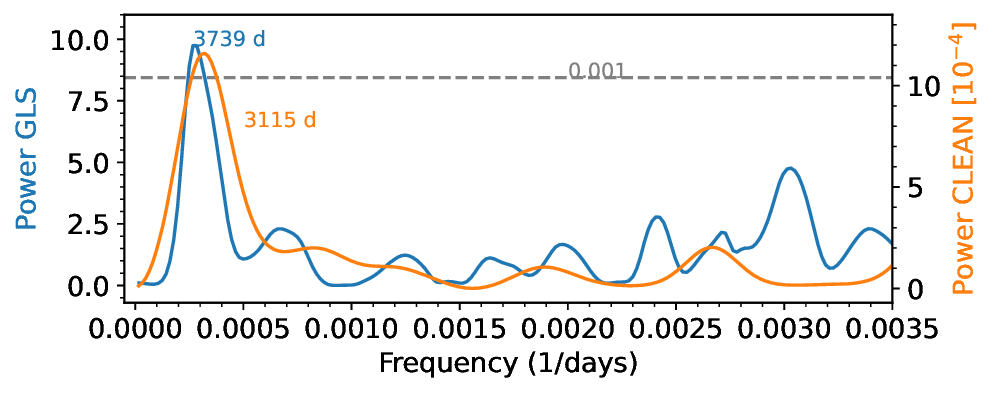}
    \includegraphics[width=0.25\textwidth]{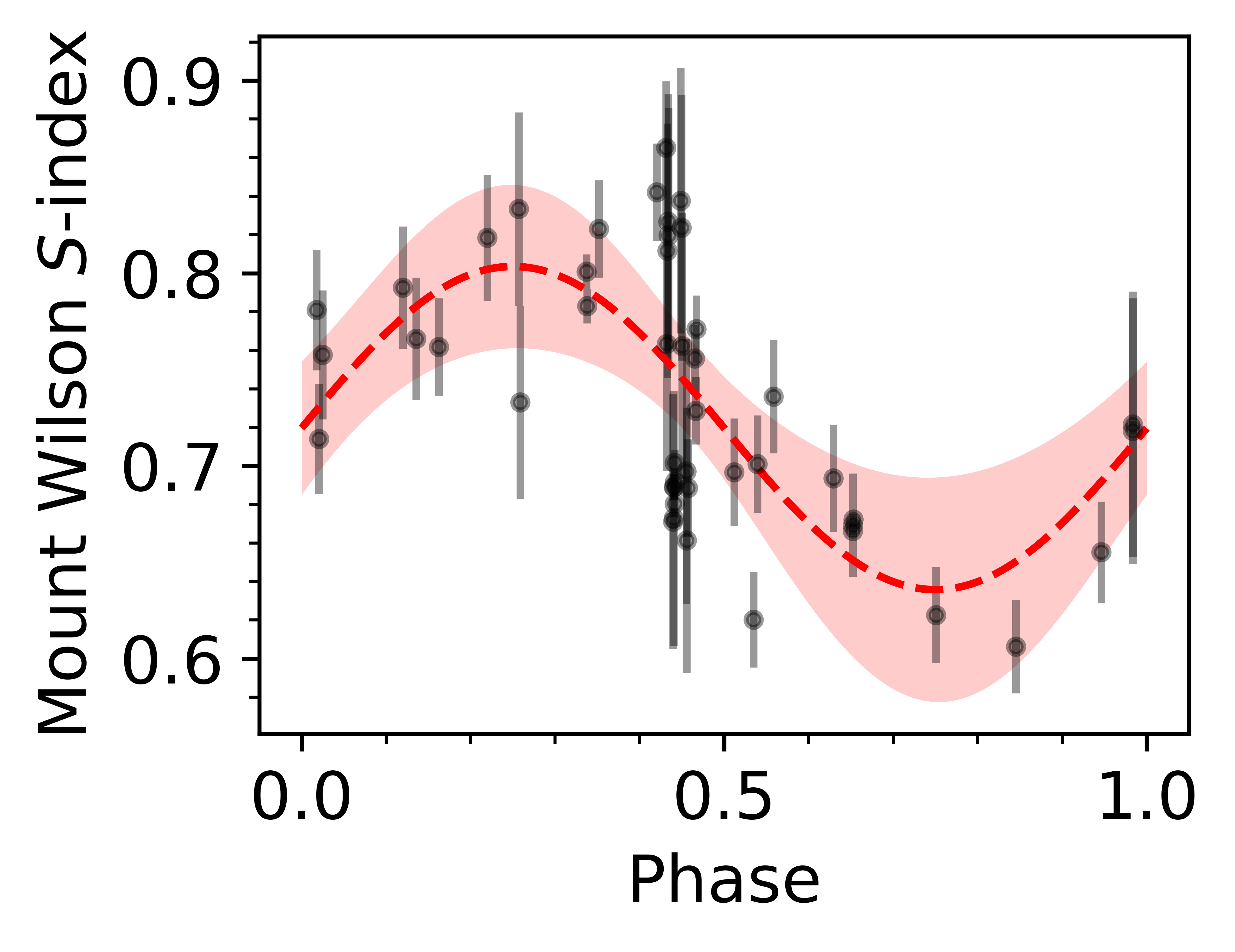}
\end{center}
\caption{Same as Fig. \ref{per_gj205} for \textbf{GJ 526.} $P_{GLS} = (3739 \pm 288)$ days with FAP of 0.013\% and $P_{CLEAN} = (3115 \pm 1336)$ days. \textit{Bottom.} Phase folded time series with a $\sim3700$ day period.} 
\label{per_gl526}
\end{figure}

\begin{figure}[htb!]
\begin{center}
    \includegraphics[width=0.4\textwidth]{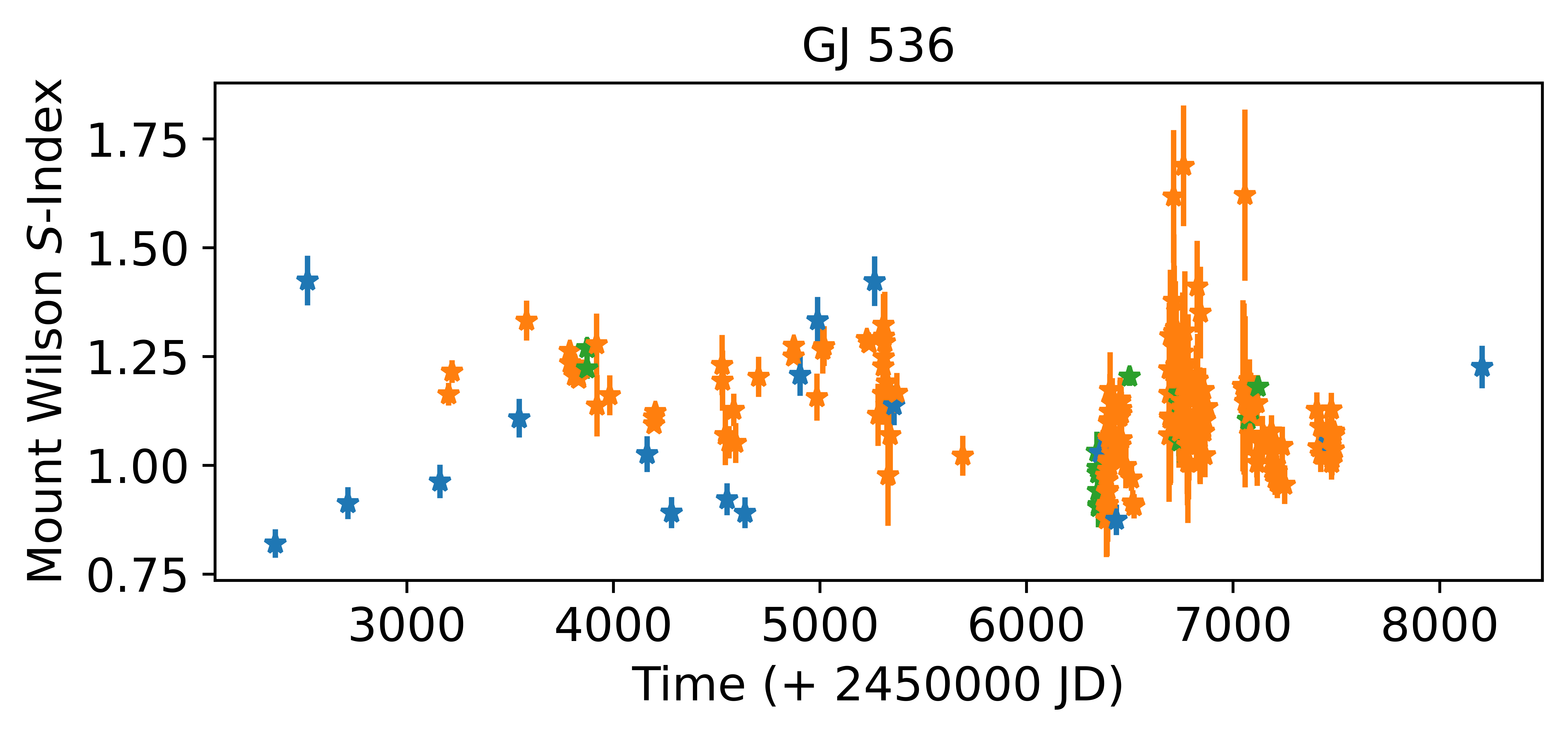}
    \includegraphics[width=0.45\textwidth]{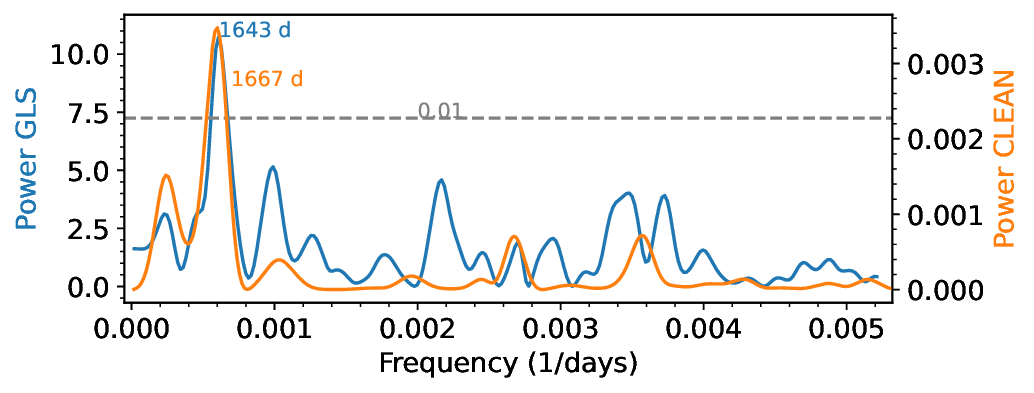}
    \includegraphics[width=0.25\textwidth]{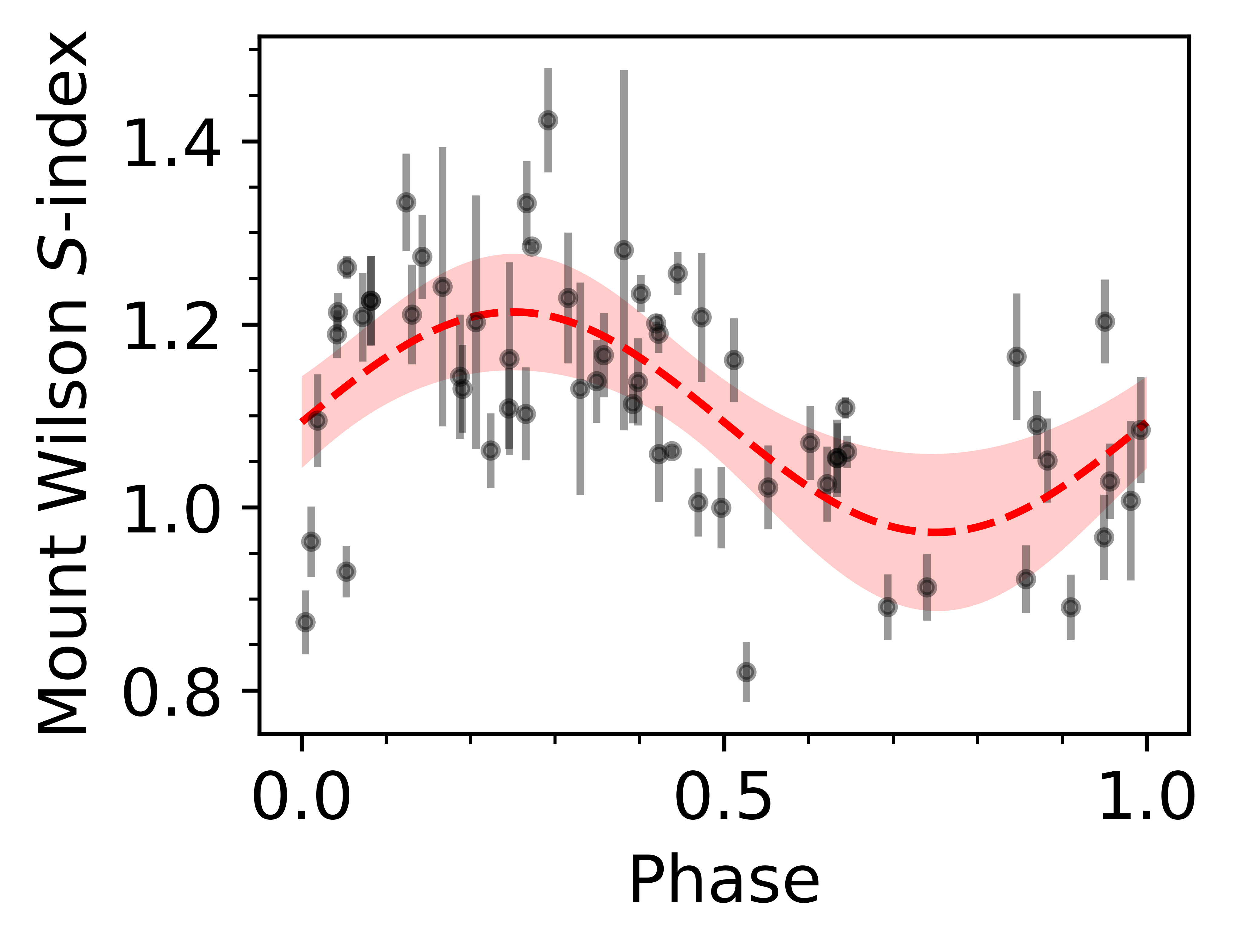}
    \includegraphics[width=0.45\textwidth]{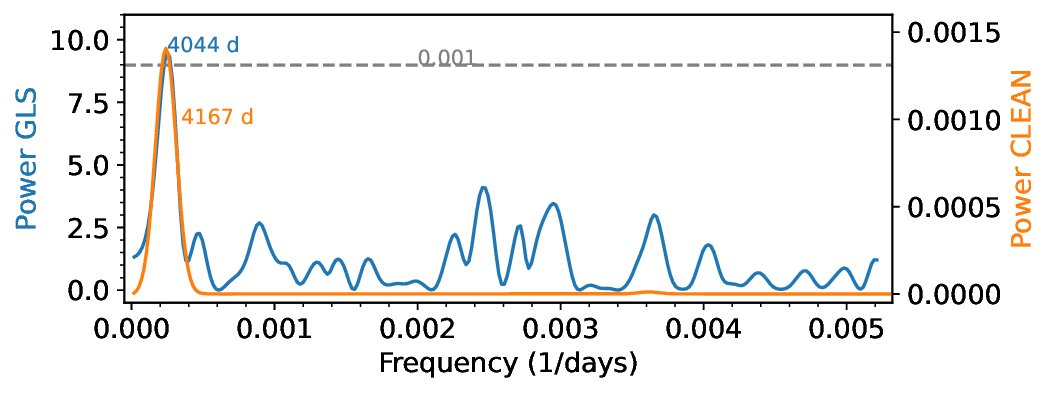}
    \includegraphics[width=0.25\textwidth]{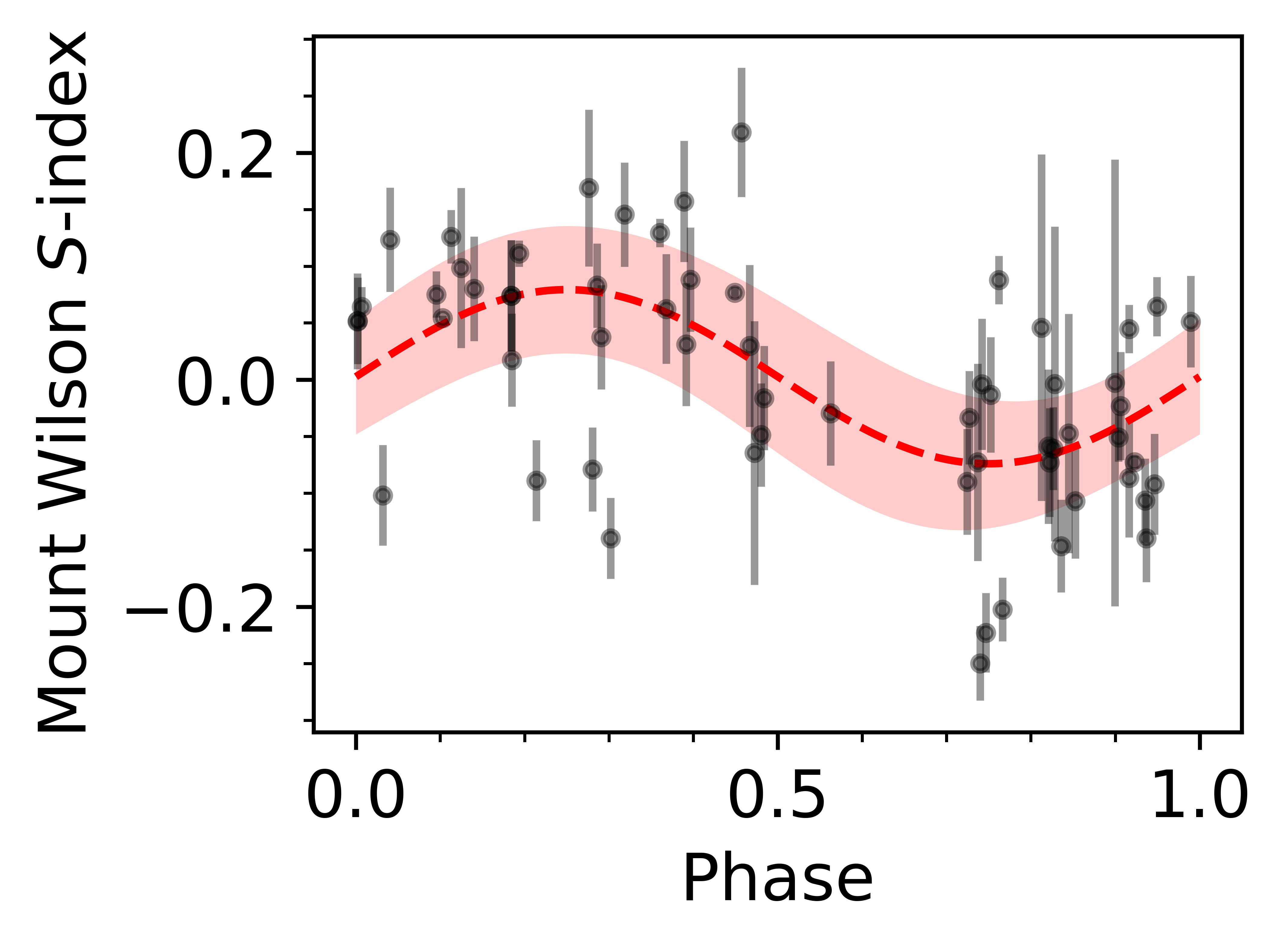}
\end{center}
\caption{Same as Fig. \ref{per_gj205} but for \textbf{GJ 536.} $P_{GLS} = (1643 \pm 58)$ days with a FAP of $8\times10^{-5}$; CLEAN periodogram: $(1667 \pm 24)$ days. \textit{Third row.} Phase folded time series with a $\sim1600$ day period.
\textit{Fourth row.} GLS and CLEAN periodograms after substract the signal of $\sim$1600-day peak, $P_{GLS} = 4044 \pm 443$ days with FAP of 0.05\% and $P_{CLEAN} = 4167 \pm 149$ days. \textit{Bottom.} Phase folded time series with a $\sim4100$ day period.} 
\label{per_gl536}
\end{figure}

\begin{figure}[htb!]
\begin{center}
    \includegraphics[width=0.4\textwidth]{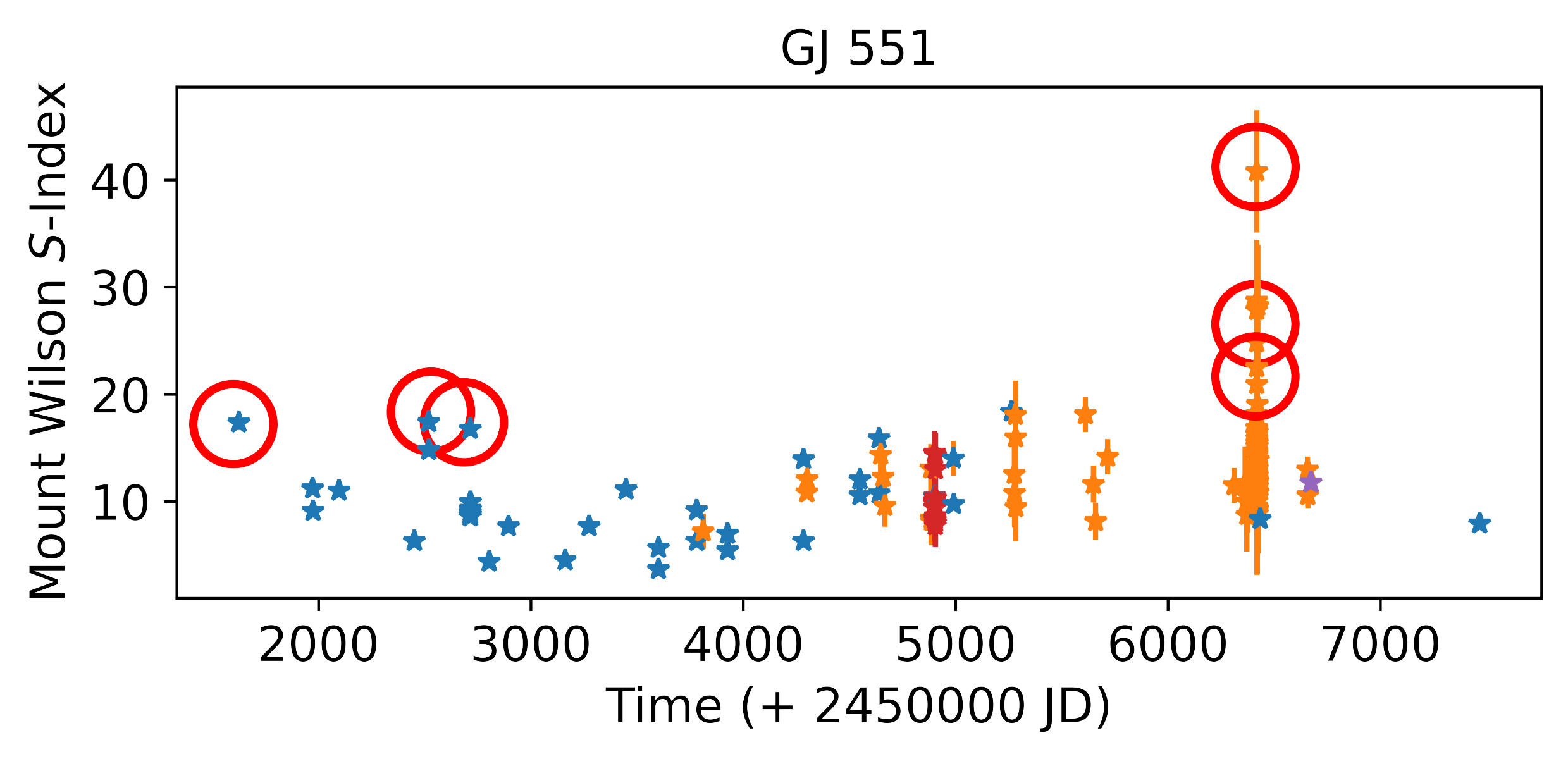}
    \includegraphics[width=0.45\textwidth]{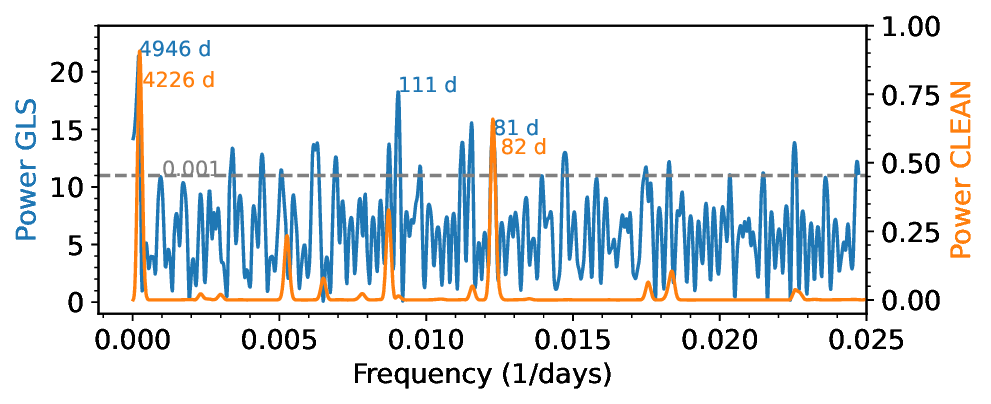}
    \includegraphics[width=0.25\textwidth]{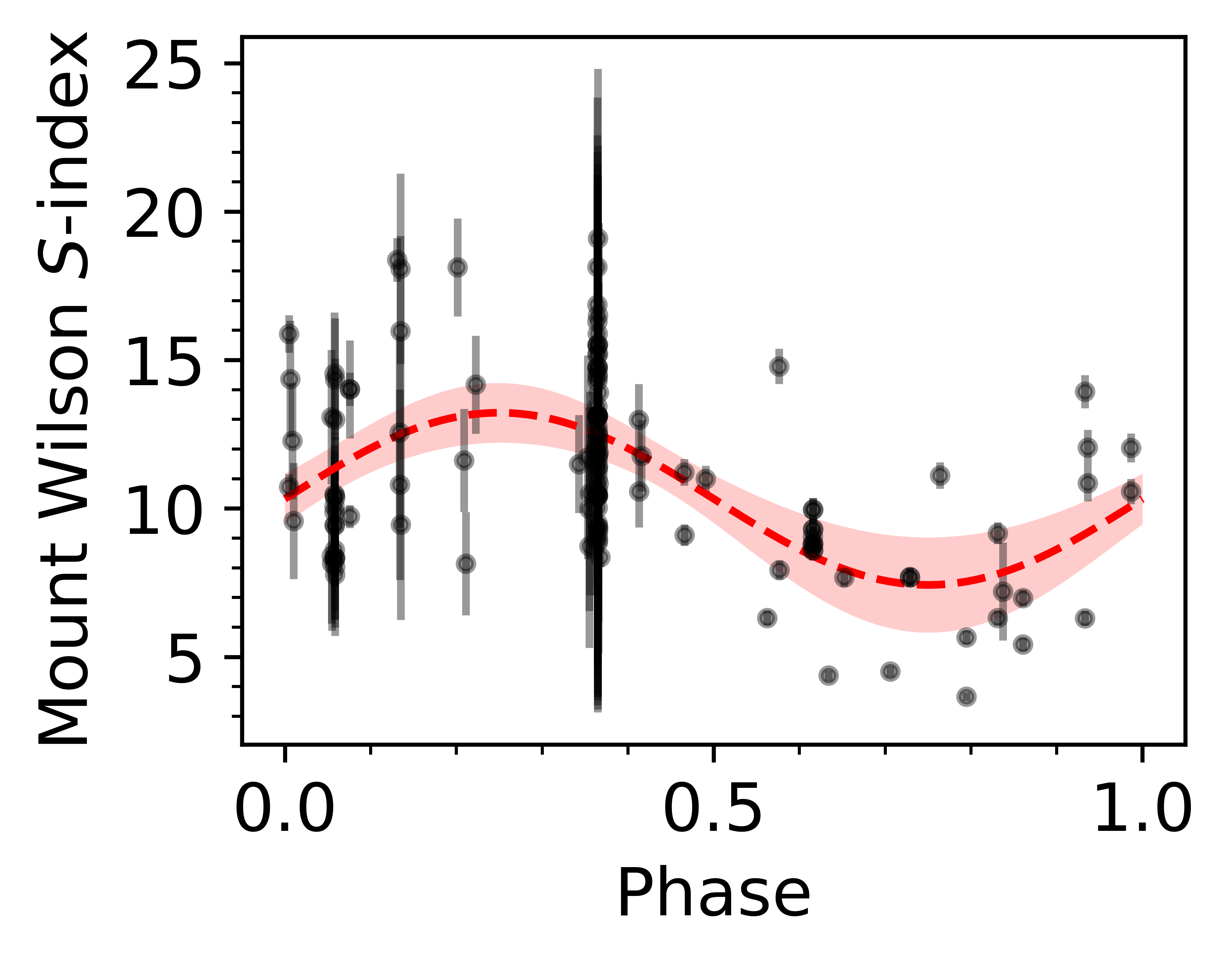}
    \includegraphics[width=0.45\textwidth]{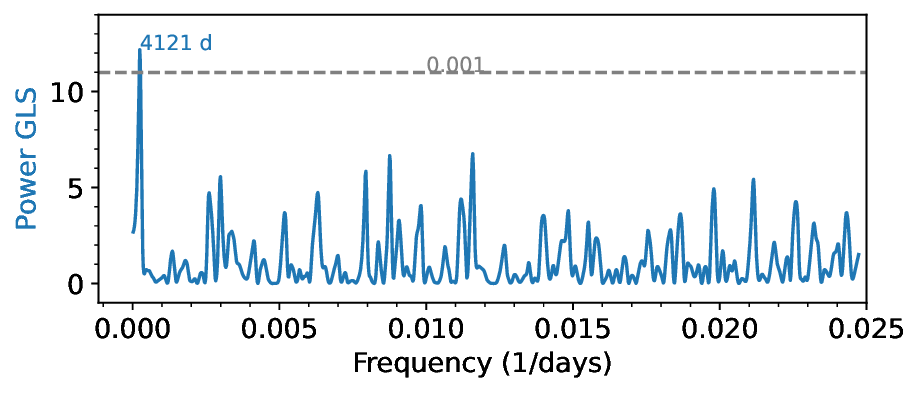}
    \includegraphics[width=0.25\textwidth]{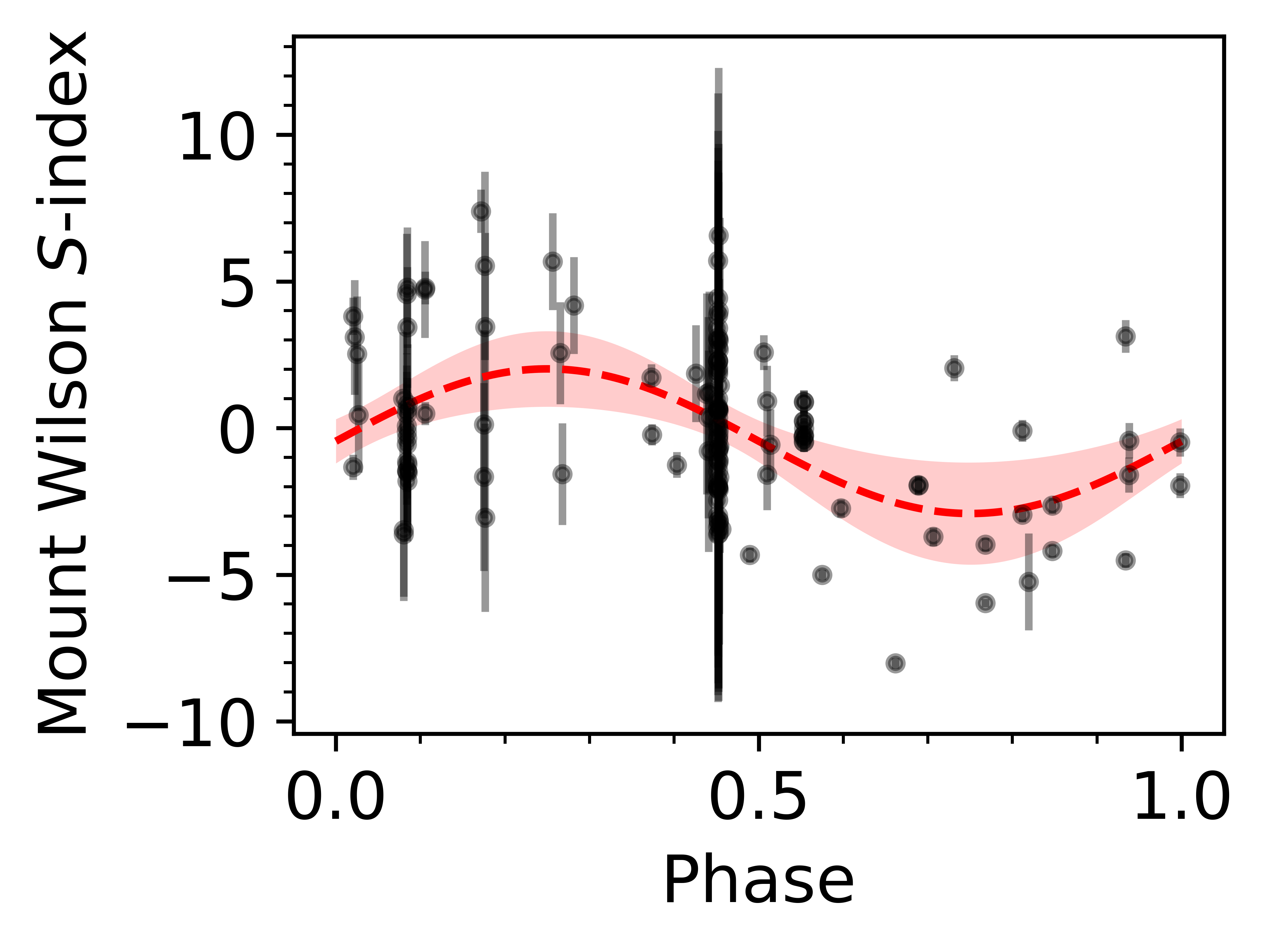}
\end{center}
\caption{Same as Fig. \ref{per_gj205} but for \textbf{GJ 551.} GLS: $P_{GLS; 1} = (4946 \pm 522)$ d, $P_{GLS; 2} = (110.6 \pm 0.3)$ d, $P_{GLS; 3} = (81.5 \pm 0.3)$ d con FAPs $<0.01$ \%). CLEAN: $P_{CLEAN; 1} = (4226 \pm 325)$ d and $P_{CLEAN; 2} = (81.5 \pm 0.2)$ d. \textit{Third row.} Phase folded time series with a $\sim4900$ day period. 
\textit{Fourth row.} GLS and CLEAN periodograms after substract the signal of $\sim$82-day peak, $P_{GLS} = 4121 \pm 321$ days with FAP of 0.0002. \textit{Bottom.} Phase folded time series with a $\sim4100$ day period.} 
\label{per_gj551}
\end{figure}

\begin{figure}[htb!]
\begin{center}
    \includegraphics[width=0.4\textwidth]{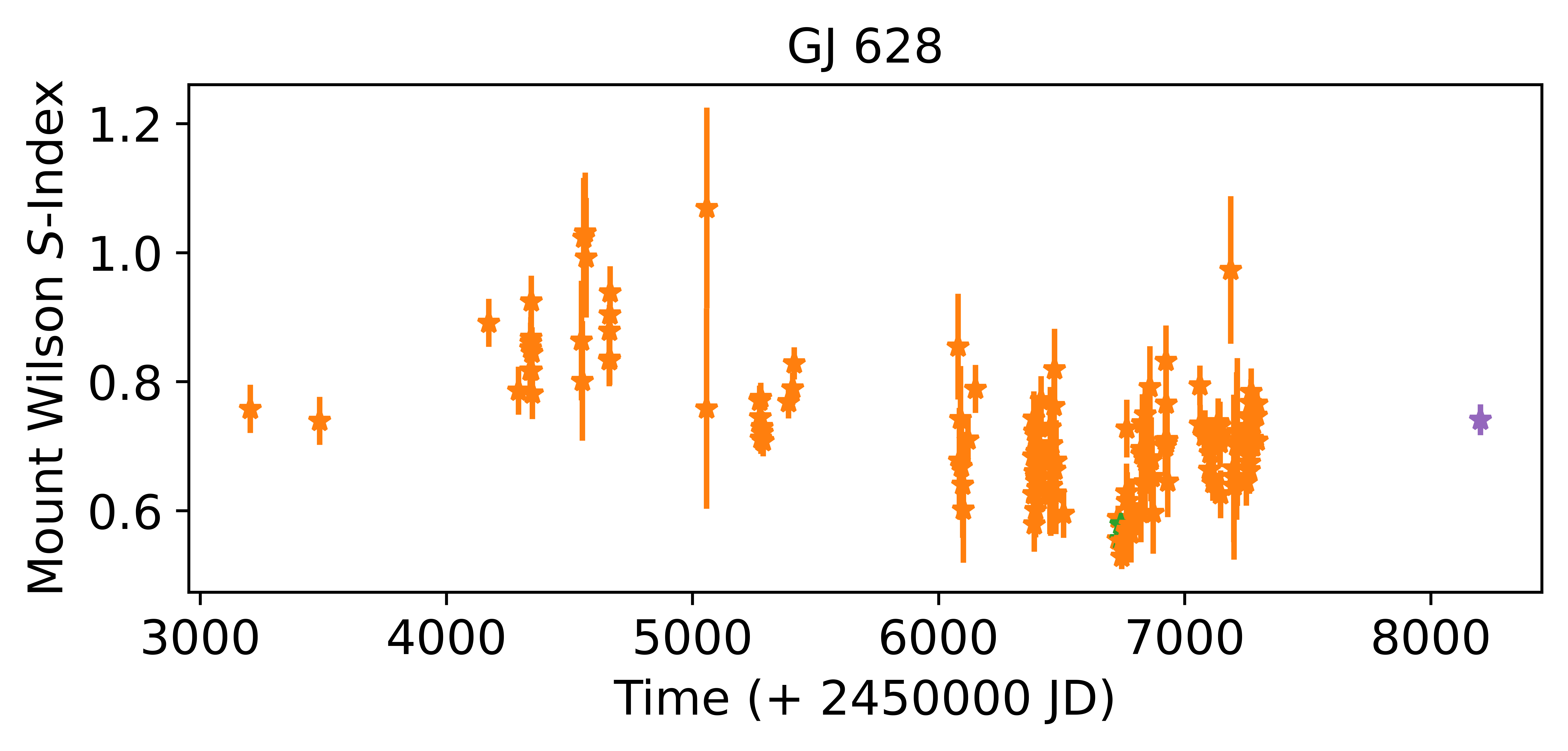}
    \includegraphics[width=0.45\textwidth]{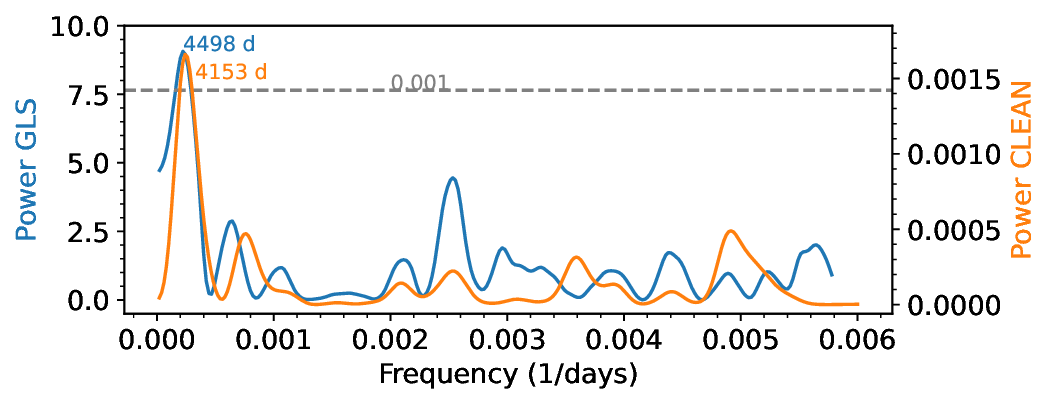}
    \includegraphics[width=0.25\textwidth]{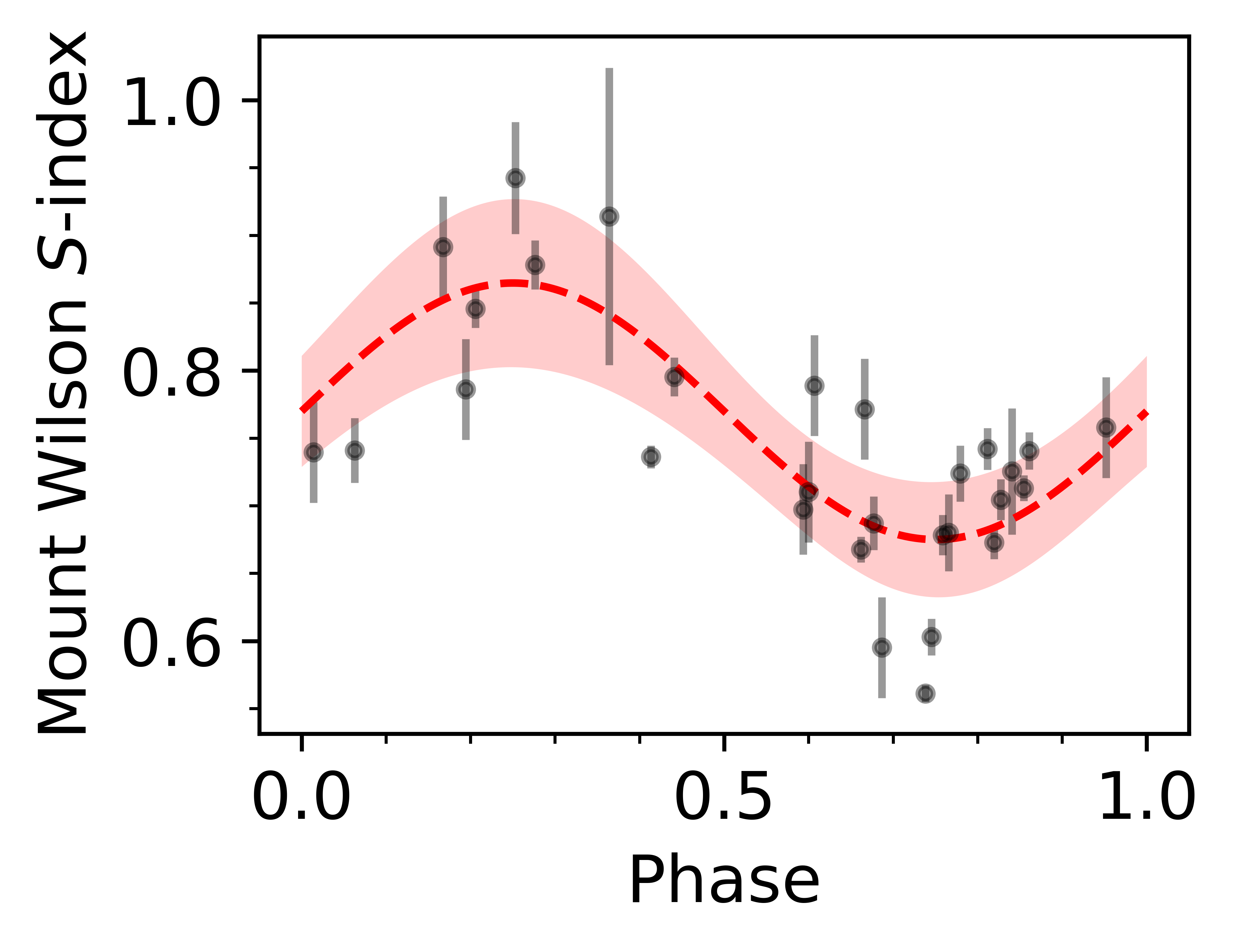}
\end{center}
\caption{Same as Fig. \ref{per_gj205} but for \textbf{GJ 628.} GLS and CLEAN periodograms obtained for the time series binned every 30 days, $P_{GLS}$ = (4498 $\pm$ 607) d with FAP $\sim 0.004$ \%. From the CLEAN periodogram we obtained a significant period of $P_{CLEAN}$ = (4153 $\pm$ 173) d. \textit{Bottom.} Phase folded time series with a $\sim4400$ day period.}
\label{per_gl628}
\end{figure}

\begin{figure}[htb!]
\begin{center}
    \includegraphics[width=0.4\textwidth]{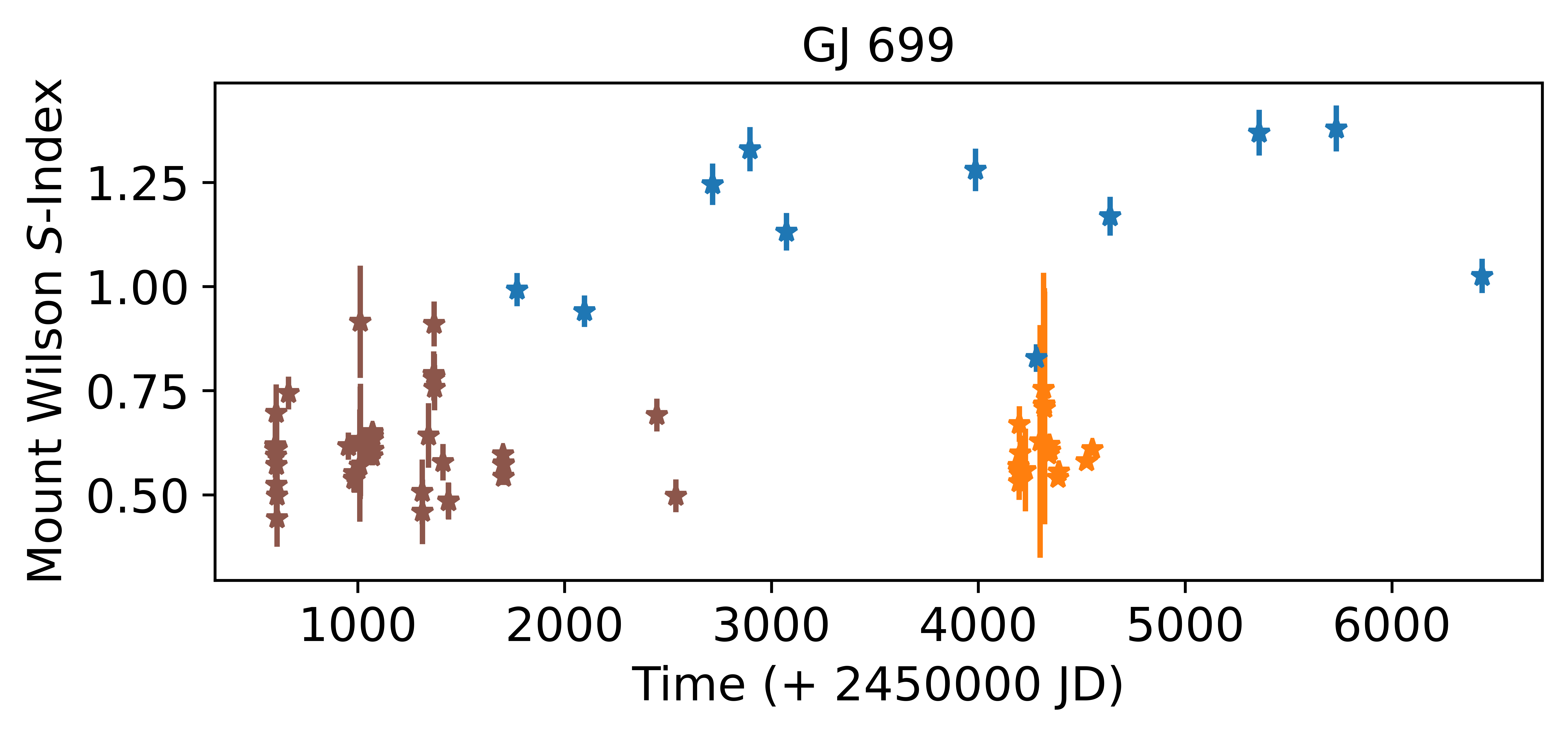}
    \includegraphics[width=0.45\textwidth]{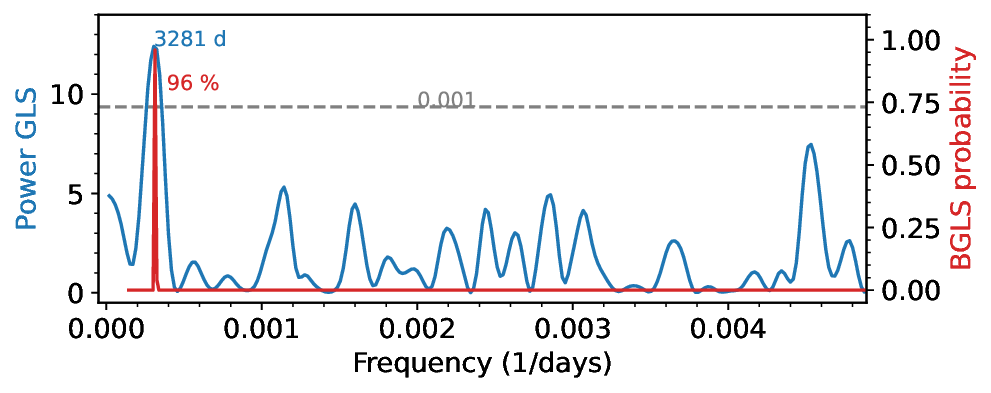}
    \includegraphics[width=0.25\textwidth]{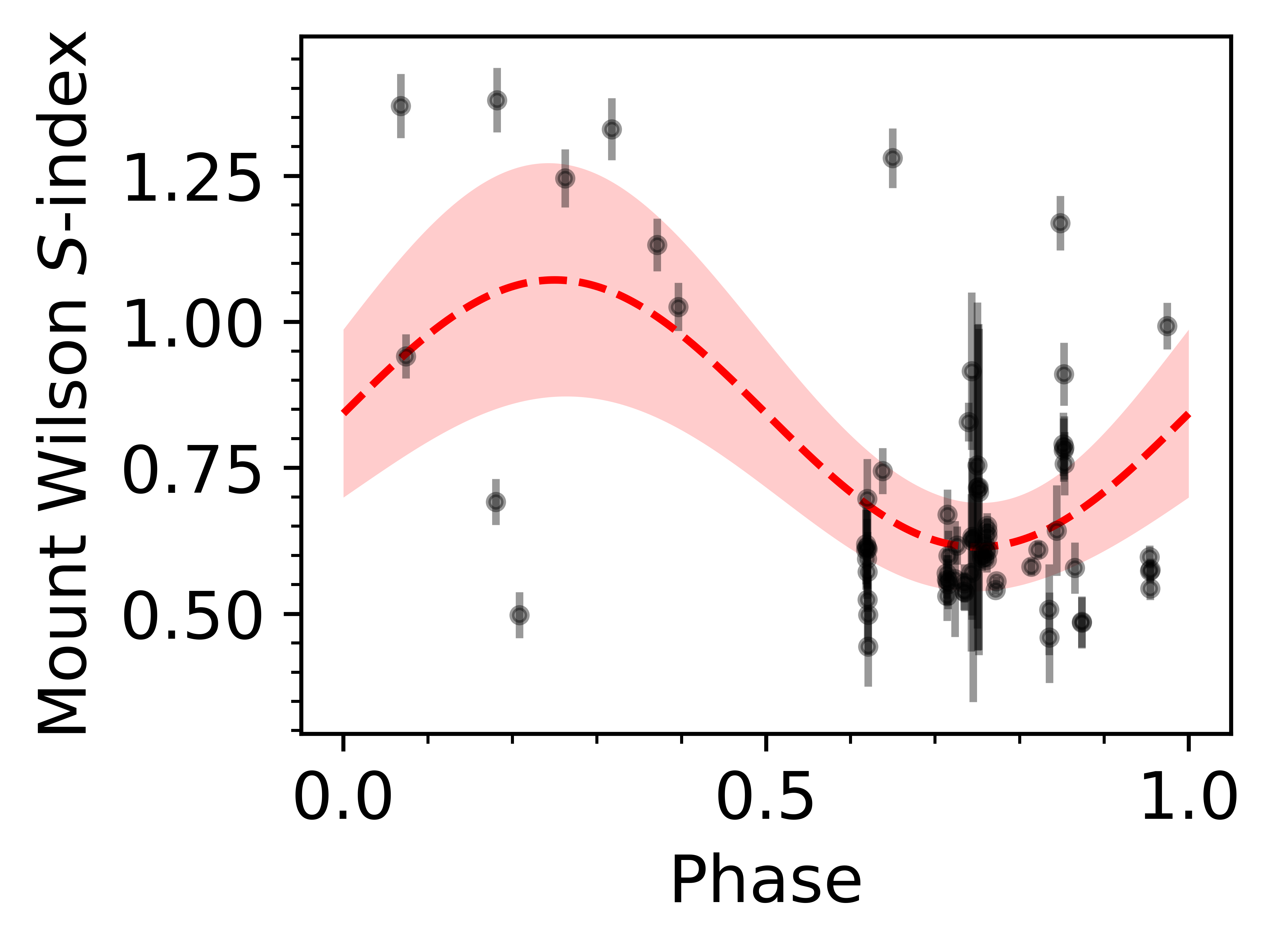}
\end{center}
\caption{Same as Fig. \ref{per_gl1} but for \textbf{GJ 699.} $P_{cyc} = 3281 \pm 255$ days with FAP of $1\times10^{-5}$ and a probability of 96\% that the period correspond with the activity cycle for Gl 699. \textit{Bottom.} Phase folded time series with a $\sim3300$ day period.}
\label{per_gl699}
\end{figure}

\begin{figure}[htb!]
\begin{center}
    \includegraphics[width=0.4\textwidth]{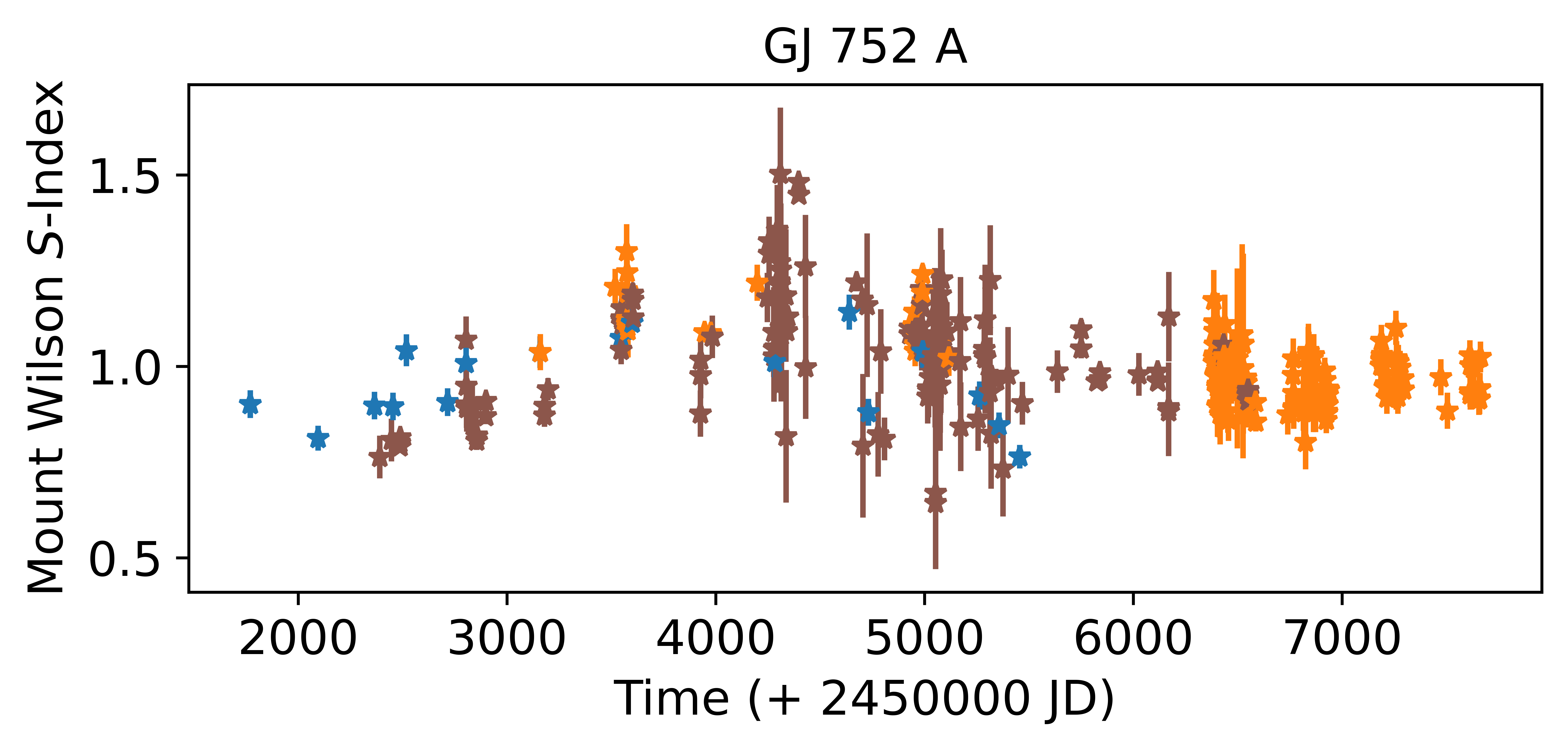}
    \includegraphics[width=0.45\textwidth]{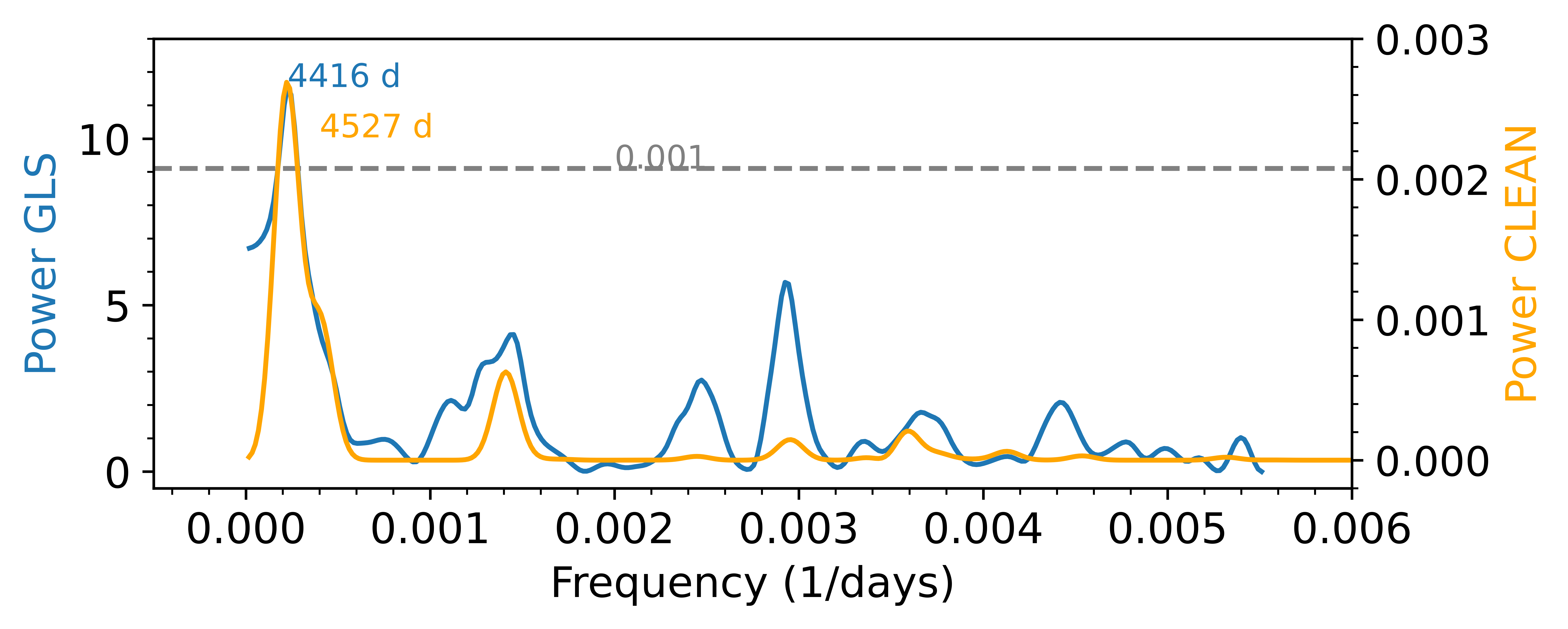}
    \includegraphics[width=0.25\textwidth]{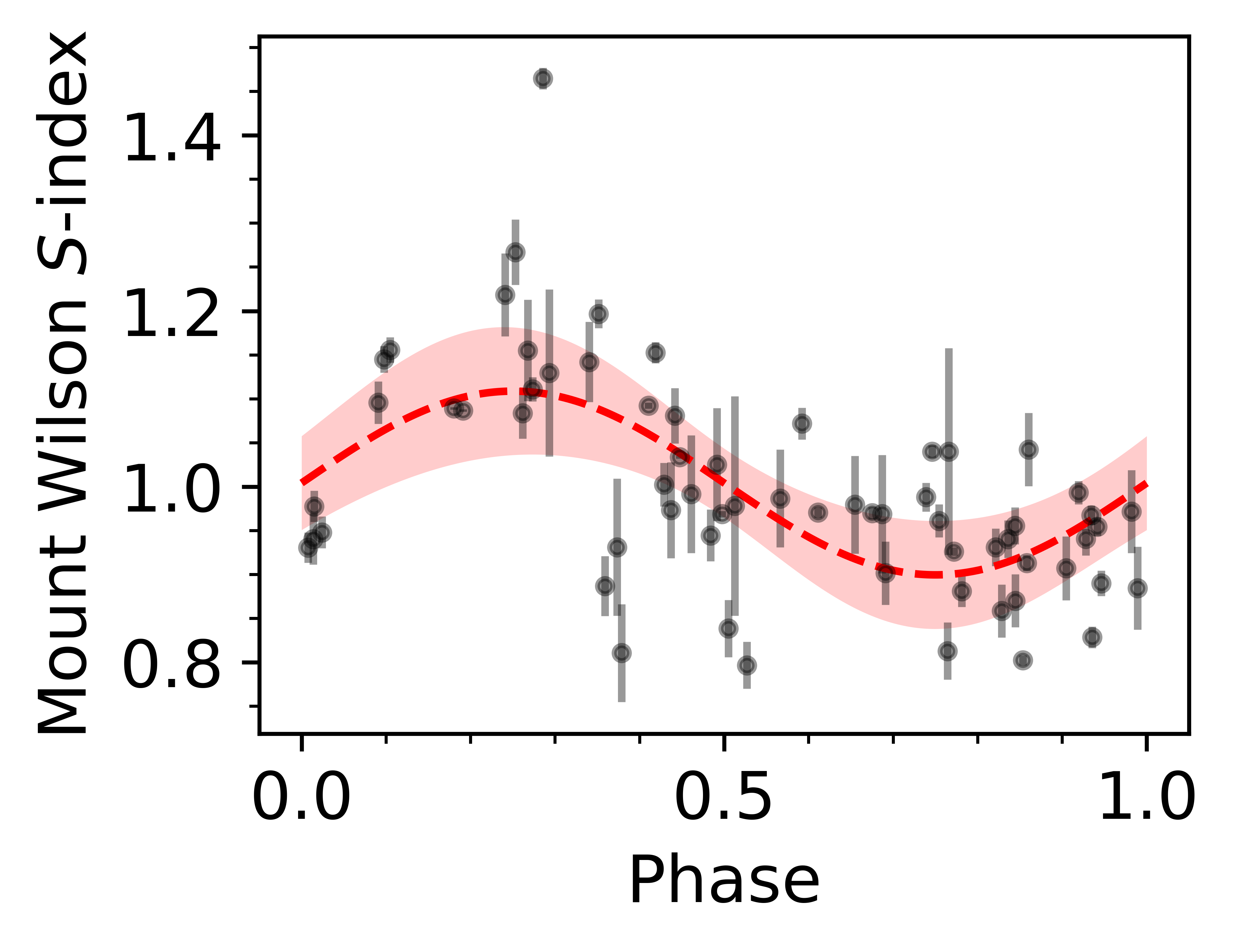}
\end{center}
\caption{Same as Fig. \ref{per_gj205} but for \textbf{GJ 752 A.} $P_{GLS} = (4416 \pm 462)$ d with FAP < 0.01\% and  $P_{CLEAN} = (4527 \pm 174)$ d. \textit{Bottom.} Phase folded time series with a $\sim4400$ day period.}
\label{per_gj752A}
\end{figure}

\begin{figure}[htb!]
\begin{center}
    \includegraphics[width=0.4\textwidth]{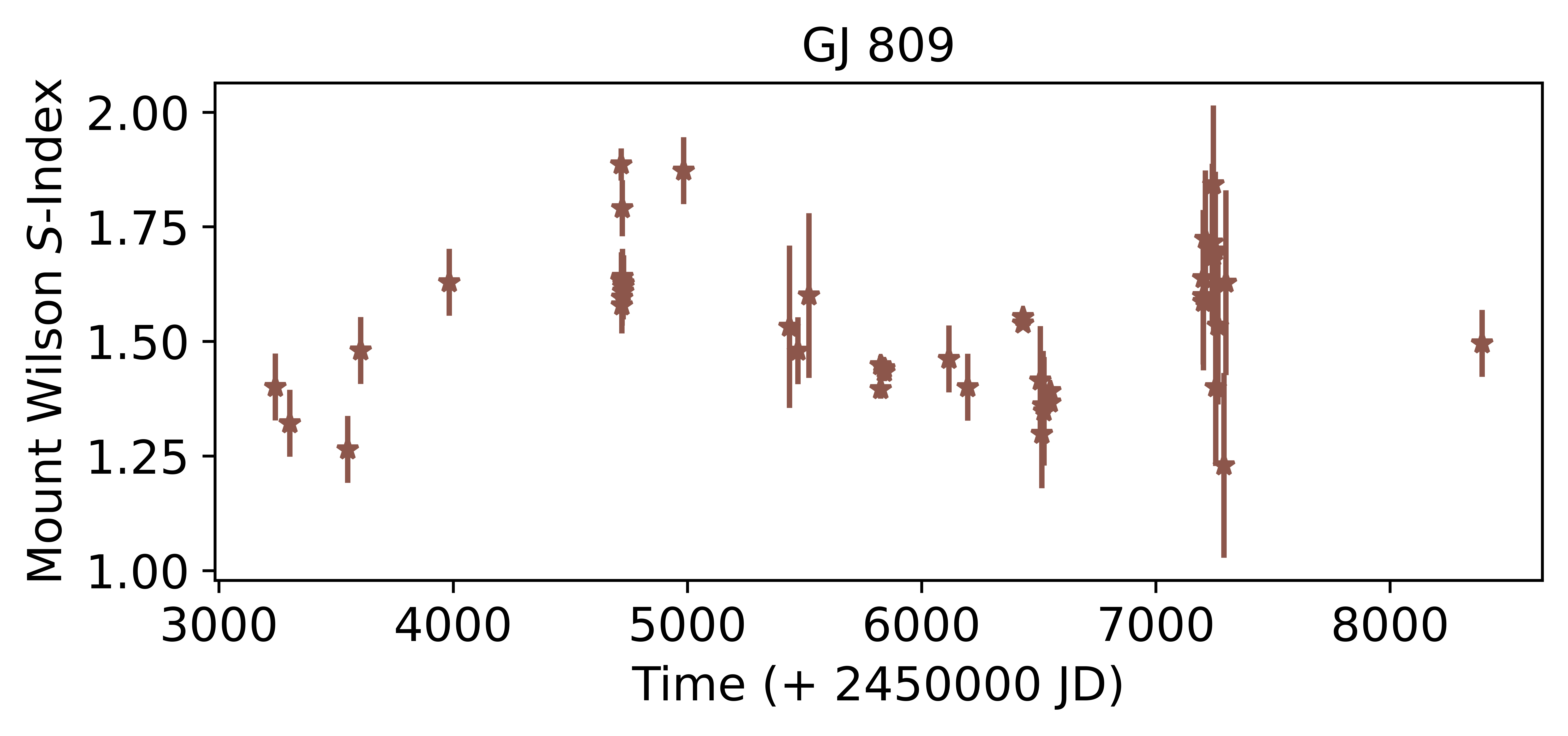}
    \includegraphics[width=0.45\textwidth]{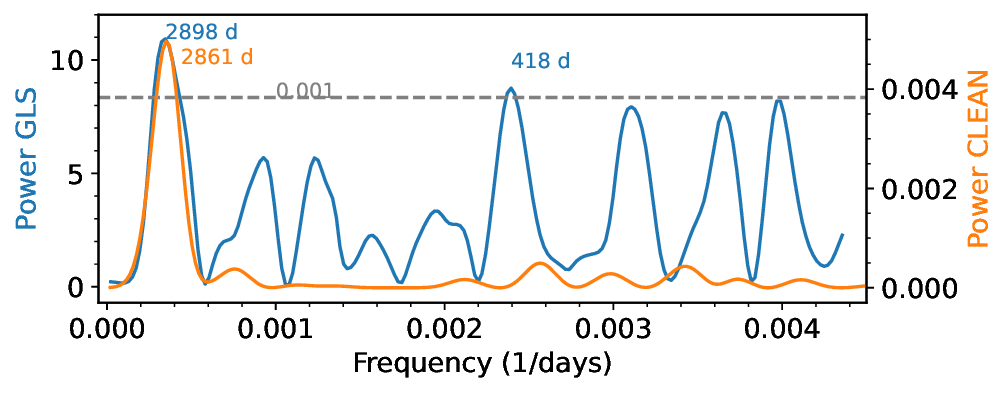}
    \includegraphics[width=0.25\textwidth]{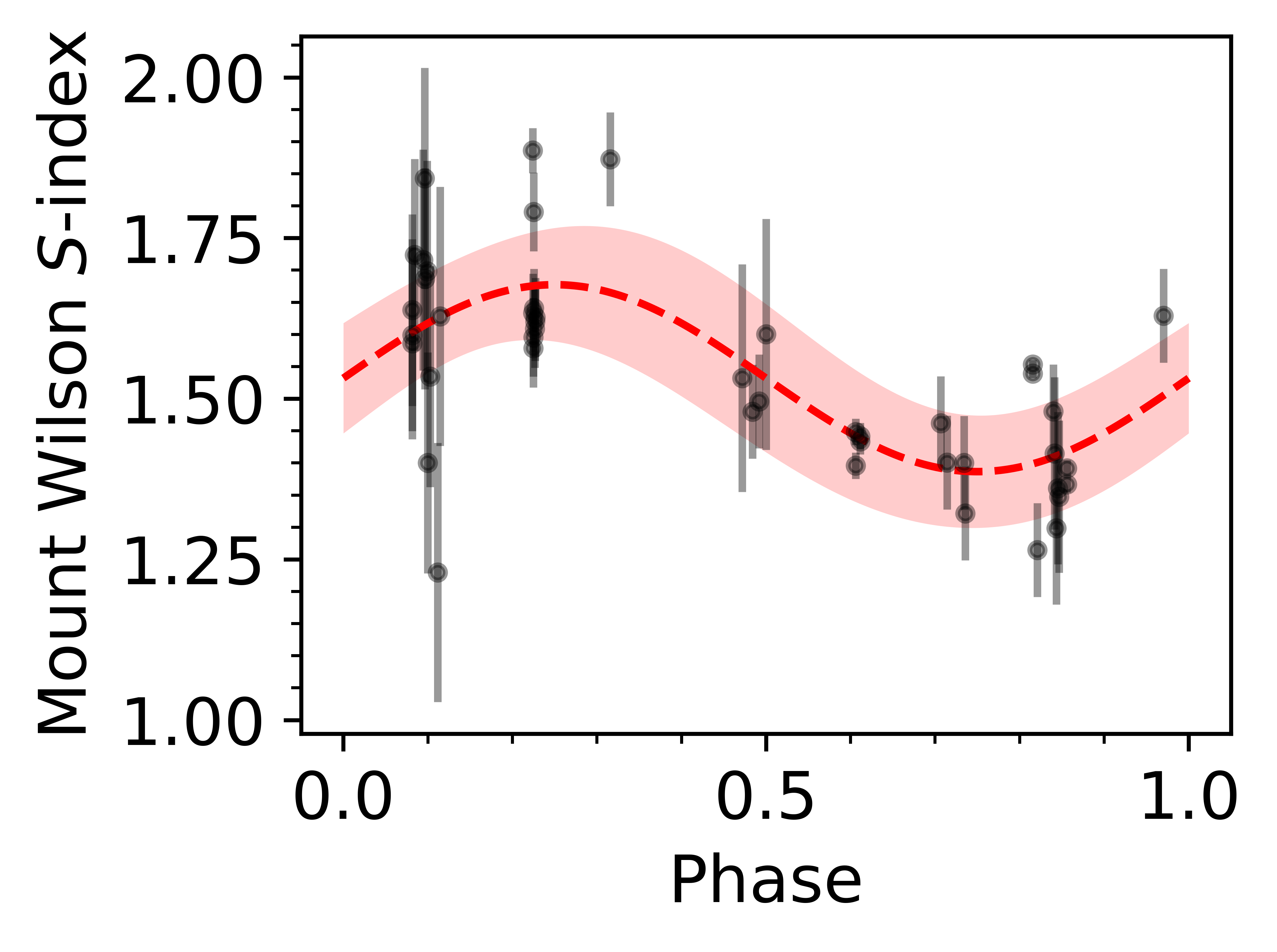}
\end{center}
\caption{Same as Fig. \ref{per_gj205} but for \textbf{GJ 809.} $P_{GLS; 1}$ = (2898 $\pm$ 273) d and $P_{GLS; 2}$ = (418 $\pm$ 6) d with FAPs $\sim 0.001$ \% and $\sim 0.05$ \%, respectively. From the CLEAN periodogram we obtained a significant period of $P_{CLEAN}$ = (2861 $\pm$ 79) d. \textit{Bottom.} Phase folded time series with a $\sim2900$ day period.} 
\label{per_gl809}
\end{figure}

\begin{figure}[htb!]
\begin{center}
    \includegraphics[width=0.4\textwidth]{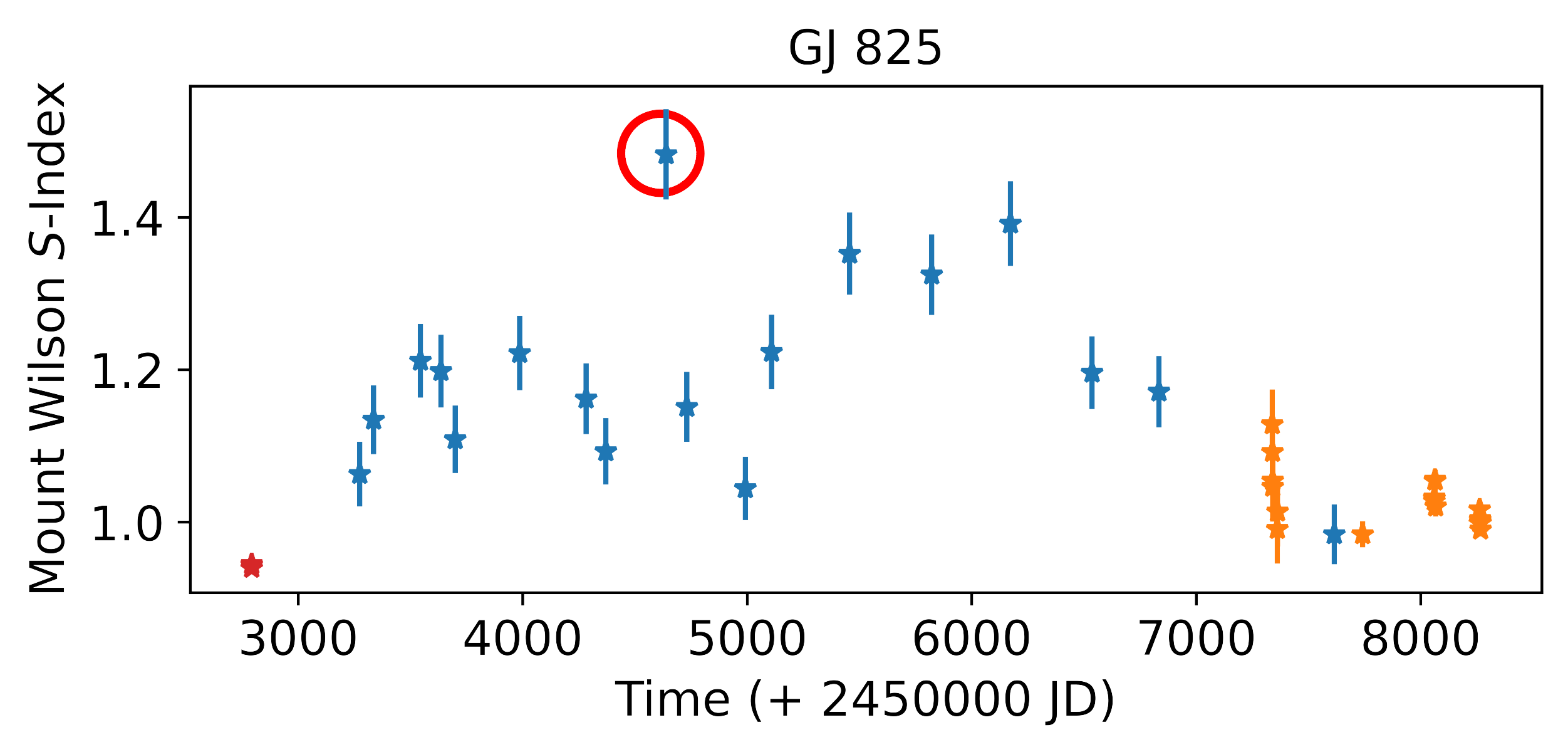}
    \includegraphics[width=0.45\textwidth]{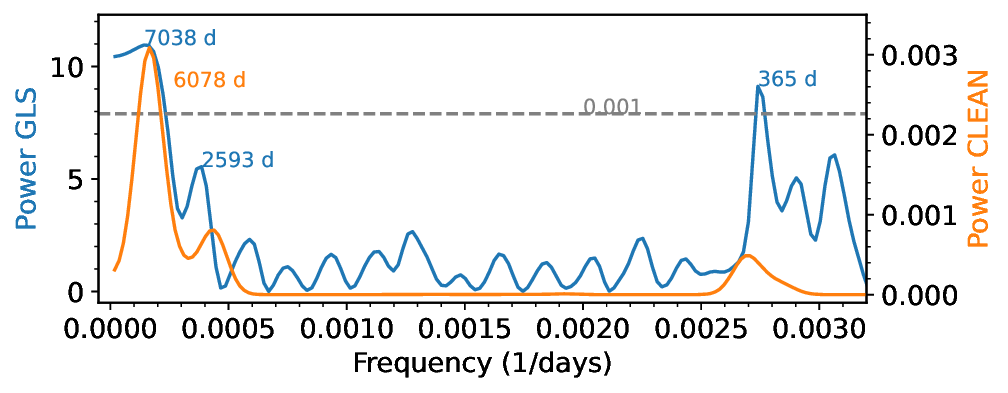}
    \includegraphics[width=0.25\textwidth]{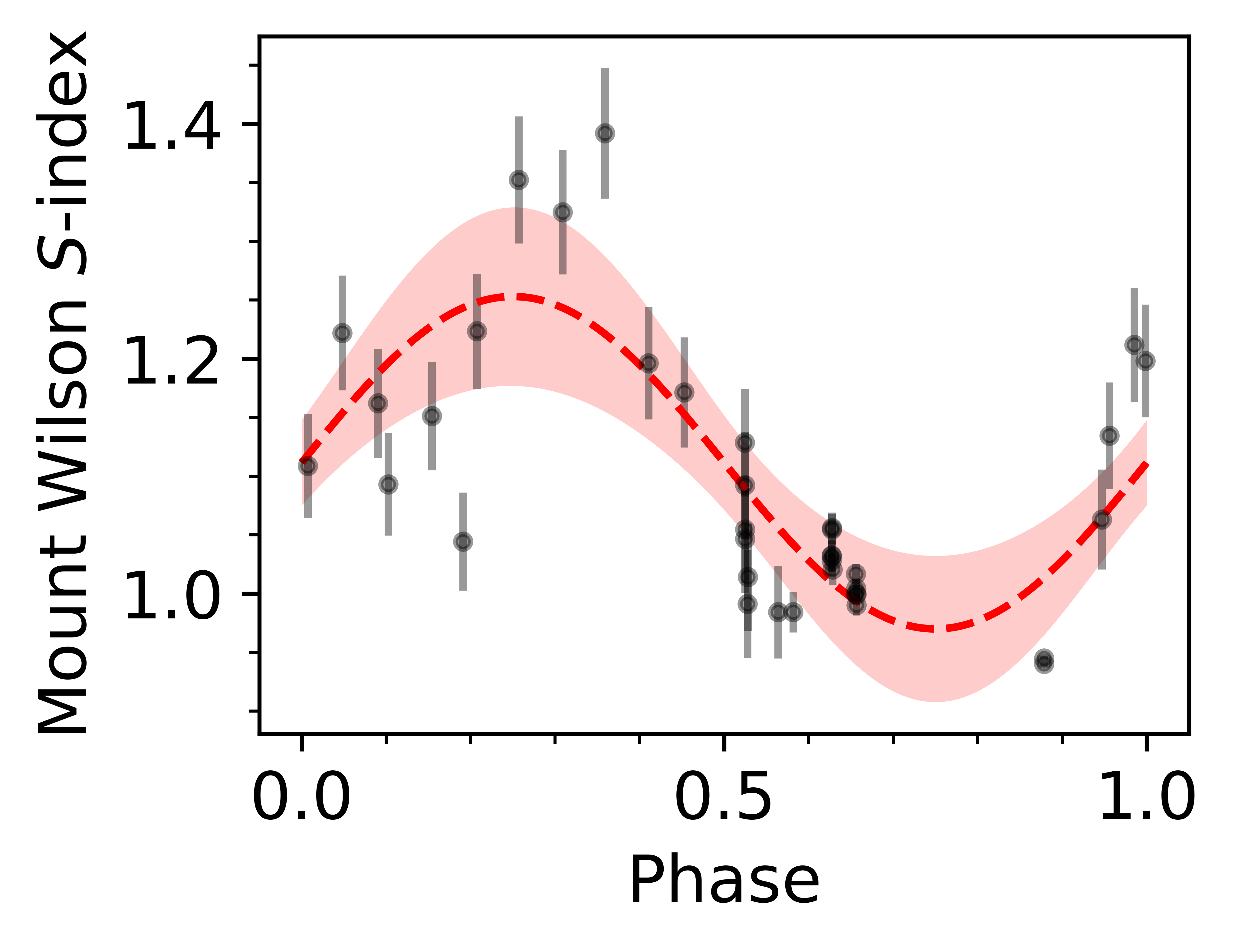}
    \includegraphics[width=0.45\textwidth]{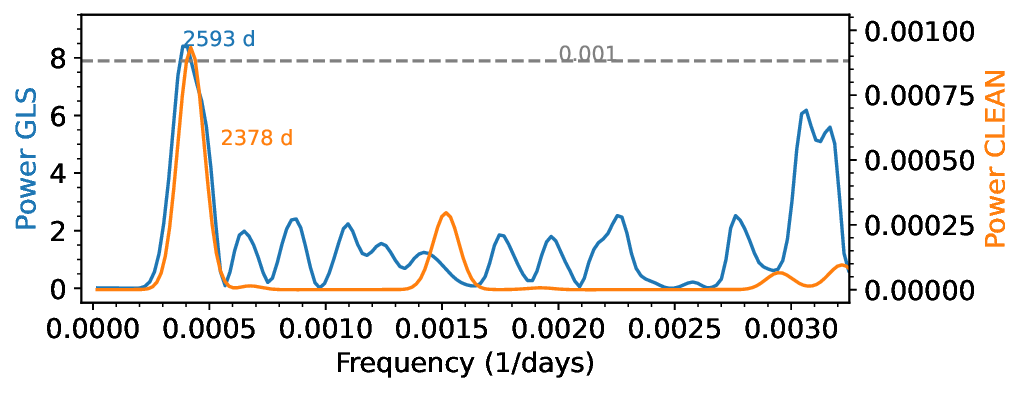}
    \includegraphics[width=0.25\textwidth]{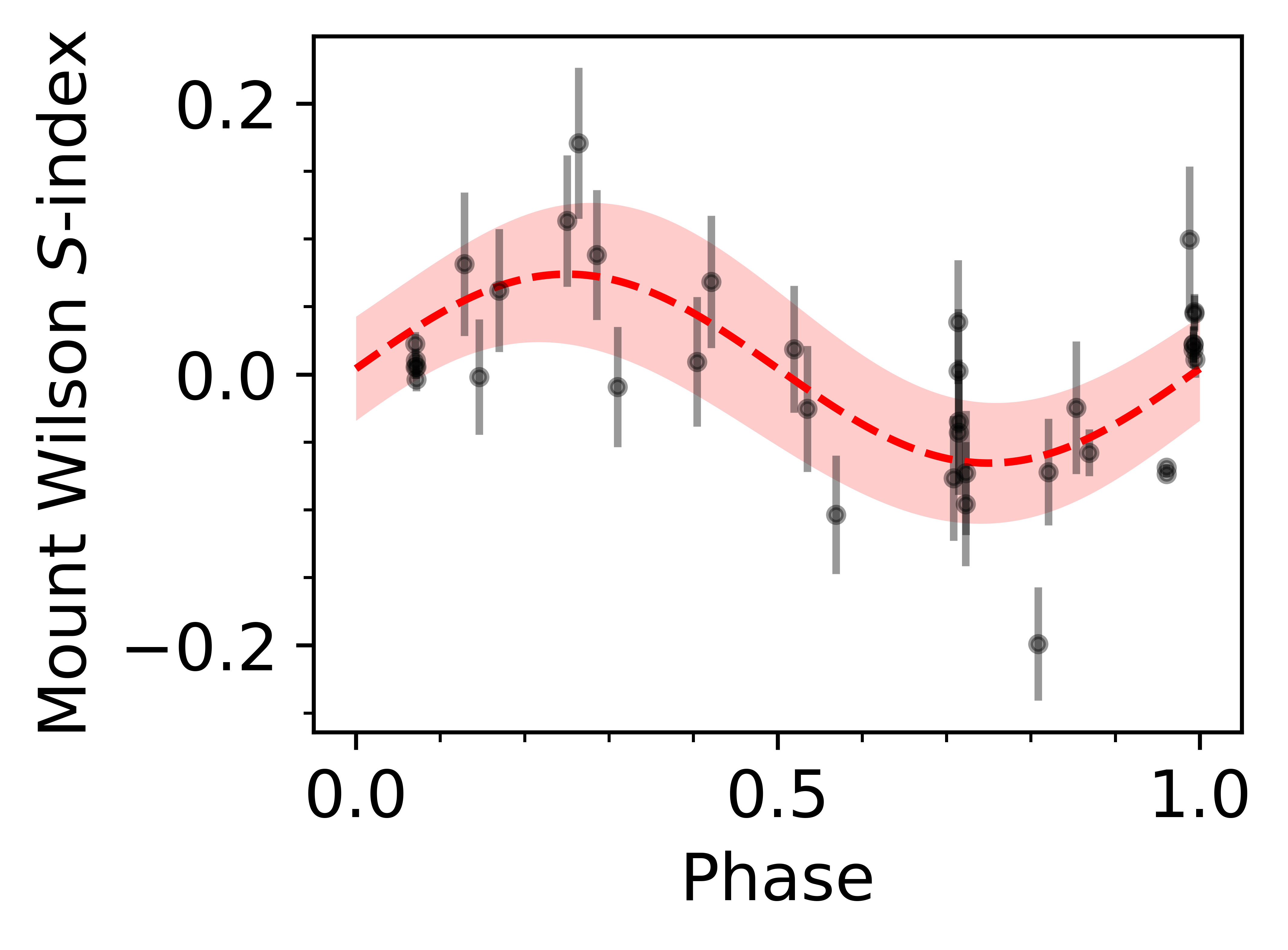}
\end{center}
\caption{Same as Fig. \ref{per_gj205} but for \textbf{GJ 825.} $P_{GLS;1}$ = (7038 $\pm$ 2700) d and  $P_{GLS;2}$ = (365 $\pm$ 1) d with FAPs of $2\times10^{-6}$  and $1\times10^{-4}$, respectively;  CLEAN periodogram: $P_{CLEAN}$ = (6078 $\pm$ 338) d. \textit{Third row.} Phase folded time series with a $\sim7000$ day period.
\textit{Fourth row.} GLS and CLEAN periodograms after substract the $\sim$7000-days peak, $P_{GLS}$ = (2593 $\pm$ 137) d with FAP  0.04\% and $P_{CLEAN}$ = (2378 $\pm$ 104) d. \textit{Bottom.} Phase folded time series with a $\sim2500$ day period.}
\label{per_gl825}
\end{figure}

\begin{figure}[htb!]
\begin{center}
    \includegraphics[width=0.4\textwidth]{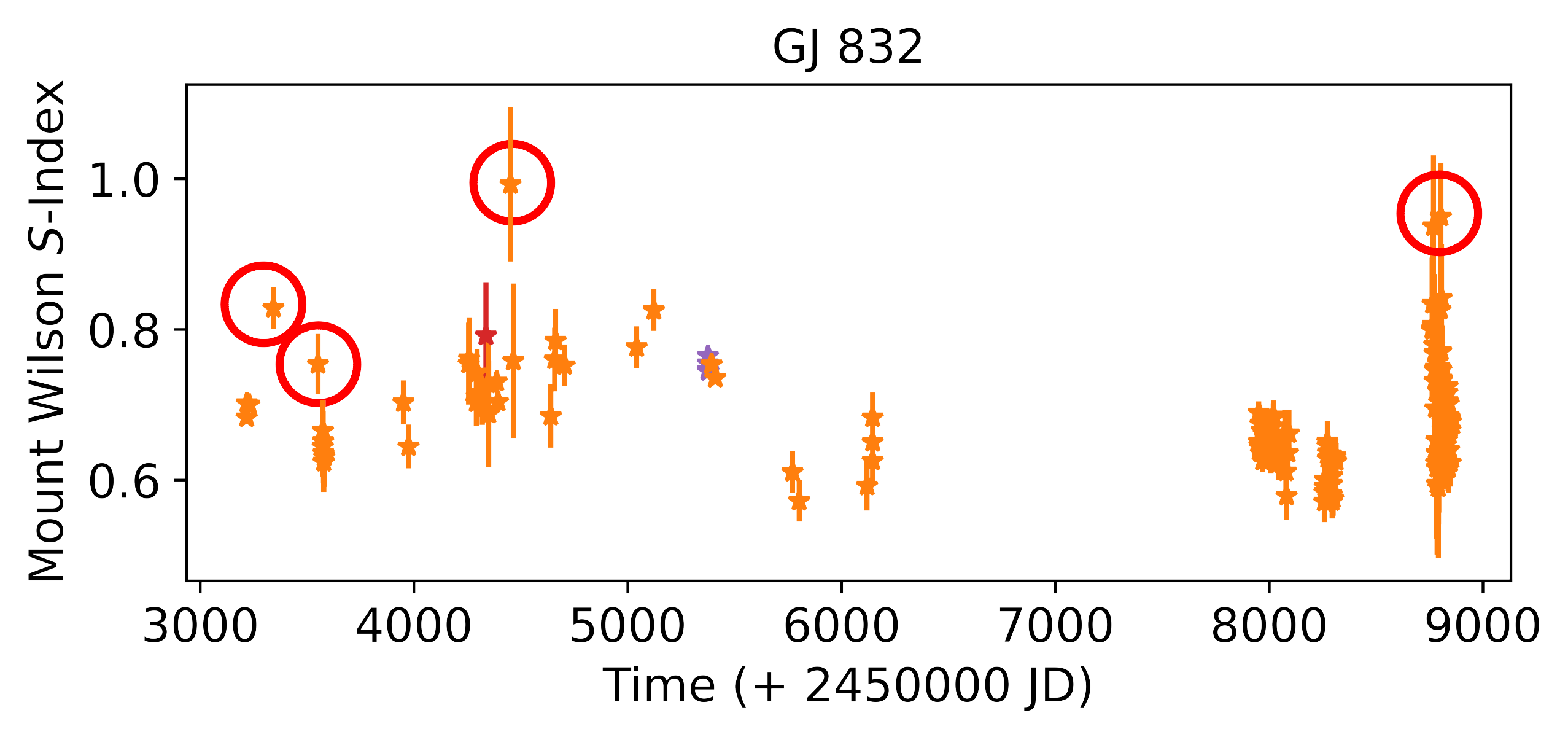}
    \includegraphics[width=0.45\textwidth]{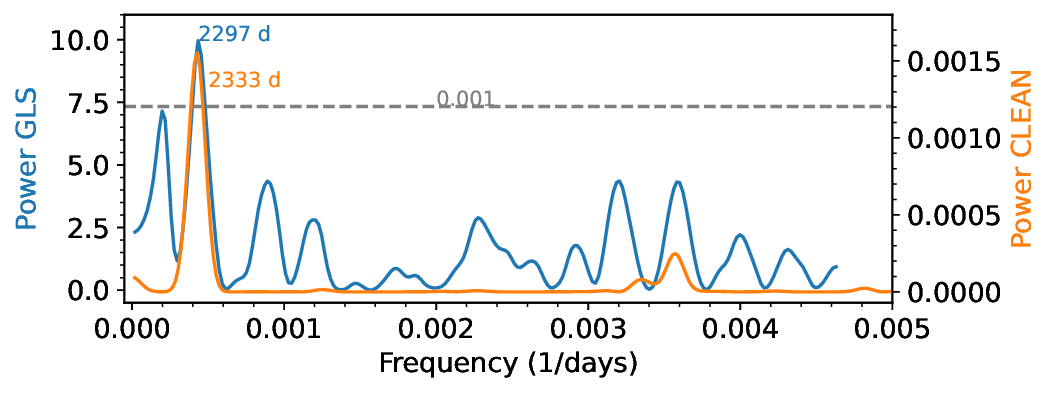}
    \includegraphics[width=0.25\textwidth]{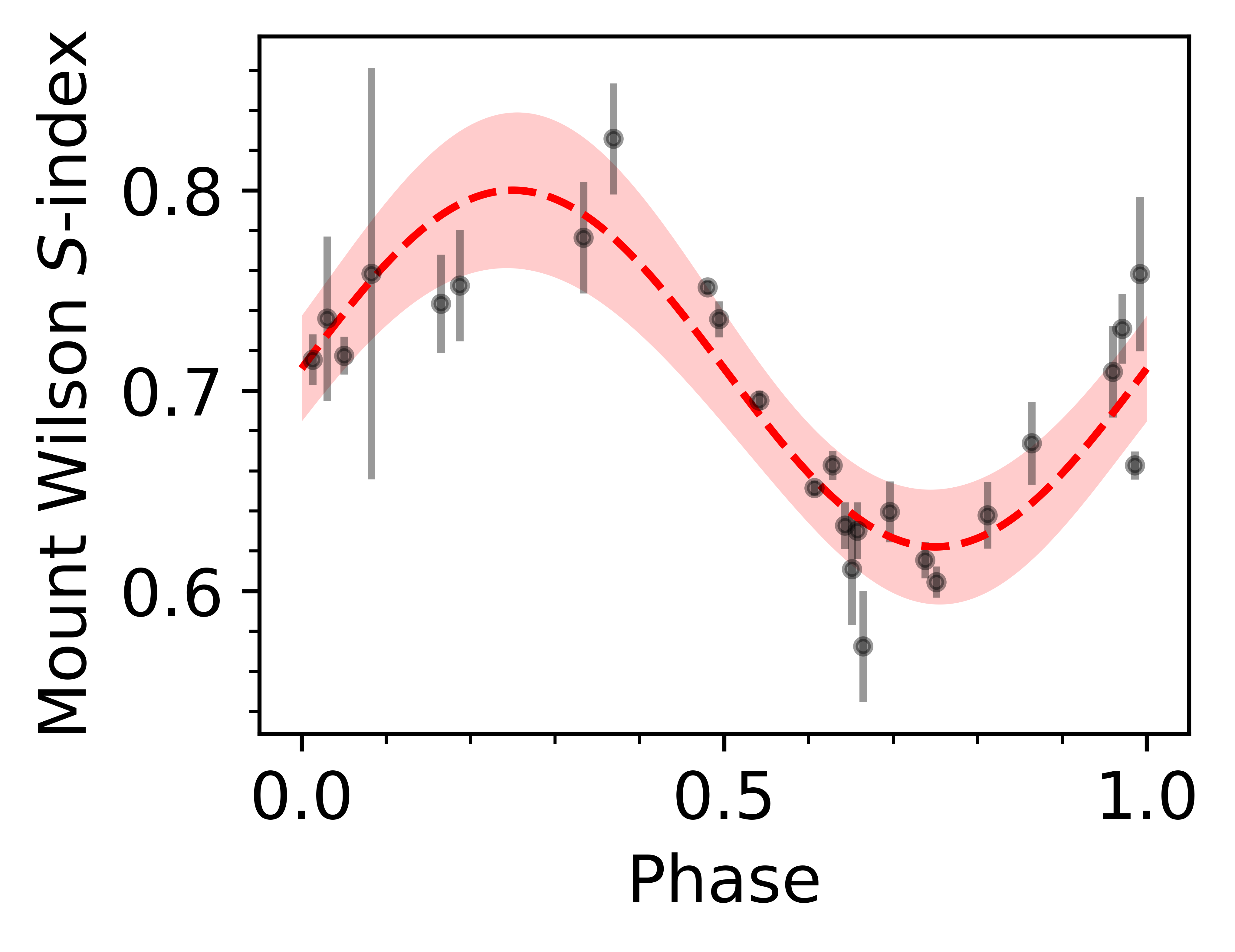}
\end{center}
\caption{Same as Fig. \ref{per_gj205} but for \textbf{GJ 832.} $P_{GLS}$ = (2297 $\pm$ 54) d with FAP of $3\times10^{-7}$ and $P_{CLEAN}$ = (2333 $\pm$ 49) d. \textit{Bottom.} Phase folded time series with a $\sim2300$ day period.}
\label{per_gl832}
\end{figure}

\FloatBarrier

\subsection{Other M dwarfs} \label{ss.noncycles}

\begin{figure}[htb!]
\begin{center}
    \includegraphics[width=0.45\textwidth]{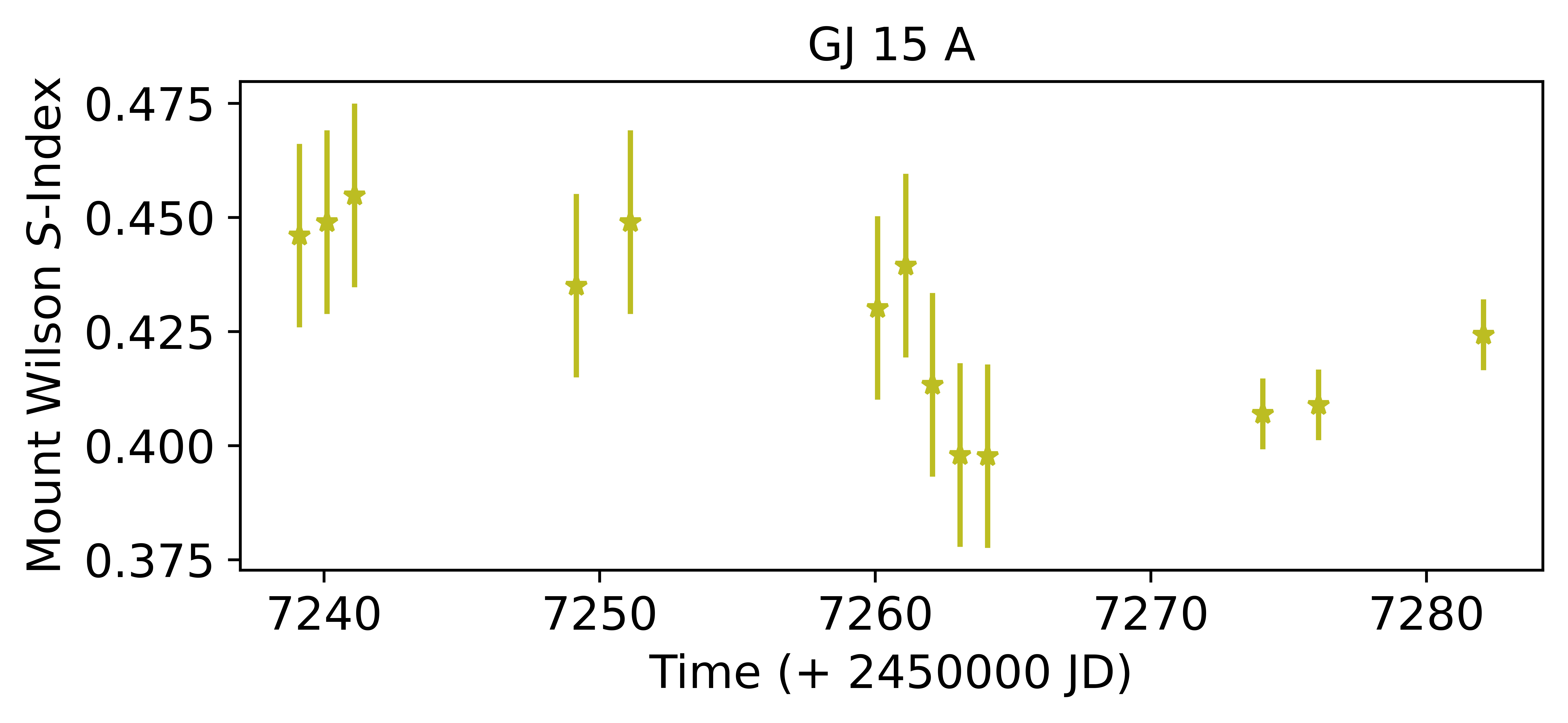}
    \includegraphics[width=0.45\textwidth]{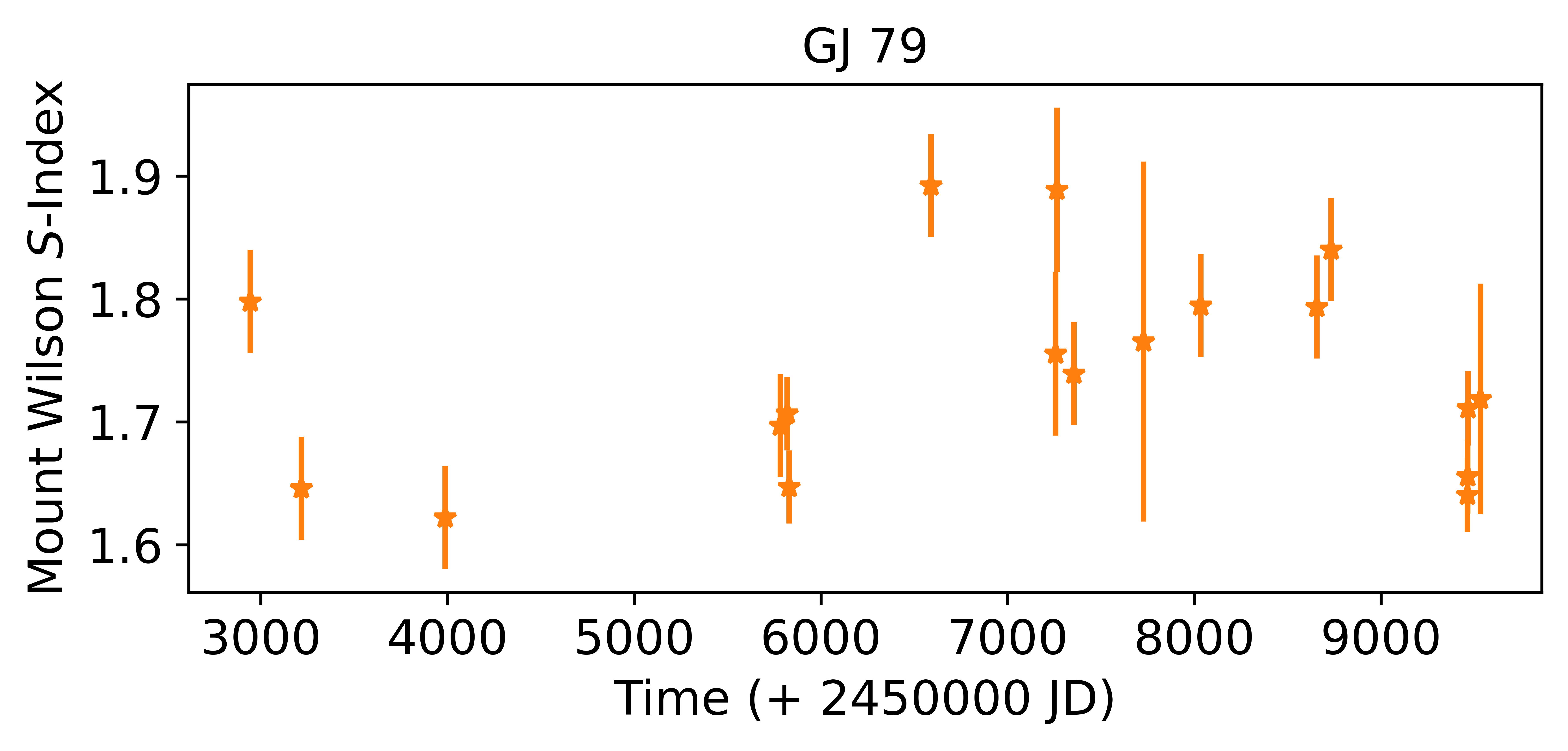}
    \includegraphics[width=0.45\textwidth]{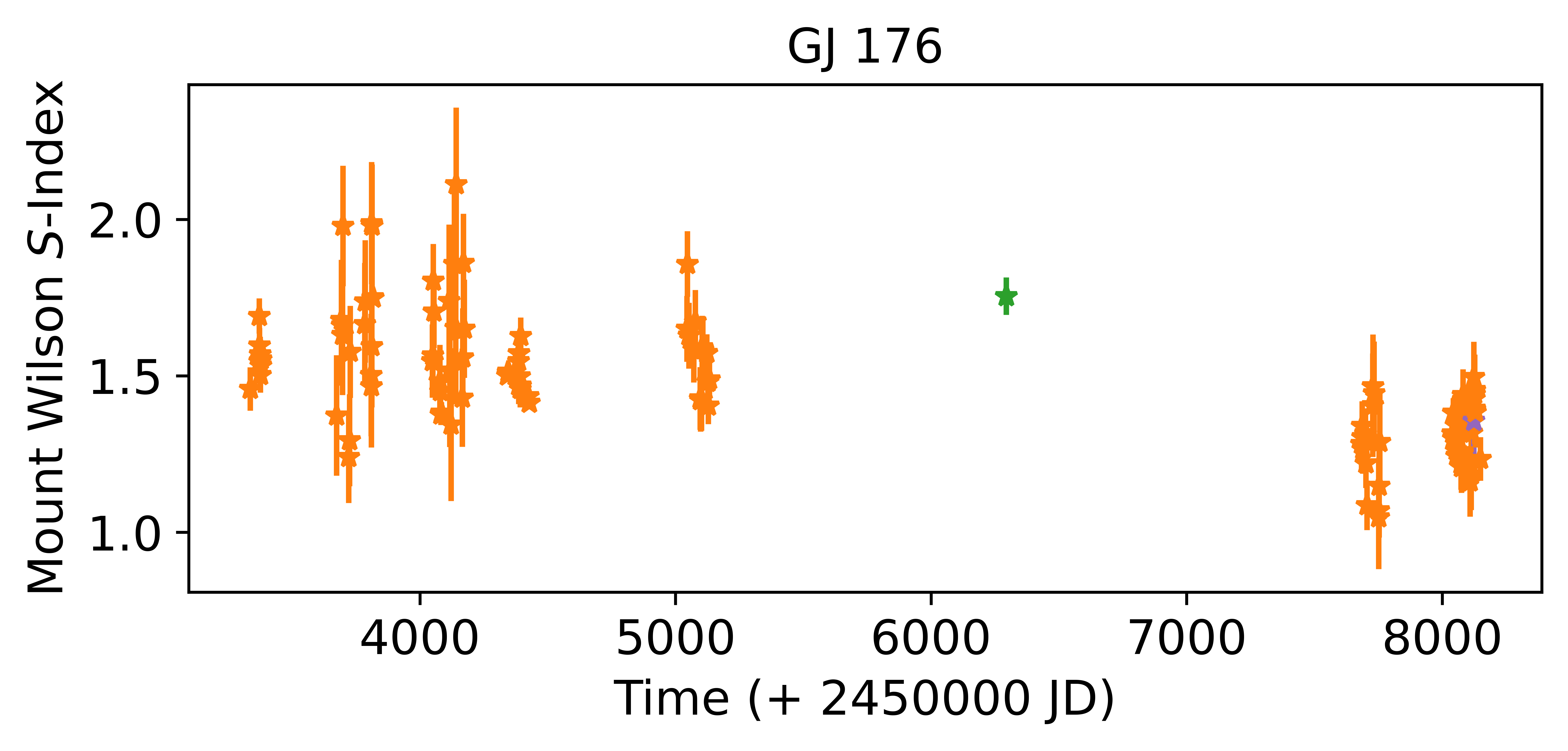}
    \includegraphics[width=0.45\textwidth]{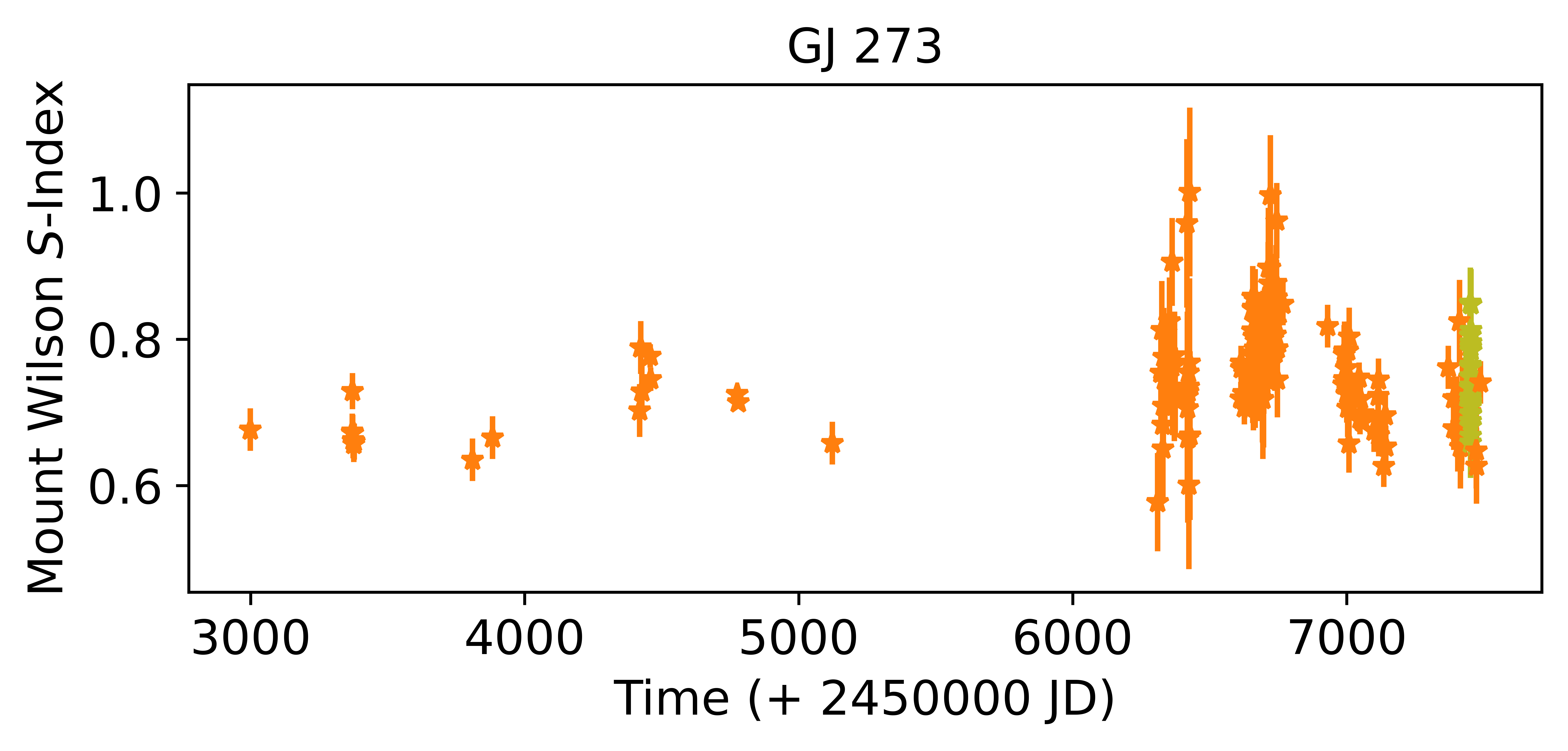}
\end{center}
\caption{$S$-Index time series.}
\label{st_gl1}
\end{figure}

\begin{figure}[htb!]
\begin{center}
    \includegraphics[width=0.45\textwidth]{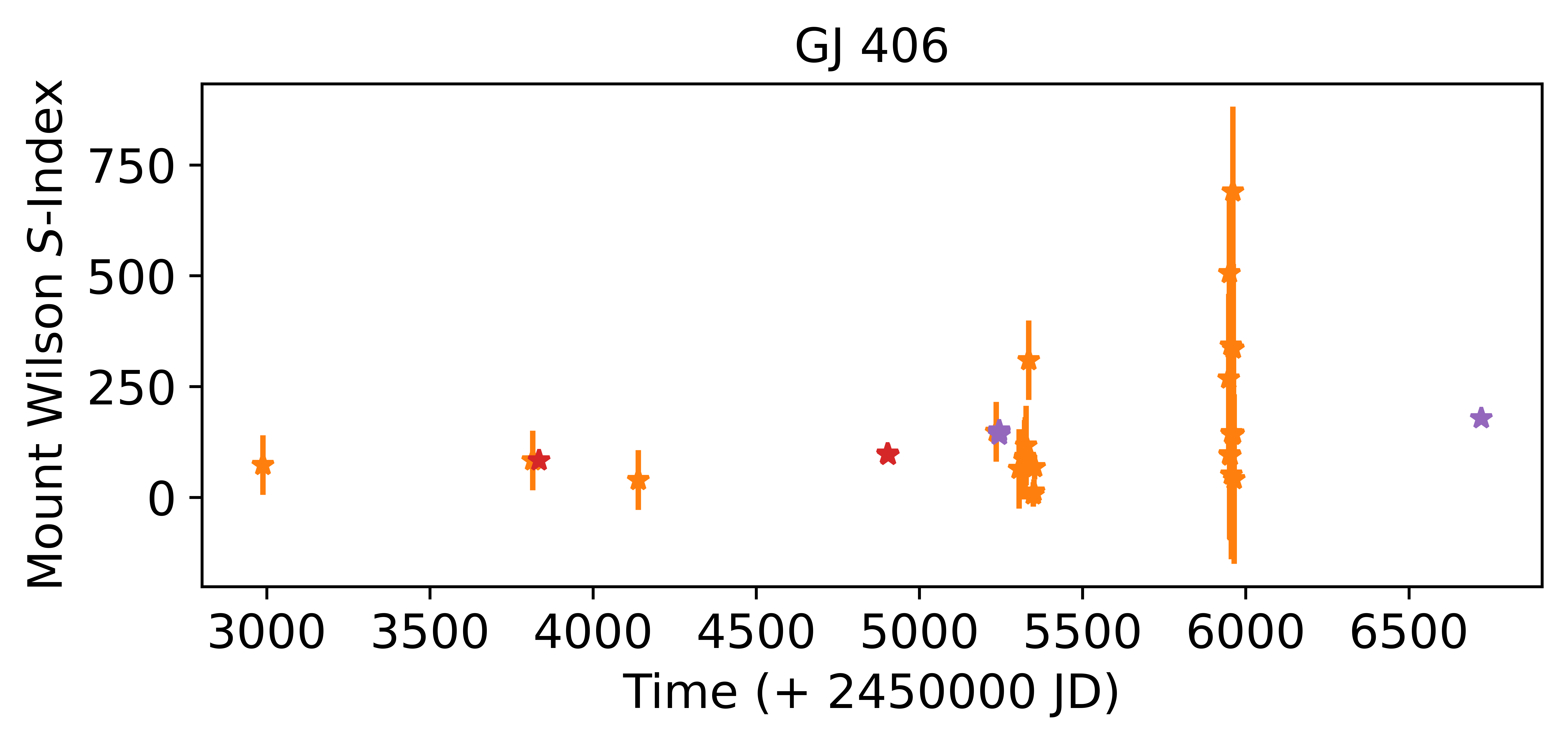}
    \includegraphics[width=0.45\textwidth]{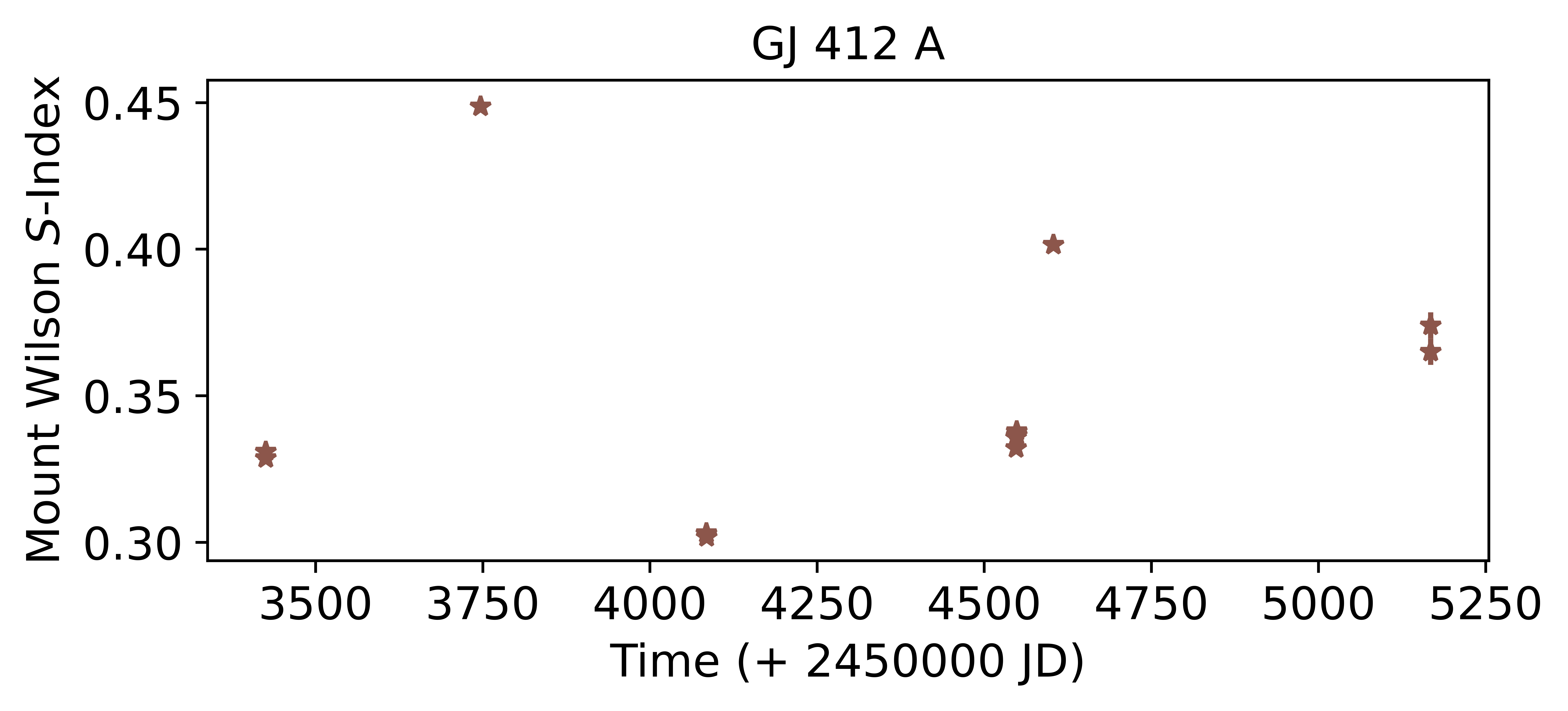}
    \includegraphics[width=0.45\textwidth]{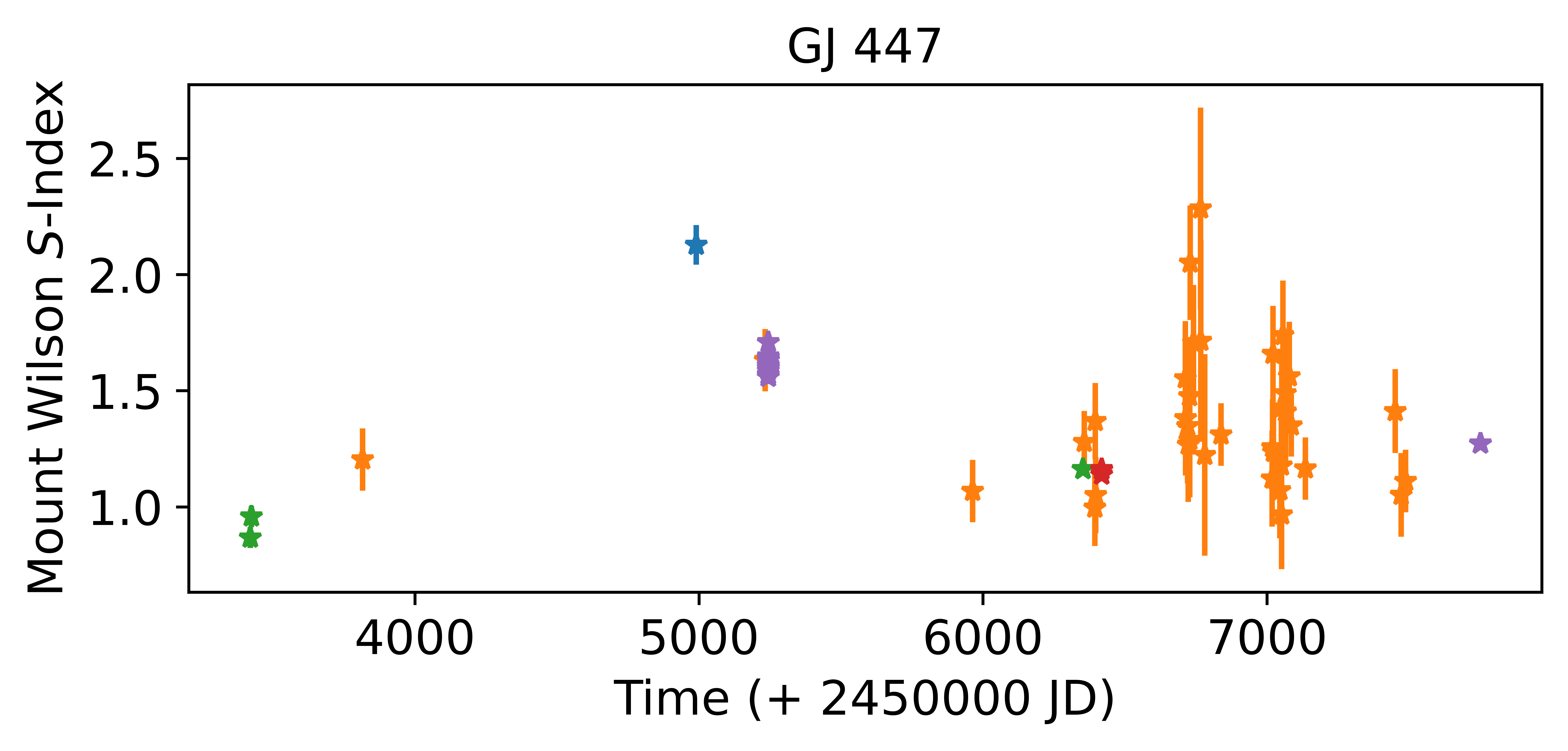}
    \includegraphics[width=0.45\textwidth]{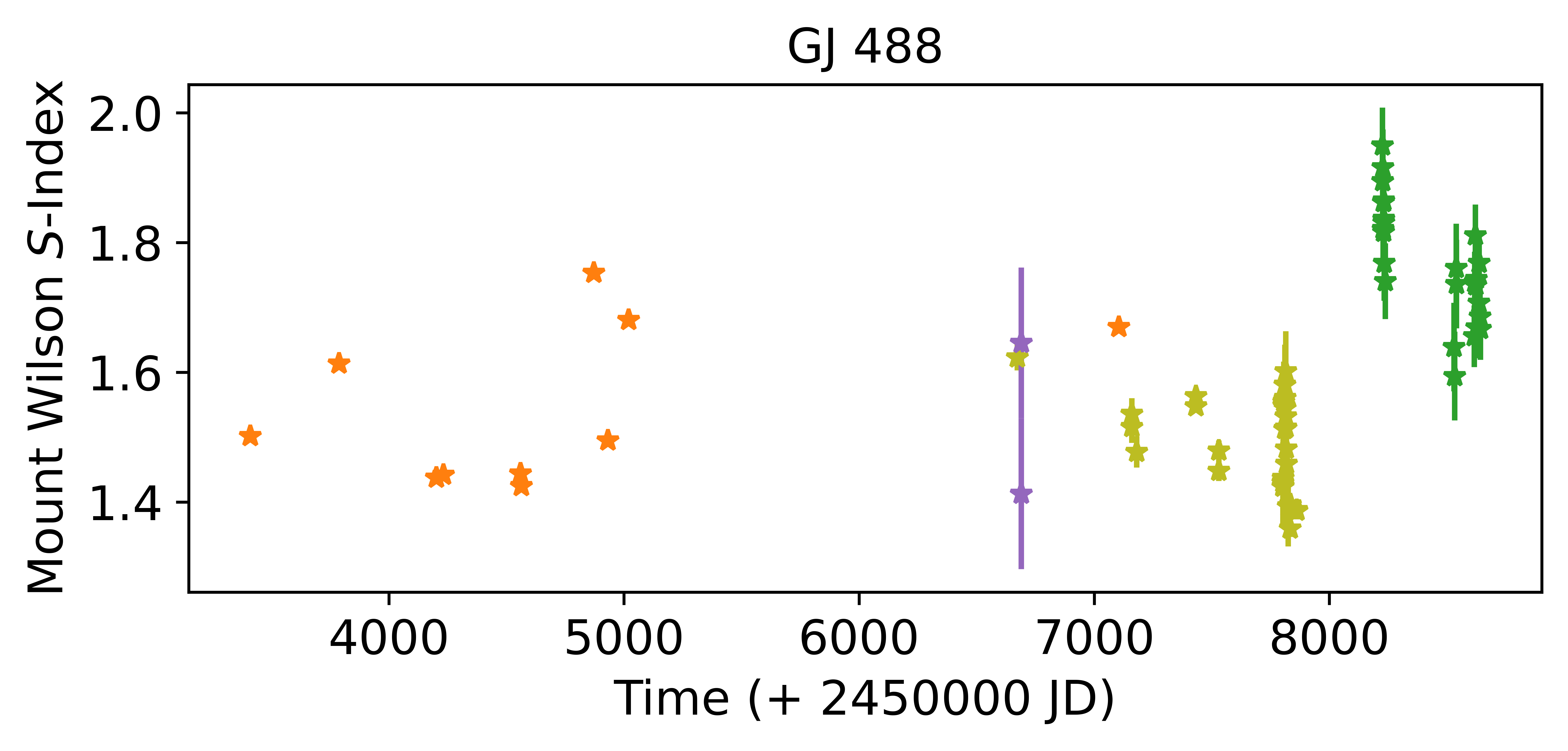}
    \includegraphics[width=0.45\textwidth]{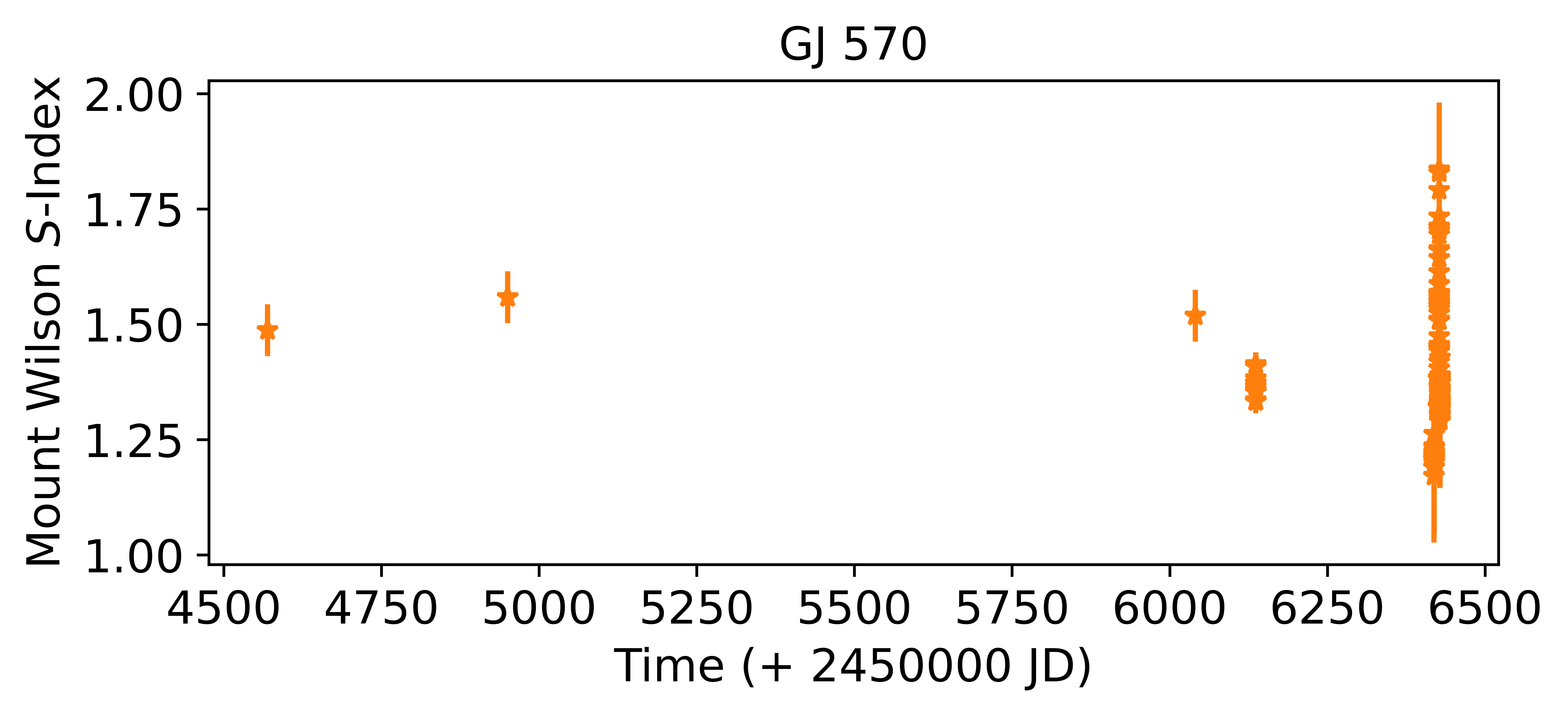}
\end{center}
\label{st_gl536}
\end{figure}

\begin{figure}[htb!]
\begin{center}
    \includegraphics[width=0.45\textwidth]{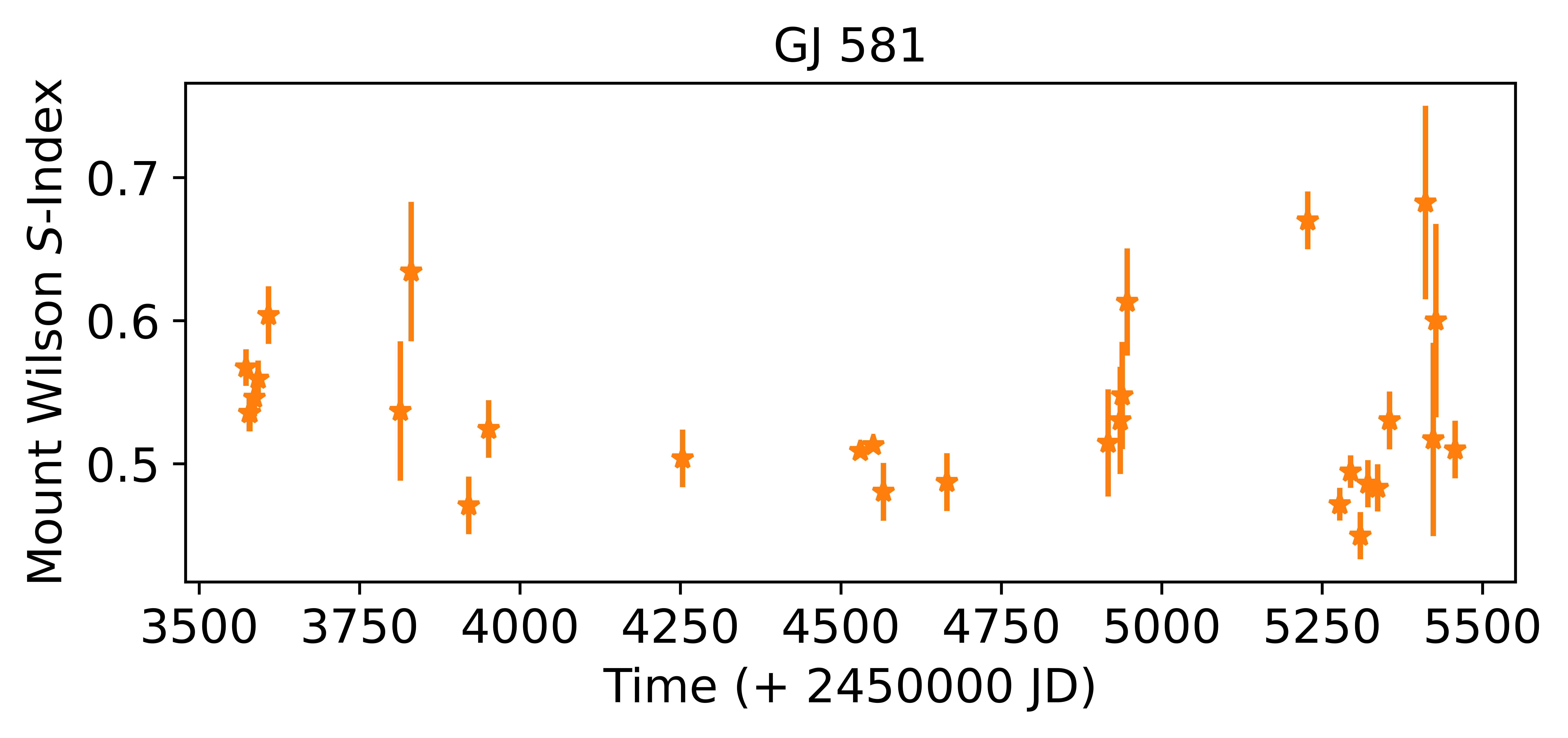}
    \includegraphics[width=0.45\textwidth]{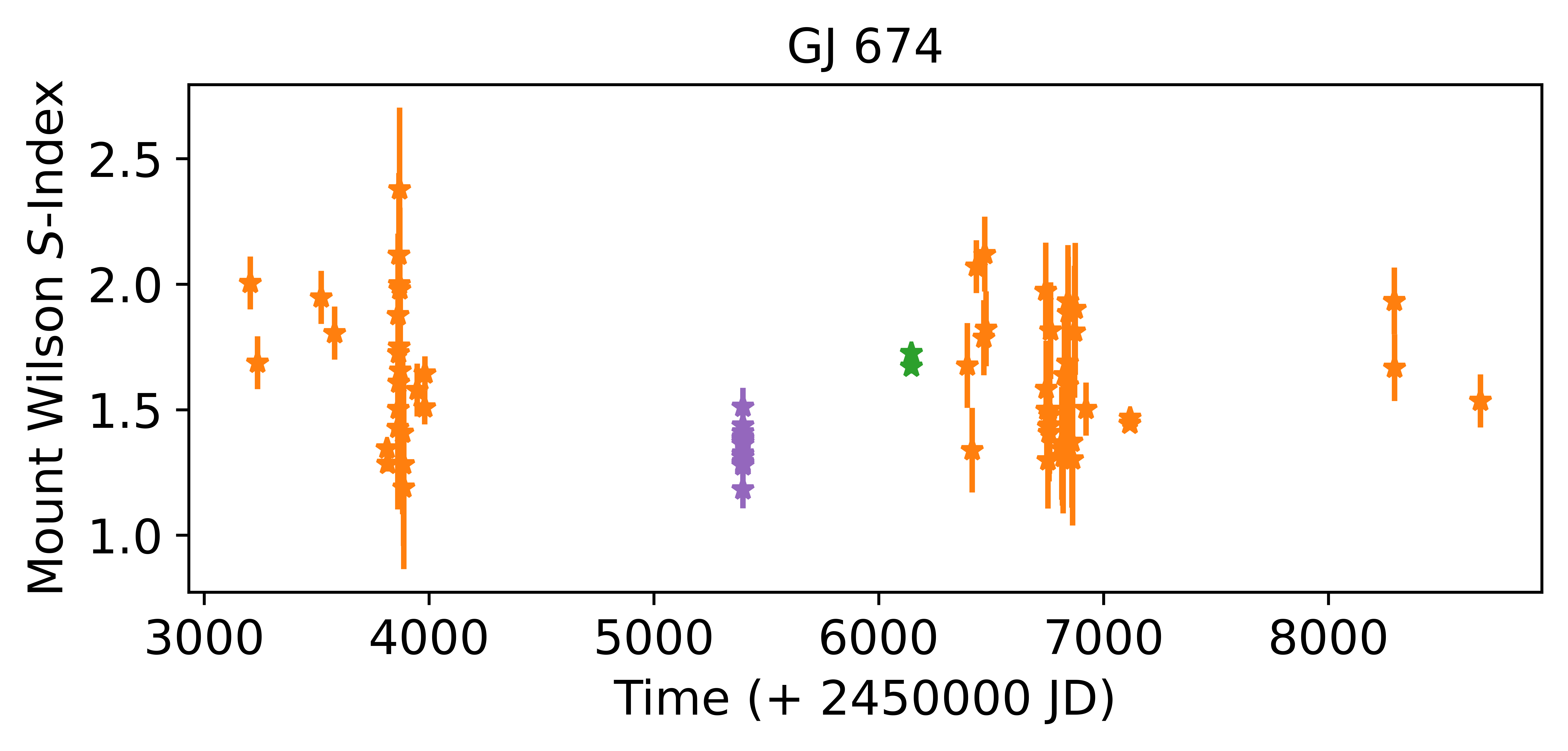}
    \includegraphics[width=0.45\textwidth]{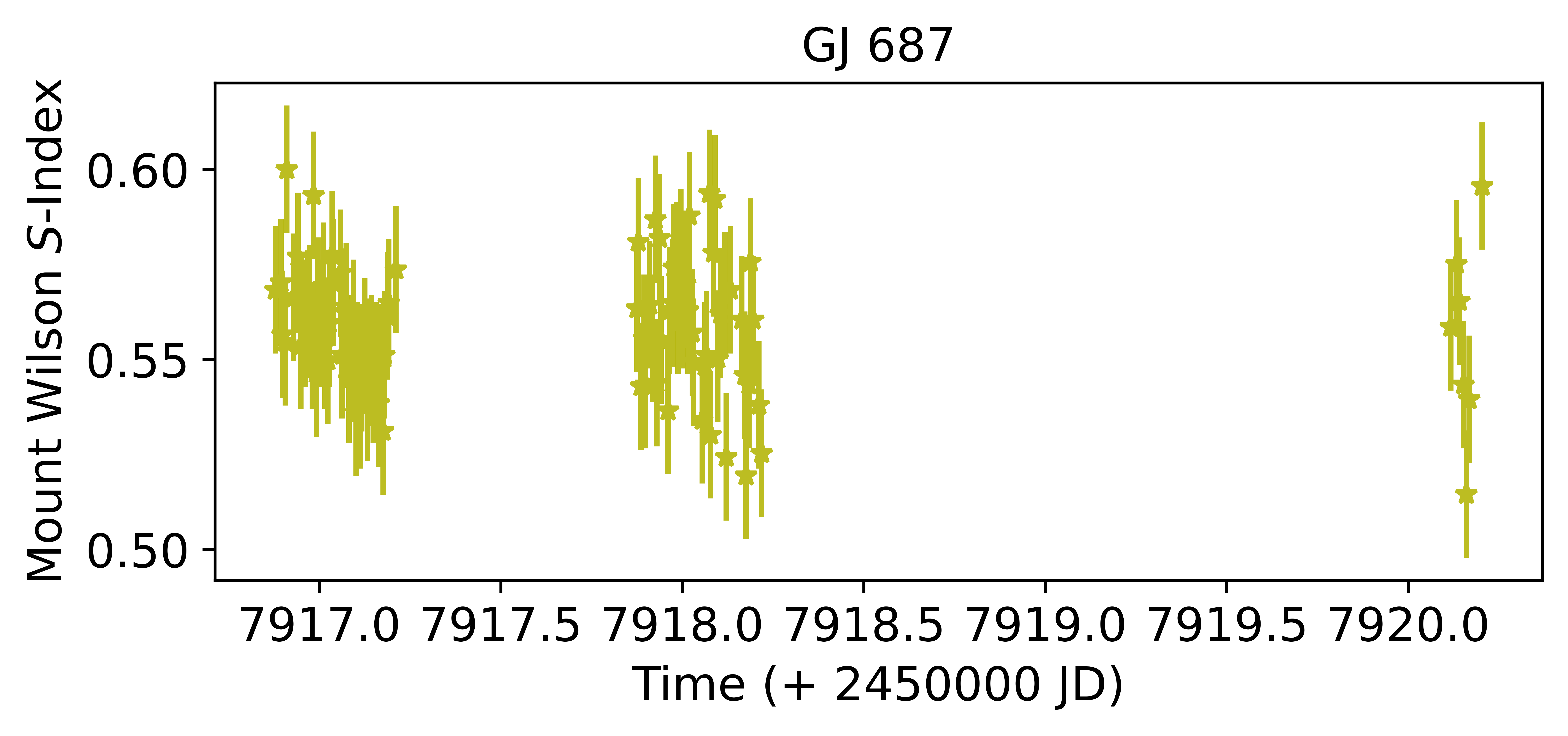}
    \includegraphics[width=0.45\textwidth]{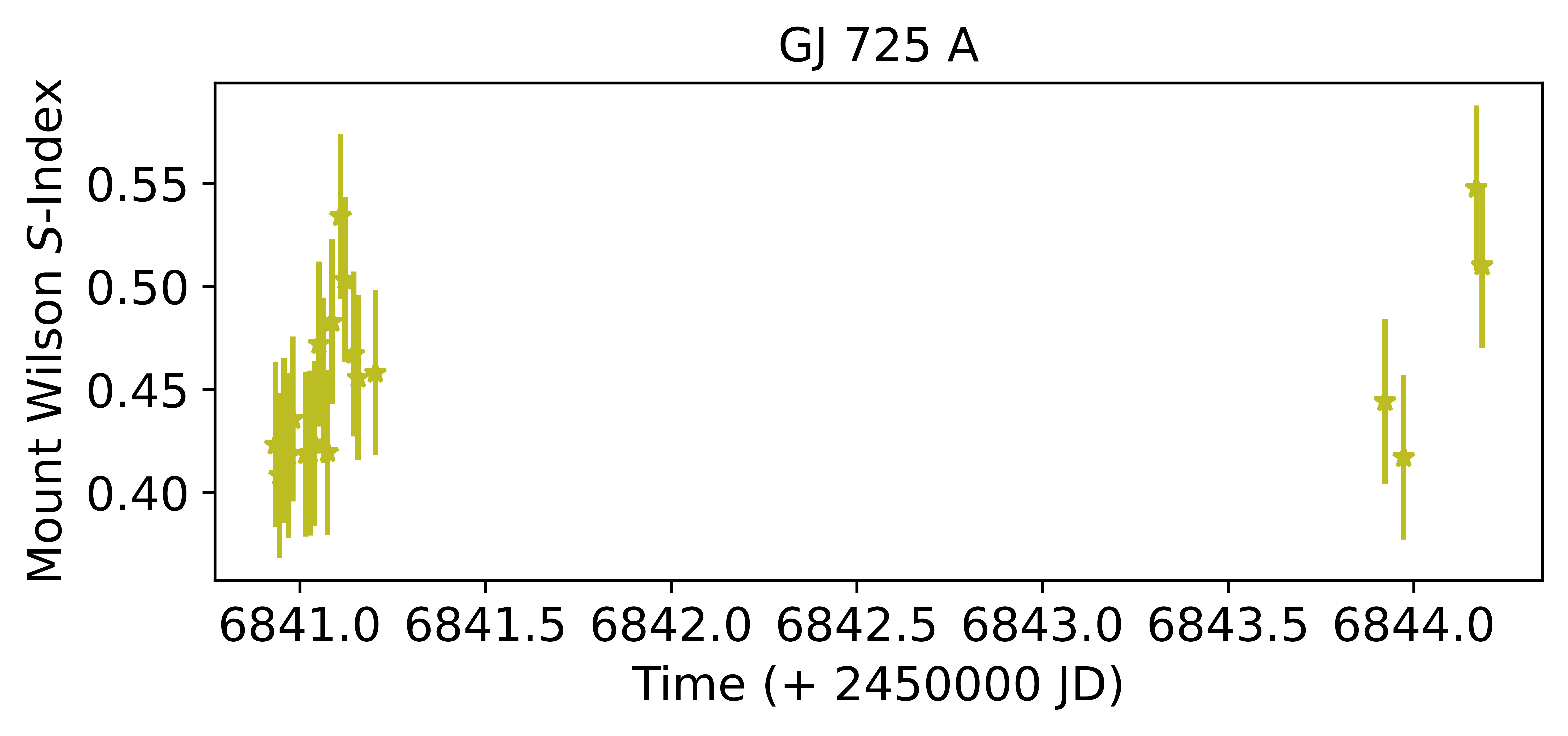}
    \includegraphics[width=0.45\textwidth]{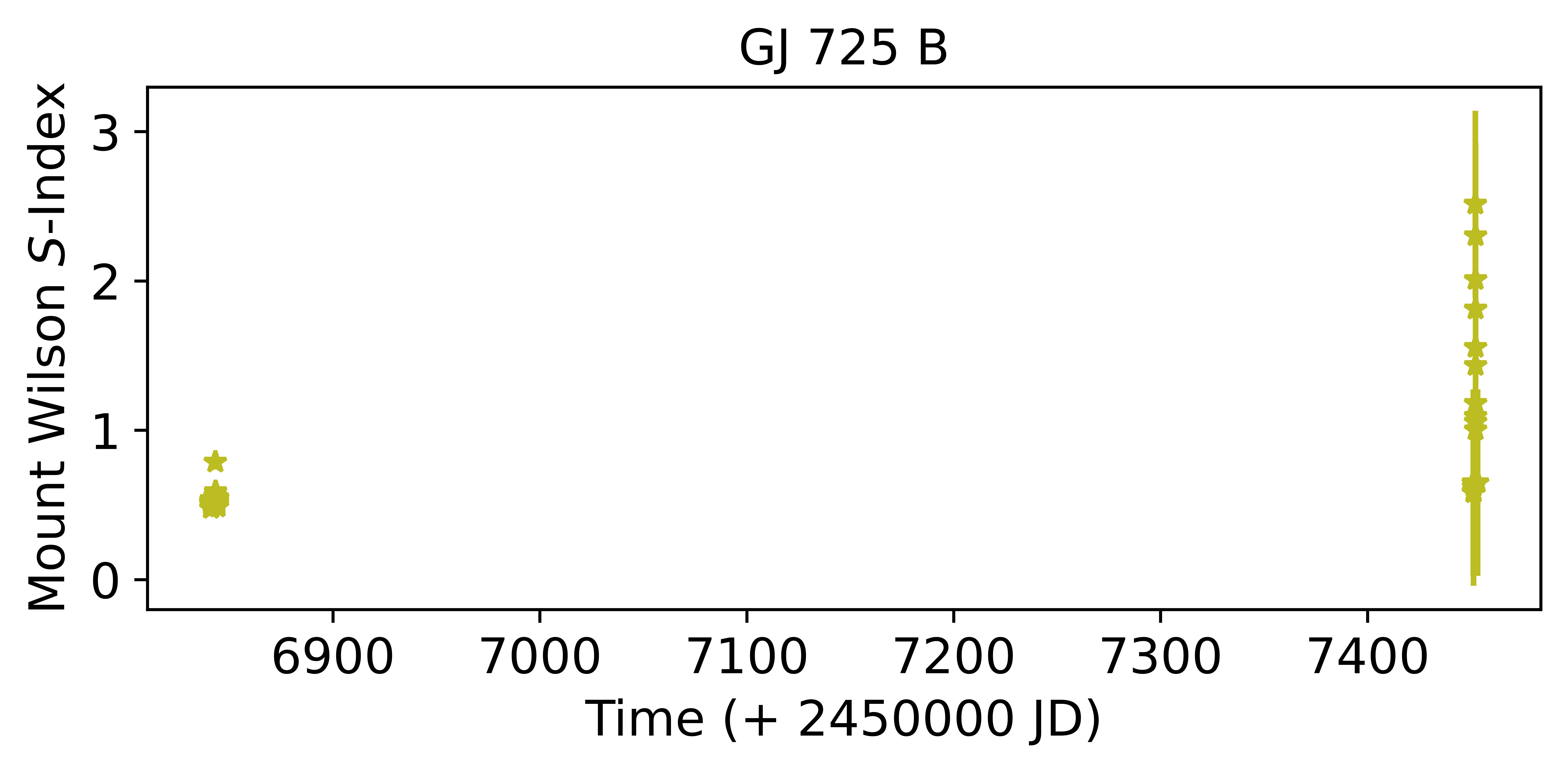}
\end{center}
\label{st_gl699}
\end{figure}

\begin{figure}[htb!]
\begin{center}
    \includegraphics[width=0.45\textwidth]{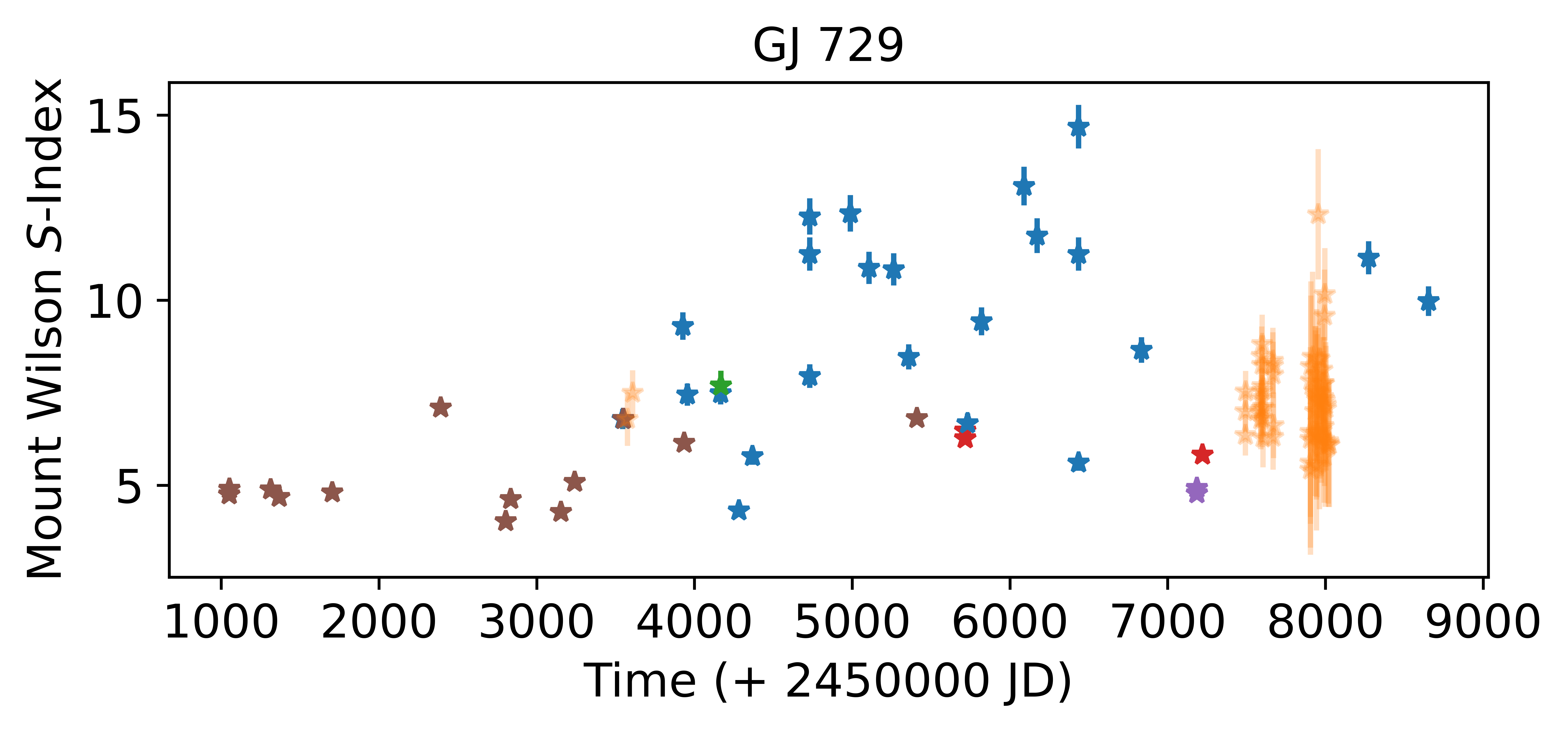}
    \includegraphics[width=0.45\textwidth]{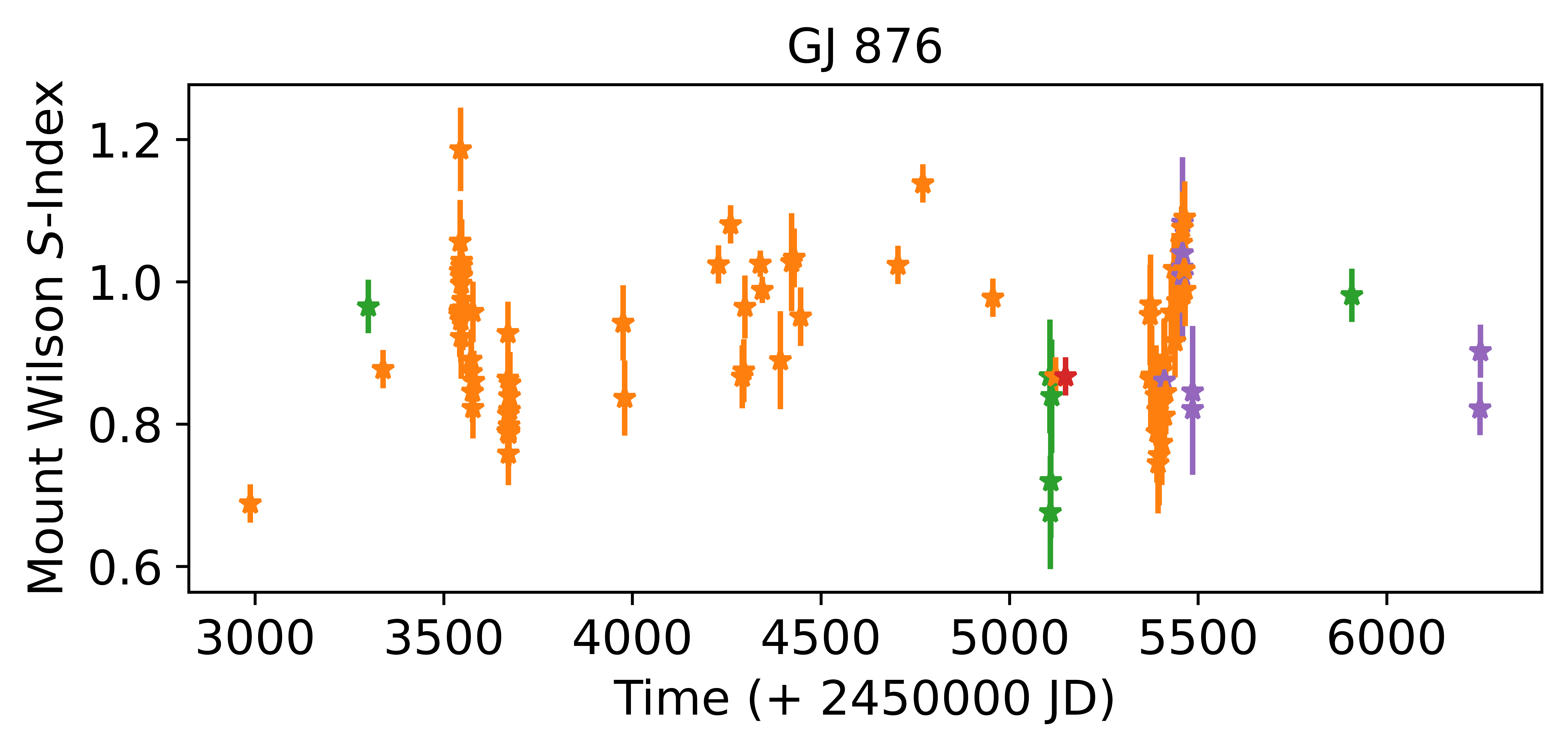}
    \includegraphics[width=0.45\textwidth]{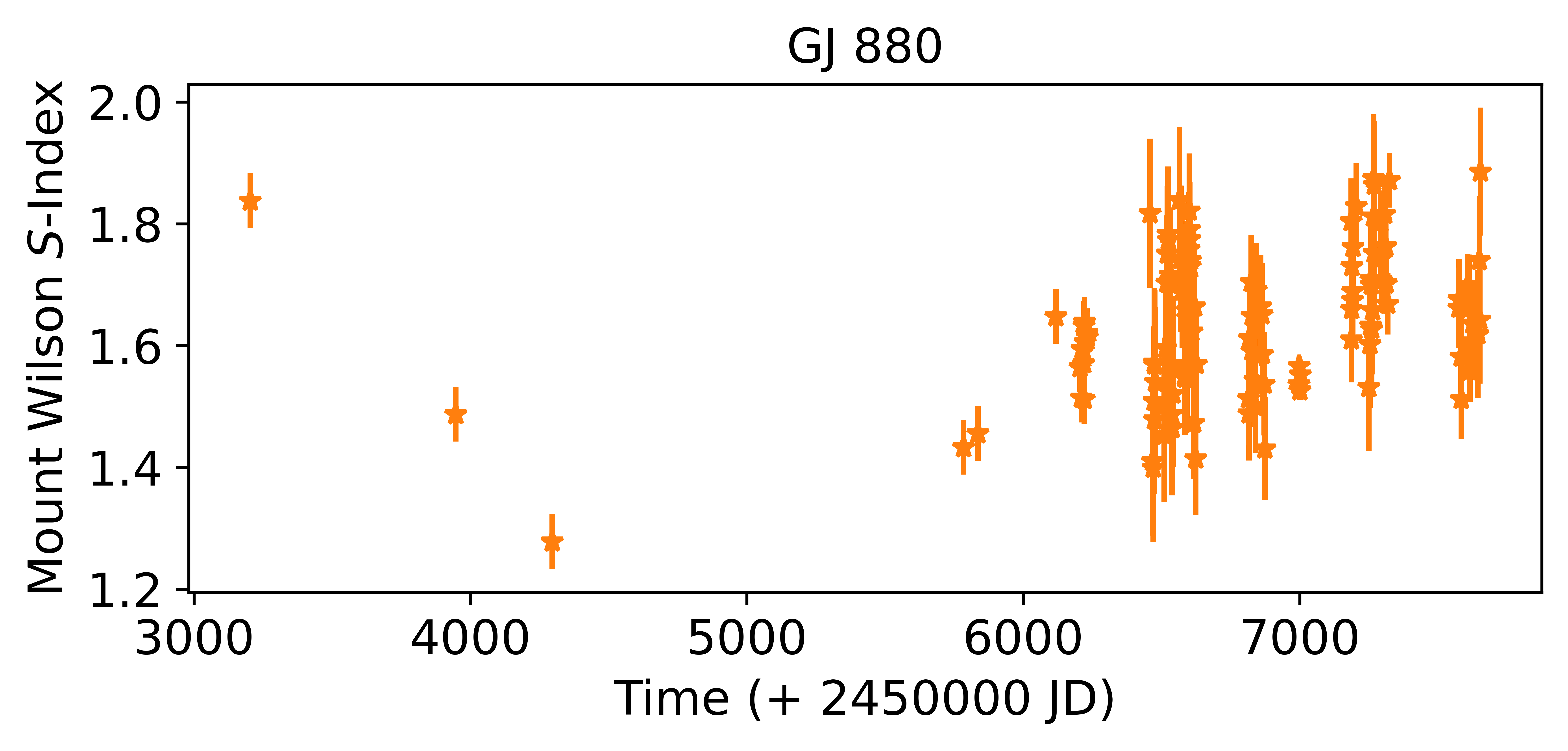}
    \includegraphics[width=0.45\textwidth]{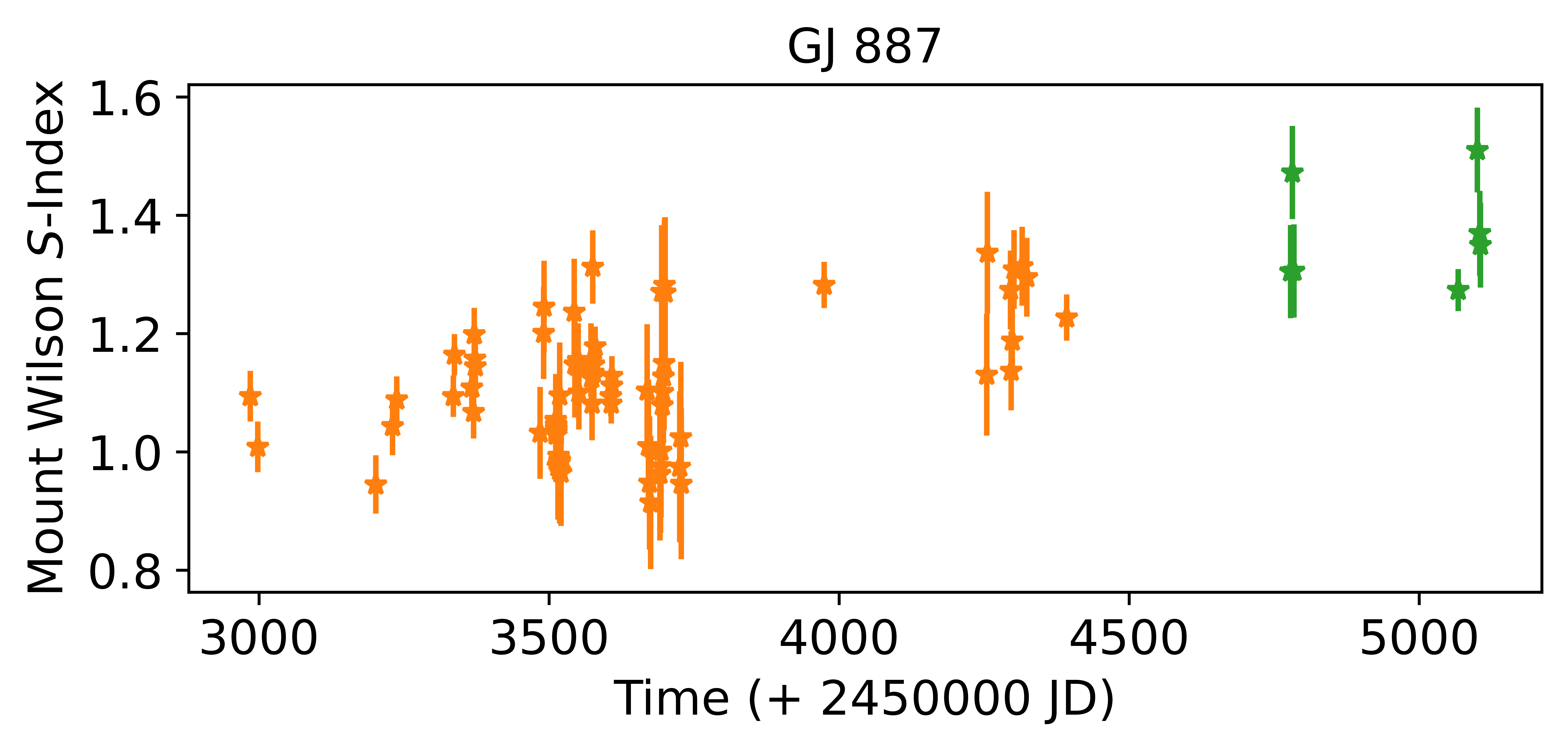}
    \includegraphics[width=0.45\textwidth]{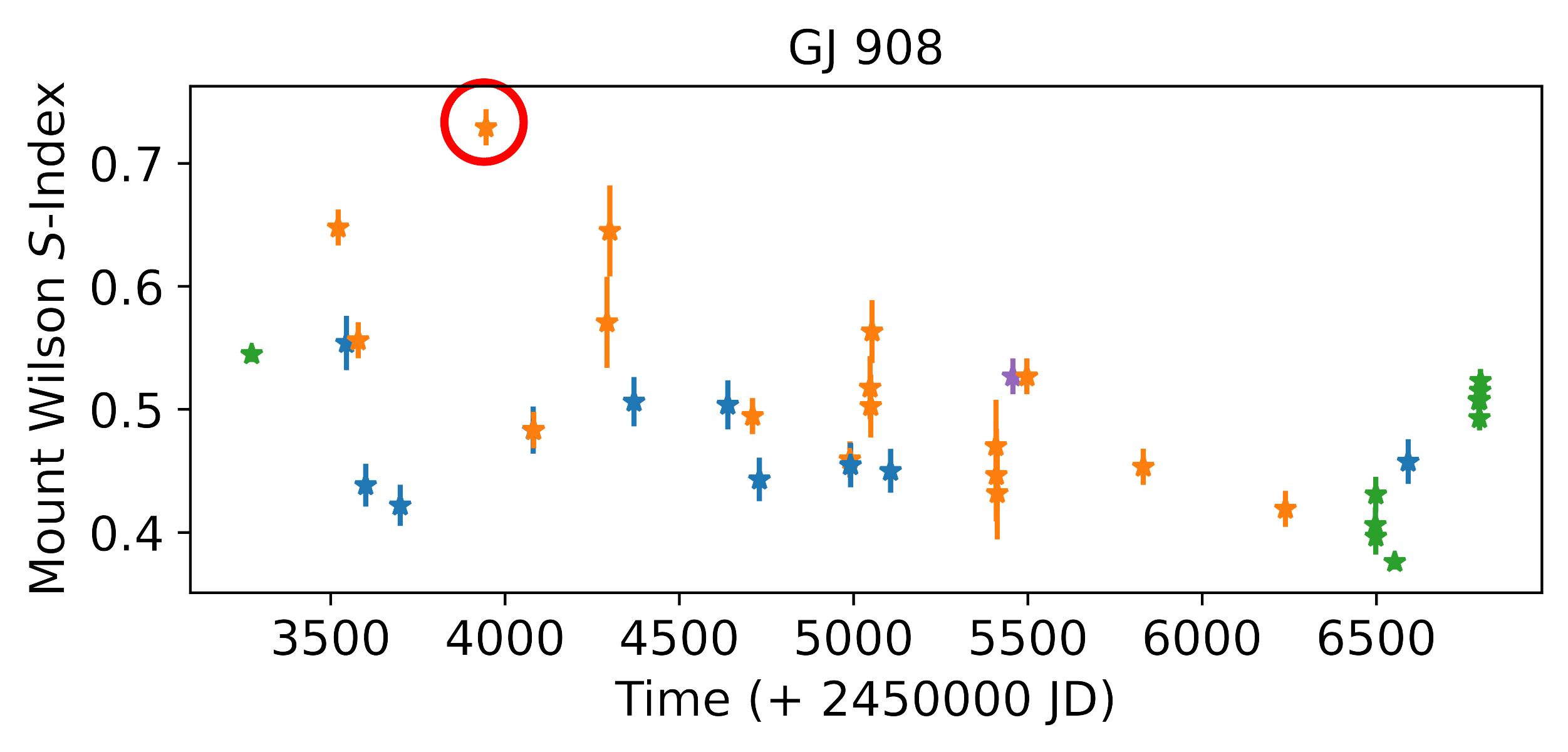}
\end{center}
\label{st_gl825}
\end{figure}

\end{appendix}

\end{document}